\documentclass{aa}
\usepackage[varg]{txfonts}

\usepackage[normalem]{ulem}

\usepackage[switch]{lineno}

\usepackage[usenames,dvipsnames]{color}
\usepackage[caption=false]{subfig}
\usepackage{graphicx,url,twoopt,natbib}
\usepackage{tikz} 
\usetikzlibrary{shapes,arrows} 

\usepackage[breaklinks=true]{hyperref} 
\makeatletter

\bibpunct{(}{)}{;}{a}{}{,}    
\makeatletter
  \protected\def\stonyslink{%
     \def\hyper@linkstart##1##2{}\let\hyper@linkend\@empty}
  \newcommandtwoopt{\citeads}[3][][]{%
   \href{http://ui.adsabs.harvard.edu/abs/#3/abstract}%
        {\stonyslink \citealp[#1][#2]{#3}}
   \biblink{#3}{\href{http://ui.adsabs.harvard.edu/abs/#3/abstract}{ADS}}}
 \newcommandtwoopt{\citepads}[3][][]{%
   \href{http://ui.adsabs.harvard.edu/abs/#3/abstract}%
        {\stonyslink \citep[#1][#2]{#3}}
   \biblink{#3}{\href{http://ui.adsabs.harvard.edu/abs/#3/abstract}{ADS}}}
 \newcommandtwoopt{\citetads}[3][][]{%
   \href{http://ui.adsabs.harvard.edu/abs/#3/abstract}%
        {\stonyslink \citet[#1][#2]{#3}}
  \biblink{#3}{\href{http://ui.adsabs.harvard.edu/abs/#3/abstract}{ADS}}}
 \newcommandtwoopt{\citeyearads}[3][][]{%
   \href{http://ui.adsabs.harvard.edu/abs/#3/abstract}%
        {\stonyslink \citeyear[#1][#2]{#3}}
   \biblink{#3}{\href{http://ui.adsabs.harvard.edu/abs/#3/abstract}{ADS}}}
\makeatother

\begin{document}

\newcommand{\annot}[1]{\textbf{\textcolor{OrangeRed}{{{#1}}}}}

\title{Covering factor in AGNs: evolution versus selection}


\author{Mateusz Ra{\l}owski\inst{1,}\inst{3} 
    \and Krzysztof Hryniewicz\inst{2}
  \and Agnieszka Pollo\inst{1,}\inst{2}
  \and {\L}ukasz Stawarz\inst{1}}

\offprints{M. Ra{\l}owski, \email{mralowski@oa.uj.edu.pl}}

\institute{Astronomical Observatory of the Jagiellonian University, Faculty of Physics, Astronomy and Applied Computer Science, ul. Orla 171, 30-244 Cracow, Poland; \email{mralowski@oa.uj.edu.pl}
  \and National Centre for Nuclear Research, ul. Pasteura 7, 02-093 Warsaw, Poland;
  \and Jagiellonian University, Doctoral School of Exact and Natural Sciences, Astronomy}

\date{A\&A Accepted 26.10.2023}

\abstract {} {In every proposed unification scheme for Active Galactic Nuclei (AGN), an integral element is the presence of circumnuclear dust arranged in torus-like structures, partially obscuring the nuclear (accretion-associated) radiation. A crucial model parameter in this context is the covering factor (CF), which can be defined as the ratio between the infrared luminosity of the dusty torus, $L_{\rm IR}$, and the accretion disk bolometric luminosity, $L_{\rm agn}$. Recent research has discussed the potential redshift evolution of the CF. Our study aims to determine whether this observed evolution is genuine or if selection effects significantly influence it.}
{Based on cross-matched multiwavelength photometrical data from the five major surveys (SDSS, GALEX, UKIDSS, WISE, SPITZER), a sample of over 17,000 quasars was derived. The main parameters of quasars, such as black hole masses and the Eddington ratios, were calculated based on the spectroscopic data. The data were divided into two redshift bins: Low-$z$ (redshift from 0.7 to 1.1) and High-$z$ (from 2.0 to 2.4) quasars. The associated smaller data sets with higher quality data were constructed from a) the WISE W3\&W4 detections with SNR\,$>5$, and b) the SPITZER MIPS 24\,$\mu$m photometry. The CF was determined by computing the ratio of integrated luminosities, $L_{\rm IR}$ and $L_{\rm agn}$, using two methods: 1) power-law fitting, and 2) the area between all photometric points. We explored different selection effects and their influence on CF estimates. Finally, statistical tests were employed to assess the hypothesis of CF evolution within the higher-quality datasets.}
{We identified an issue with the accuracy of the WISE W4 filter. Whenever feasible, it is recommended to utilize SPITZER MIPS 24\,$\mu$m data. Luminosities obtained through direct integration of all photometric data points exhibit higher accuracy compared to values derived from a power-law approximation. The Efron \& Petrosian test confirmed the presence of luminosity evolution with redshift for both $L_{\rm IR}$ and $L_{\rm agn}$. Both the Low-$z$ and High-$z$ samples exhibit a similar correlation between $L_{\rm agn}$ and $L_{\rm IR}$. The calculated median CF values are comparable within errors: $\log$\,CF$_{\textrm{low}-z} = -0.18 \pm 0.11$ and $\log$\,CF$_{\textrm{high}-z} = -0.01 \pm 0.13$. Additionally, the SPITZER photometry dataset reinforces this consistency with $\log$\,CF$_{\textrm{low}-z} = -0.19\pm 0.11$ and $\log$\,CF$_{\textrm{high}-z}= -0.18\pm 0.11$.}
{No discernible evolution of the CF was observed in the subsample of quasars with high SMBH mass bin or high luminosities, as the CF values for Low-$z$ and High-$z$ quasars have the same distributions. The relationship between $L_{\rm IR}$ and $L_{\rm agn}$ deviates slightly from the expected 1:1 scaling, suggesting a more intricate connection between CF and $L_{\rm agn}$. However, no statistically significant dependence of CF on luminosities could be claimed across the entire dataset (merged redshifts). It's worth noting that the Low-$z$/low-luminosity portion of the CF distribution is influenced by contamination, possibly due to polar dust, as suggested in the literature, while the High-$z$/high-luminosity segment is affected by observational biases.}

\maketitle

\section{Introduction}

According to the unification schemes for Active Galactic Nuclei (AGNs; \citealt{antonucci1985, urry1995,  2015ARA&A..53..365N, Hickox2018}), dusty tori are a crucial component in our understanding of the AGN properties. Initial models \citep{antonucci1985}, describe such tori as axisymmetric structures made of dust, and placed around central Super Massive Black Holes (SMBHs), with the equivalent hydrogen column densities large enough to completely obscure the central engine in some directions. The inner radius of a torus can be constrained by the dust sublimation temperature, beyond which dust grains evaporate. Given the UV emission of the innermost segments of the AGN accretion disks, this corresponds to scales ofn the order of parsecs and sub-parsecs (e.g., \citealt{suganuma2006, koshida2014, 2015ARA&A..53..365N}). Other unification models explore the connection between the circumnuclear dust in AGNs and the disk winds \citep{Proga_2000, Elvis2000, Murray2005, Ricci2017}. Recent studies highlight the potential for a more complex, notably highly inhomogeneous or clumpy, structure of a torus \citep{Nenkova2008}. Furthermore, these studies suggest a direct connection between the torus, the Broad Line Region (BLR), and the accretion disk itself \citep{2011A&A...525L...8C, Czerny2019MgII}. The study by \cite{2011A&A...525L...8C} specifically demonstrates that dust grains can survive in the accretion disk within a radial zone situated between the BLR and the inner surface of the torus. In this region, the effective temperature of the disk remains below the dust sublimation temperature.

A ``covering factor'' (CF) describes the fraction of obscuration of a SMBH by a dusty torus. The initial studies defined CF as the ratio of the solid angle $\Omega$ between the torus inner boundaries and a central SMBH, to $4\pi$, denoted as CF\,$=\Omega/4\pi$ \citep{1993ApJ...415..541H}. This original definition, while accurate and straightforward, poses challenges in practical implementation using photometric observations. Currently, a widely adopted approach for CF calculation involves the ratio between the nuclear infrared luminosity, ${L_{\rm IR}}$, and the bolometric AGN luminosity, ${L_{\rm agn}}$, expressed simply as CF\,$= L_{\rm IR}/L_{\rm agn}$ \citep{Maiolino2007, Treister2008ApJ...679..140T, 2013ApJ...773..176G, toba2021}. Three assumptions are necessary for this luminosity-based definition to correspond directly to the primary, angle-based definition of the CF: (1) $L_{\rm IR}$ is dominated by the emission of the circumnuclear hot dust and depends on the amount of radiation captured from the accretion disc and reprocessed within the torus; (2) $L_{\rm agn}$ accounts for the bulk of the luminosity of the accretion disk in the active nucleus; (3) $L_{\rm agn}$ and $L_{\rm IR}$ are both intrinsically isotropic. Hence, the CF defined through the luminosity ratio should be directly proportional to the original definition of the factor. It should be noted that the values calculated from the angle-based definition are always in the range between 0 and 1. This cannot be said for the luminosity-based definition. The relation between the luminosity-based and the angle-based definitions of the CF, was addressed by \cite{Treister2008ApJ...679..140T} among the others.

The CF can also be calculated through various alternative methods employing different observables. One such approach involves modeling the observed X-ray continua of AGNs \citep{Steffen2008, 2012MNRAS.423..702B}. Another option is to calculate the CF based as a population mean covering factor. In this approach, the CF value is derived as a ratio of type II and type I AGNs in a sample, expressed as CF\,$= N(\text{type-II})/N(\text{type-I})$ \citep{Hasinger2008, Toba2014}. The mean CF method is suitable for a sizable source sample but is not applicable on an individual object basis. For an in-depth comparison of different methods, we refer the reader to \cite{2015ARA&A..53..365N}.

The CF parameter has been widely discussed and analyzed in the literature \citep[e.g.,][]{Maiolino2007,  Treister2008ApJ...679..140T, Hasinger2008, Lawrence_2010, 2013ApJ...773..176G, toba2021}. Several studies have addressed a possible evolution of the CF with redshift \citep{LaFranca2005, Treister2008ApJ...679..140T, Hasinger2008,2013ApJ...773..176G, 2015ARA&A..53..365N, toba2021}. An anticorrelation between CF and $L_{\rm agn}$ was also reported in several works \citep{2013ApJ...773..176G, toba2021}. In particular, \cite{2013ApJ...773..176G}, gathered a sample of nearly 6,000 quasars from optical data in the Sloan Digital Sky Survey (SDSS) Data Releases 7 and 9 (hereafter DR7 and DR9, respectively), infrared data from the Wide-field Infrared Survey Explorer (WISE), ultraviolet data from Galaxy Evolution Explorer (GALEX), and near-infrared data from the UKIRT Infrared Deep Sky Survey (UKIDSS). They estimated $L_{\rm IR}$ and $L_{\rm agn}$ through spectral energy distribution (SED) integrals for targets in two redshift bins ($2.0\leq z \leq 2.4$ and $0.7\leq z \leq 1.1$). Their analysis revealed significant anticorrelations between CF and $L_{\rm agn}$ in both redshift bins, a finding also supported by \cite{toba2021}. However, explaining the evolution of CF with redshift within the basic AGN unification scheme remains challenging \cite{2015ARA&A..53..365N}. 
Indeed, other studies suggested no significant evolution of the CF with redshift \citep{ Gilli2007, Toba2014, Vito2018MNRAS.473.2378V}.

In this study, we investigate the CF in a general population of quasars with available broad-band coverage and explore the potential evolution of CF with redshift. 
The central question that we attempt to answer is: is the evolution of the CF real, or do the selection effects have a major influence on the CF estimates? In our study, we use the largest up-to-date photometrical data sample covering the IR--to--UV spectral range. Based on the work of \cite{Kozlowski}, we estimate the masses of the SMBHs, $M_{\rm BH}$, as well as the corresponding Eddington ratios, $\lambda_{\rm Edd}$, for the objects in the sample. We explore correlations between CF and $M_{\rm BH}$, or $\lambda_{\rm Edd}$. In this context, we thoroughly investigate multiple potential selection effects, signal-to-noise ratio (SNR) challenges in the infrared domain, as well as various approaches for calculating $L_{\rm IR}$ and $L_{\rm agn}$.

The paper is organized as follows. Section\,\ref{sec:Data} introduces the data and data reduction, Section\,\ref{sec:Luminosity} contains a discussion on luminosity estimates, and Section,\ref{sec:methods} reviews the methods used for the statistical analysis; Section\,\ref{sec:results} presents the main analysis results, further discussed in Section\,\ref{sec:Discussion}, and concluded in Section\,\ref{sec:Conclusions}. Cosmological parameters $H_{0}= 70$\,km\,s$^{-1}$\,Mpc$^{-1}$, $\Omega_{m}=0.3$, and $\Omega_{\Lambda}= 0.7$ are used throughout the article. The spectral index $\alpha$ is defined as $f_{\nu} \propto \nu^{-\alpha}$, where $f_{\nu}$ is the flux spectral density at a frequency $\nu$.

\section{Data}
\label{sec:Data}

The primary dataset for our analysis relies on photometric observations of quasars, specifically utilizing the SDSS Quasar Catalog, Sixteenth Data Release, was used (DR16Q\footnote{\url{https://data.sdss.org/datamodel/files/BOSS_QSO/DR16Q/DR16Q_v4.html}}; \citealt{2020ApJS..250....8L}). The DR16Q contains $\sim$\,750,000 quasars, including $\sim$\,225,000 newly identified sources in comparison to previous data releases. Quasars in the SDSS DR16Q were classified based on the spectral fitting for the SDSS-IV/the Extended Baryon Oscillation Spectroscopic Survey (eBOSS) spectra \citep{Dawson2016eBOSS}. The final data catalog is estimated to have a completeness rate of 99.8\% with contamination levels ranging from 0.3 to 1.3\%; for further details, see \cite{2020ApJS..250....8L}. The SDSS photometric data are given in asinh magnitudes, or in the so-called nanomaggies\footnote{\url{https://www.sdss.org/dr16/algorithms/magnitudes/\#Fluxunits:maggiesandnanomaggies}} \citep{1999AJ....118.1406L}. Measurements in nanomaggies can be converted to AB fluxes after correcting for the shift in the \textit{u} and \textit{z} filters\footnote{\url{https://www.sdss.org/dr12/algorithms/fluxcal/\#SDSStoAB}}. 

The DR16Q data were split into two distinct redshift ranges. There are 110,563 targets in the $0.7\leq z \leq 1.1$ redshift range (hereafter the ``Low-$z$'' sample), and 128,591 quasars in the $2.0\leq z \leq 2.4$ range (hereafter the ``High-$z$'' sample). By cross-matching with the WISE \citep{wright2010} All-Sky Survey, using a cross-matching radius of 2 arcsec, we identified 97,187 objects in the Low-$z$ sample and 60,056 in the High-$z$ sample. The main selection criteria for the cross-match were significant detections in all four WISE bands W1 -- W4. Further matching with the UKIDSS \citep{2007MNRAS.379.1599L} DR8, returned 16,841 objects in the Low-$z$ sample, and 7,999 objects in the High-$z$ sample. Finally, the Low-$z$ sample was cross-matched with the 3 arcsec matching radius with the GALEX \citep{martin2005} GR6/GR7, resulting in the Low-$z$ sample of 9,024 objects. The main condition for the source selection here was that all the fluxes, including upper limits, were positive. In total, our final sample contains a set of 9,024 Low-$z$ quasars (DR16Q $\times$ WISE $\times$ UKIDSS $\times$ GALEX), and 7,999 High-$z$ quasars (DR16Q $\times$ WISE $\times$ UKIDSS). 

\begin{figure}[!htp]

  \includegraphics[clip,width=1.\columnwidth]{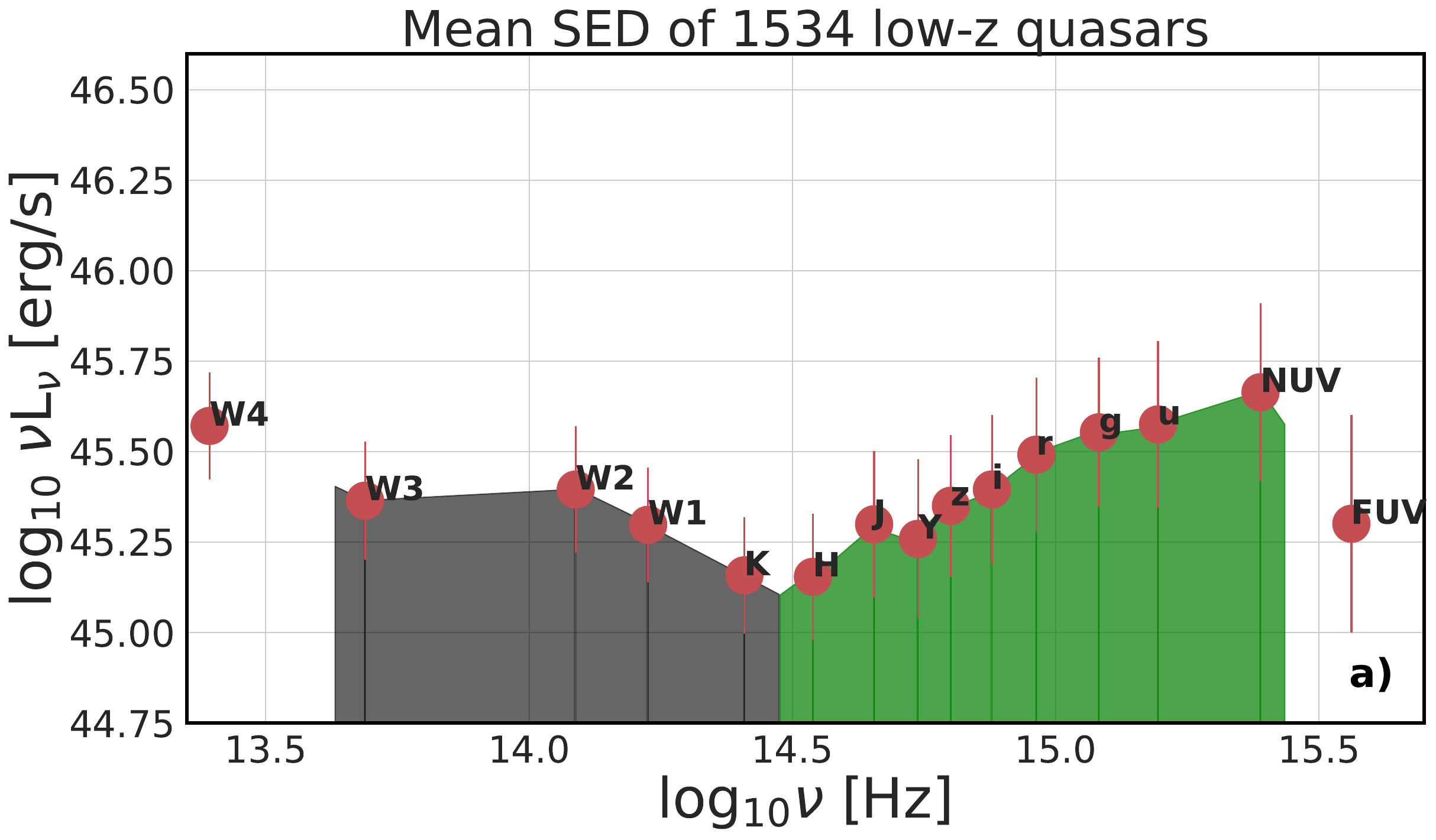}%
  \label{fig:meansed}

  \includegraphics[clip,width=1.\columnwidth]{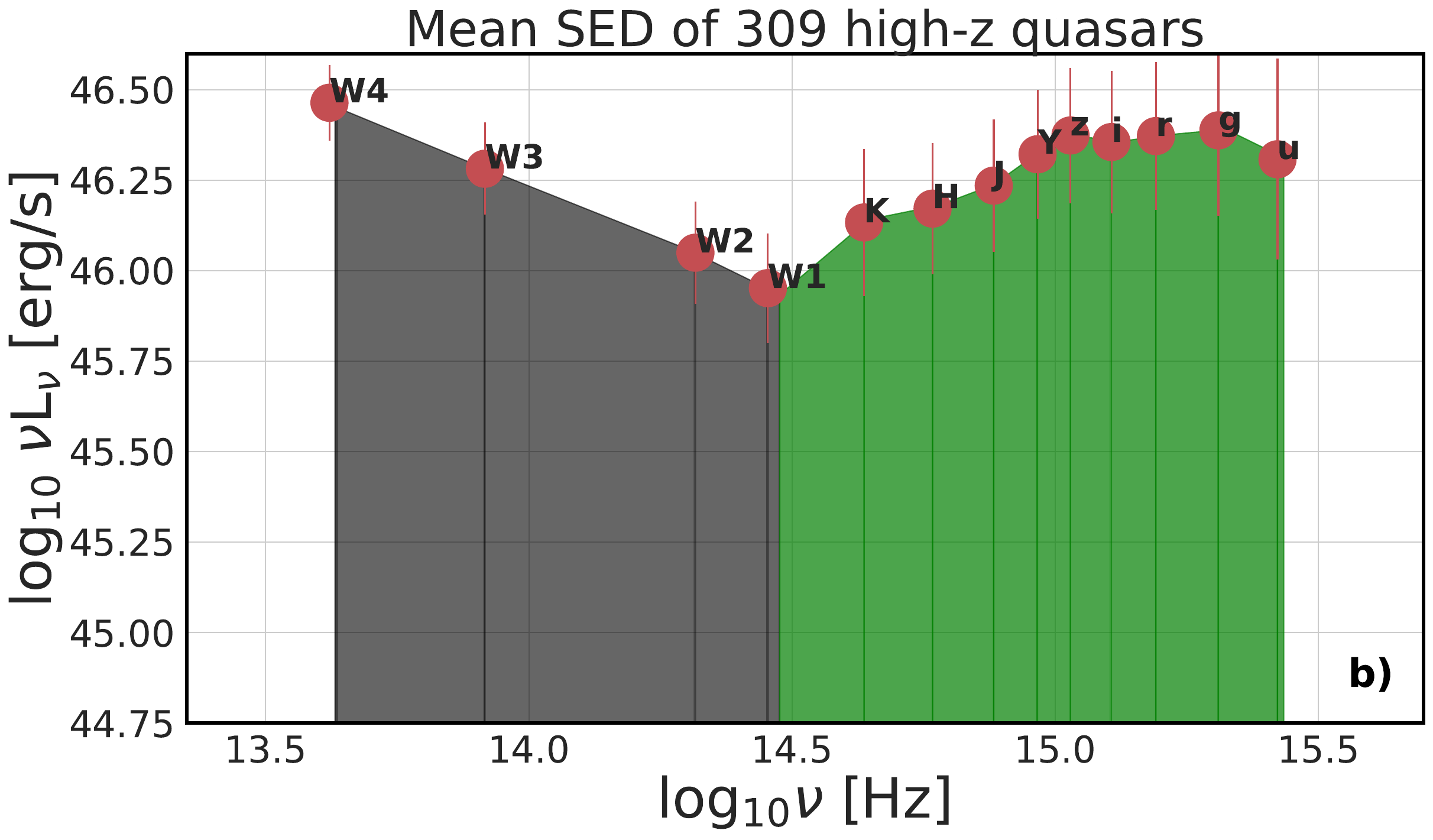}%

\caption{The top and bottom panels depict the mean rest-frame SEDs for Low-$z$ and High-$z$ quasars, respectively. Black and green shaded regions highlight the areas utilized for computing $L_{\rm IR}$ and $L_{\rm agn}$, respectively, in the all-points method.} Each photometric observation is labeled with the filter name. The errorbars denote the 1st and 3rd quartiles.
\label{fig:1}
\end{figure}

The WISE photometry was reported in previous work to have minor issues, such as redleak (see Section~\ref{sec:redleak} below). To assess the reliability of WISE W4 (hereafter W4) photometry, we used the SPITZER Multiband Imaging Photometer for SPITZER (MIPS) 24\,$\mu$m (hereafter M24) observations \citep{Spitzer_catalog2021}. The SPITZER M24 observations were available only for 290 Low-$z$ objects, and 171 High-$z$ sources. The entire data selection process is presented as a flow chart in Fig. \ref{fig:chart}.

All the photometric data underwent correction for Galactic extinction using the reddening maps of \cite{1998ApJ...500..525S}, and the extinction law of \cite{1989ApJ...345..245C}. These data were transformed into rest-frame luminosity spectral densities $L_{\nu} = 4 \pi d_{\rm L}^2 f_{\nu} / (1+z)$, where $f_{\nu}$ represents the extinction-corrected flux, and $d_{\rm L}$ signifies the luminosity distance calculated using the chosen cosmological model based on the code of \cite{Wright2006}.

The mean SEDs for the Low-$z$ and High-$z$ sources are presented in Figure~\ref{fig:1}. As shown, the slopes of the low- and high-frequency segments of the SEDs are different between the Low-$z$ and High-$z$ samples. Also, the High-$z$ quasars have higher monochromatic luminosities, especially in the IR range. Given the overall concave shape of the SEDs, hereafter we define the torus-related IR luminosity $L_{\rm IR}$ as the one corresponding to the integrated continuum emission below the rest-frame 1\,$\mu$m, and the accretion-related bolometric (optical-to-UV) luminosity $L_{\rm agn}$ corresponding to frequencies above 1\,$\mu$m (see Figure~\ref{fig:1}).


\tikzstyle{decision} = [rectangle, draw, fill=green!20, text width=4.5em, text badly centered, node distance=2cm, inner sep=5pt]
\tikzstyle{block} = [rectangle, draw, fill=teal!40, node distance=1.8cm , text width=5em, text centered, rounded corners, minimum height=4em]
\tikzstyle{block1} = [rectangle, draw, fill=magenta!20, node distance=1.8cm , text width=4.8em, text centered, rounded corners, minimum height=4em]
\tikzstyle{line} = [draw, -latex']

\tikzstyle{cloud_low_high} = [draw, rectangle, fill=gray!30, node distance=2.1cm, text width=4.5em, text centered, rounded corners, minimum height=4.em]

\tikzstyle{cloud_low_spitzer} = [draw, rectangle, fill=orange!50, node distance=2.1cm, text width=5em, text centered, rounded corners, minimum height=4.em]
\tikzstyle{cloud_high_spitzer} = [draw, rectangle, fill=red!50, node distance=2.1cm, text width=5em, text centered, rounded corners, minimum height=4.em]

\tikzstyle{cloud_low} = [draw, rectangle, fill=cyan!40, node distance=2.1cm, text width=5.8em, text centered, rounded corners, minimum height=5em]

\tikzstyle{cloud_high} = [draw, rectangle, fill=green!40, node distance=2.1cm, text width=5.8em, text centered, rounded corners, minimum height=5em]

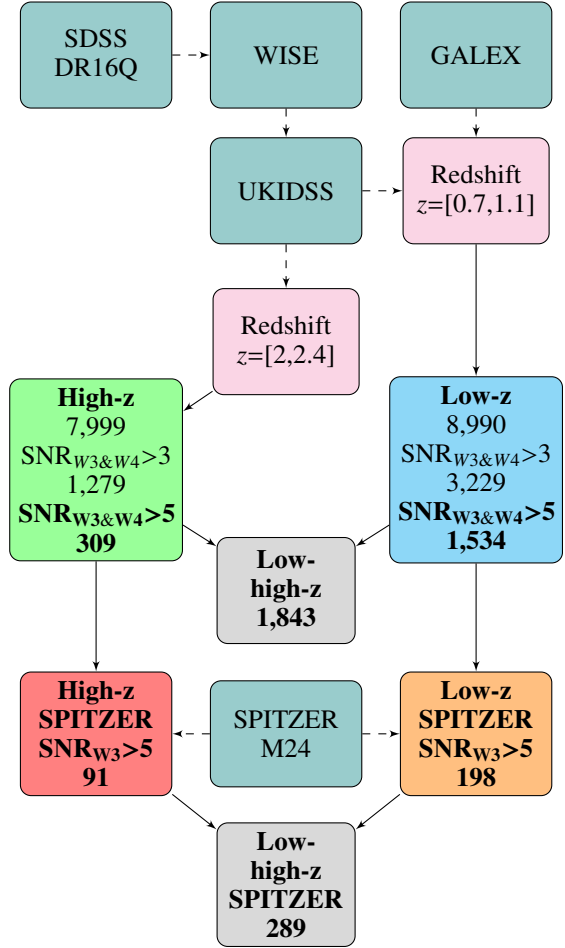
\begin{figure}[!htp]
\begin{tikzpicture}[node distance = 1cm, auto]
\node [block, node distance=2cm] (WISE) {WISE};
\node [block, below of=WISE] (UKIDSS) {UKIDSS};
\node [block, left of=WISE, node distance=2.5cm] (SDSS) {SDSS DR16Q};
\node [block, right of=WISE, node distance=2.5cm] (GALEX) {GALEX\\};
\node [block1, below of=UKIDSS, node distance=2.cm] (Redshift) {Redshift $z$=[2,2.4]};
\node [block1, right of=UKIDSS, node distance=2.5cm] (Redshift1) {Redshift $z$=[0.7,1.1]};

\node [cloud_high, below of=SDSS, node distance=5.5cm] (High) 
{\textbf{High-$\mathbf{z}$}\\ 7,999  \\ 
SNR$_{W3\&W4}$>3 \\
1,279
\textbf{SNR$_{\mathbf{W3\&W4}}$>5}\\  \textbf{309}};
\node [cloud_low, below of=GALEX, node distance=5.5cm] (Low) {\textbf{Low-$\mathbf{z}$}\\ 8,990 \\ 
SNR$_{W3\&W4}$>3\\
3,229
\textbf{SNR$_{\mathbf{W3\&W4}}$>5}\\  \textbf{1,534}};

\node [cloud_low_high, below of=Redshift, node distance=3.3cm] (Low-High) {\textbf{Low-high-$\mathbf{z}$\\ 1,843}};

\node [block, below of=Redshift, node distance=5.2cm] (SPITZER) {SPITZER \\M24};

\node [cloud_high_spitzer, below of=High, node distance=3.5cm] (High1) {\textbf{High-$\mathbf{z}$ SPITZER\\ SNR$_{\mathbf{W3}}$>5\\ 91}};
\node [cloud_low_spitzer, below of=Low, node distance=3.5cm] (Low1) {\textbf{Low-$\mathbf{z}$ SPITZER\\ SNR$_{\mathbf{W3}}$>5\\ 198}};

\node [cloud_low_high, below of=SPITZER, node distance=2.cm] (Low-High_Spitzer) {\textbf{Low-high-$\mathbf{z}$ SPITZER\\ 289}};

\path [line, dashed] (SDSS) -- (WISE);
\path [line, dashed] (WISE) -- (UKIDSS);
\path [line, dashed] (UKIDSS) -- (Redshift);
\path [line, dashed] (UKIDSS) -- (Redshift1);
\path [line, dashed] (GALEX) -- (Redshift1);
\path [line] (Redshift1) -- (Low);
\path [line] (Redshift) -- (High);
\path [line] (Low) -- (Low1);
\path [line] (High) -- (High1);
\path [line] (Low) -- (Low-High);
\path [line] (High) -- (Low-High);
\path [line, dashed] (SPITZER) -- (Low1);
\path [line, dashed] (SPITZER) -- (High1);
\path [line] (Low1) -- (Low-High_Spitzer);
\path [line] (High1) -- (Low-High_Spitzer);

\end{tikzpicture}
\caption{The flowchart showing the data selection process, as described in Section \ref{sec:Data}. The final samples are: 1,534 objects for Low-$z$ with SNR$_{W3\&W4}$>5, 309 objects for High-$z$ SNR$_{W3\&W4}$>5. The Low-$z$ and High-$z$ data-set combined create the Low-high-$z$. High precision datasets with SPITZER observations have: 198 objects for Low-$z$ SPITZER and 91 objects for High-$z$ SPITZER, both with SNR$_{\mathbf{W3}}$>5. The datasets with different SNR$_{W3\&W4}$ cuts are analyzed in Appendix \ref{sec:Relations}.}
\label{fig:chart}

\end{figure}

\subsection{WISE Data Check}
\subsubsection{Readleak}
\label{sec:redleak}

Prior research has highlighted minor issues with the WISE data, such as redleak, which have been appropriately addressed in our analysis. In particular, following the methodology outlined in \cite{redleak}, the effective wavelength of W4 was changed from 22.1 to 22.8\,$\mu$m, and the AB magnitude of the Vega zero point of W4 was changed from $m_{\rm W4} = 6.59$ to $m_{\rm W4}=6.66$. As follows, this correction is relatively minor.

\subsubsection{Signal-to-Noise Ratio}

For the faint objects in the sample, the SNR is notably low and the measured W3 and W4 fluxes have relatively large errors, making some of the observations unreliable. To address this issue, we imposed a more strict requirement for SNR to exceed $5\sigma$ for the W3 and W4 filters. It's worth noting that the equivalent restriction on W1 and W2 does not alter the amount of data after the condition SNR$_{\rm W3\&W4} >5$ is set, as the W3 and W4 data present more challenges in this context. In the end, there are 1,534 high-SNR objects in the Low-$z$ sample and 309 High-$z$ sources. Low-$z$ and High-$z$ combined together are called Low-high-$z$. This dataset has 1,843 objects. This associated subset of data was subsequently examined for possible alterations in the source parameter distributions, and a possible evolution of the CF (see Appendices\,\ref{sec:results_spitzer_check} and \ref{sec:Relations} for further details). A similar restriction is applied to the associated SPITZER dataset, utilizing only W3 from the two problematic filters. When SNR$_{\rm W3}>5$ is applied, the SPITZER Low-$z$ sample contains 198 objects, while SPITZER High-$z$ includes 91 objects. These two sets, when combined, are collectively referred to as the ``Low-High-$z$ SPITZER'' dataset, consisting of 289 objects. Going forward, unless explicitly stated otherwise, the presented data adhere SNR$_{\rm W3\&W4} >5$.

\subsection{Value Added Catalogs}

To estimate the physical parameters of the analyzed sources, such as the SMBH mass $M_{\rm BH}$, or Eddington ratios $\lambda_{\rm Edd}$, high-quality spectroscopic data are essential. For this we utilized the value-added catalog (VAC) for the SDSS by \cite{Kozlowski}. This catalog is based on spectral line fitting using the SDSS DR12Q \citep{Paris2017} data. In particular, \cite{Kozlowski} provided virial estimates for $M_{\rm BH}$, using the BLR radius--disk luminosity relations, and the FWHM of the Mg\,II and C\,IV emission lines. This approach relies on the phenomenological scaling relation $R_{\rm BLR} \propto L_{\rm disk}^{1/2}$ connecting the characteristic spatial scale of the BLR, $R_{\rm BLR}$, with the continuum disc luminosity, $L_{\rm disk}$. This relation is established through reverberation mapping \citep{McLure2002, Kaspi2007, Bentz2013}. Utilizing this estimate, the virial theorem can be used with the relation $M_{\rm BH}\propto R_{\rm BLR} \, v^2 $, where $v$ is the characteristic velocity of the BLR clouds inferred from the FWHM of the broad emission lines \citep{Vestergaard2006, Shen2013}. We note that single-epoch viral $M_{\rm BH}$ estimates come with methodological biases, which are discussed in \cite{Shen2016} and \cite{Kozlowski}. 

Based on the obtained values of $M_{\rm BH}$, the Eddington luminosities were calculated as
\begin{equation}
L_{\rm Edd} =  \frac{4 \pi G M m_p c}{\sigma_T} \simeq 1.26 \times 10^{38} \,\left(\frac{M_{\rm BH}}{M_{\odot}}\right) \, \rm erg\,s^{-1} \, ,
   \label{eq:2}
\end{equation}
and the corresponding logarithmic Eddington ratios as
\begin{equation}
   \lambda_{\rm Edd} = \log \left(\frac{L_{\rm agn}}{L_{\rm Edd}}\right) \, .
    \label{eq:3}
\end{equation}
\newline

We opted to utilize exclusively Mg II based estimations of $M_{\rm BH}$ from \cite{Kozlowski}. The cross-match process with our SNR$_{\rm W3\&W4}>5$ WISE sample returned 229 Low-$z$ and 125 High-$z$ sources, and with the SPITZER SNR$_{\rm W3}>5$ sample 41 and 33 targets, respectively. Subsequently, we derived the Eddington luminosities and Eddington ratios from these data.

\section{Luminosity Estimates}
\label{sec:Luminosity}

With the comprehensive multiwavelength data we gathered, we precisely estimated the infrared and bolometric luminosities of quasars using the `all-points method'. This method offers an alternative to the commonly employed `power-law method' in the literature  \citep[e.g.,][]{2013ApJ...773..176G}. Specifically, we utilized the trapezoid method to integrate the quasar SEDs. This involved fitting a line in log-log space between consecutive photometry data points and then integrating the area beneath it, as opposed to using a fixed wider frequency range. Furthermore, we incorporated interpolated luminosity densities at wavelengths of 0.11\,$\mu$m ($L_{0.11\,\mu \textrm{m}}$), 1\,$\mu$m ($L_{1\,\mu  \textrm{m}}$), and 7\,$\mu$m ($L_{7\,\mu \textrm{m}}$). This allowed us to cover similar ranges of the rest-frame continua in both the Low-$z$ and High-$z$ samples. For $L_{7\,\mu \textrm{m}}$, we interpolated values from the W4 and W3 WISE filter detections in both Low-$z$ and High-$z$ samples. $L_{1\,\mu \textrm{m}}$ was calculated based on interpolation between the W1 and K filters for Low-$z$ or W2 and W1 for High-$z$. $L_{0.01\,\mu \textrm{m}}$ was calculated using interpolation between NUV and FUV for Low-$z$ or g and u for High-$z$.

We defined $L_{\rm IR}$ as the integrated area between $L_{7\mu m}$ and $L_{1\mu m}$, while $L_{\rm agn}$ was defined as the integrated area between $L_{1\mu m}$ and $L_{0.11\mu m}$. The rationale for this definition is that in the AGN unification scheme, the dusty torus is illuminated by the AGN disk radiation and captures a fraction of the total energy proportional to the CF. The absorbed optical, UV, and soft X-ray radiation is then reemitted as a thermal IR radiation. In Appendix \ref{sec:integration_comparison} we discuss in more detail the applied luminosity integration procedure, in particular confronting it with the standard (power-law) method.

Let us reiterate our choice of estimating quasar infrared and bolometric luminosities using a model-independent method: simply integrating the observed SEDs of the selected sources. An alternative approach could involve using templates to initially fit quasar SEDs and then deriving the corresponding IR and bolometric luminosities based on the best-fit values of the model parameters obtained. One clear advantage of the template fitting method is its consideration of intrinsic source extinction. However, it introduces systematic model uncertainties, such as those related to the specific extinction law assumed. A more detailed discussion of template fitting for the analyzed samples of quasars will be provided in a follow-up article.

The CF values were next calculated in the same manner for all the objects in the sample, as simply
\begin{equation}
   \textrm{CF} =\frac{L_{\rm IR}}{L_{\rm agn}} \, .
   \label{eq:4}
\end{equation}

The resulting CF range is characterized by a widespread, with the maximum values exceeding unity; we note again that such values are not allowed within the framework of the angle-based definition CF\,$=\Omega/4\pi$.

\section{Methods}
\label{sec:methods}
\subsection{Bayesian Fitting}

To quantify the differences between the Low-$z$ and High-$z$ quasars in our sample, Bayesian regression analysis was performed. We used the Python library \textit{emcee}\footnote{\url{https://emcee.readthedocs.io/en/stable/}}, which is the implementation of the Monte Carlo Markov Chain (MCMC) ensemble sampler \citep{emcee2013}. The statistical model, $M$, in the form $y=m\times x + b$ was used in the likelihood function, with the model parameters $\Theta=(m,b)$.  For the response variable we choose $y = \log L_{\rm IR}$, or $y=\log L_{\rm IR}/L_{\rm agn}$, and for the prediction variable $x = \log L_{\rm agn}$. With this approach, we determined the posterior probability density function (PDF) via the Bayes theorem
\begin{equation}
    P(\Theta | D, M) = \frac{P(D|\Theta,M)P(\Theta|M)}{P(D|M)} \, ,
\label{eq:Bayes}
\end{equation}
where $D$ stands for data, $P(D|\Theta,M)$ is the likelihood, $P(\Theta|M)$ is the prior, and $P(D|M)$ is the model evidence. The prior space was defined between $-2<m<2$, and $-35<b<35$. The priors starting values are based on the ordinary least-squares regression fitting (OLS). We have chosen the $\log$ likelihood in the form
\begin{equation}
    \begin{aligned}[b]
    & P(D|\Theta,M) = \ln p(y_n|x_n,s_{n},m,b) =\\
    &\quad -\sum_{n}\left[\frac{(y_n-mx_{n}-b)^2}{s^{2}_{n}} \times \ln\left(\frac{(y_n-mx_{n}-b)^2}{s^{2}_{n}}\right) \right] \, ,
    \end{aligned}
    \label{eq:LIKELIHOOD}
\end{equation}
where $s_n$ is the data error (see in this context Appendix\,\ref{sec:likelihood}).

 The resulting best-fit regression lines and parameters are given as the mean relations, at the 68\% confidence levels, in Figures~\ref{fig:SNR5}a), \ref{fig:Regression_Lir_Lagn}, \ref{fig:Toba}, and \ref{fig:Gu}, along with the relations obtained in the literature for comparison.

\subsection{Dispersion Statistical Analysis}
\label{sec:MC_IS}

\subsubsection{MCMC Errors}

The errors in the estimates of $L_{\rm IR}$ and $L_{\rm agn}$ (as discussed in Section \ref{sec:Luminosity}) were calculated using Monte Carlo (MC) simulations based on the observed fluxes and their uncertainties. Due to the asymmetric distribution of simulated luminosities, two-sided errors were defined as the 16th and 84th percentiles. These MC errors were considered throughout the entire analysis, especially in the regression analysis.

\subsubsection{Intrinsic Scatter}

Relations between physical quantities are hardly scatter-free. Inside each relation, there are perturbations involved (e.g. in our case dependence of other physical properties) that cause the scatter of such relation to some degree. That means that when one wants to measure the physical quantity $\theta$ in reality the $\phi=\theta+ \tilde\theta$ is measured, where $\tilde \theta$ is the perturbation.

To quantify the \emph{intrinsic} scatter in the analyzed correlation, denoted as $\sigma_{i}$, we followed the procedure described in \citet[][see Section 2 therein]{intrinsic_scatter}. Specifically, within the Bayesian framework, the PDF of $\sigma_{i}$ can be calculated as:
\begin{equation}
P(\sigma_{i} | \phi, \sigma_{\phi}) = \frac{P(\phi|\sigma_{i},\sigma_{\phi})P(\sigma_{i})}{P(\phi)} \, .
\end{equation}
Here, $P(\sigma_{i})$ represents the prior distribution of the intrinsic scatter, while $P(\phi)$ is the normalization corresponding to the measured quantity $\phi$ with its associated systematic and statistical error, $\sigma_{\phi}$.

\subsection{Efron \& Petrosian Test}
\label{sec:EP}

When working with truncated data, one can never be sure whether the dependency between the analyzed parameters is real or results from selection biases in the sample. Our data sample is built from several surveys, each with different limitations and selection criteria. The major of these effects is the flux-limitation of the sample. Handling such effects requires a statistical test to estimate how much the truncation (flux-limitation, in our case) affects the final dependence between the luminosities and the redshift. \citet[][hereafter EP]{EP1992} developed a nonparametric statistical test to determine the existence of a correlation or independence of variables from the truncated data set, with the special case being flux-limited data. For a detailed description of each step, see \cite{EP1999}, here we only briefly describe the most important assumptions and steps leading to the calculation of the EP test.

If one deals with one-sided truncation of the data, as in the luminosity-redshift case in a flux-limited survey, the following steps are required by the EP test.
First, for every object $i$ in the data sample, one should create the so-called associated set $J_{i} = \{j: L_{j} > L_{i}, L^{-}_{j} < L_{i} \}$, where the $L^{-}_{j}$ is the limiting luminosity (calculated for each object based on the chosen limiting flux), $L_{i}$ is the luminosity of the considered object, and $L_{j}$ is the luminosity of the object $j$ to which we compare the object $i$. The $J_{i}$ contains all the objects which have (a) luminosity above the chosen flux limit, and (b) luminosity higher than the luminosity of the object $i$. Based on the set $J_{i}$, the rank $R_{i}$ for each object $i$ is calculated following the equation $R_{i}={\rm number}\{j \in J_{i}: z_{j} \leq z_{i} \}$, where $z_{i}$ is the redshift of the object $i$ and $z_{j}$ is the redshift of the object $j$ from the set $J_i$. If luminosity and redshift are independent, then the rank $R_{i}$ should have a uniform distribution between 1 and $N_{i}$, where $N_{i}$ is the number of objects in $J_{i}$. 

The test statistic is then defined as
\begin{equation}
    \tau = \frac{\sum_{i} (R_{i} - E_{i})}{\sqrt{\sum_{i}V_{i}}} \, ,
    \label{eq:EP}
\end{equation}
where $E_{i}=\frac{1}{2}(N_{i}+1)$ is the mean of the uniform distribution, and $V_{i}=\frac{1}{12}(N_{i}^2-1)$ is the variance. One rejects the hypothesis of independence between luminosity and redshift if $|\tau| \geq n\sigma$, where $n$ can be chosen as the $\sigma$ multiplier for a given confidence level.

\section{Results}
\label{sec:results}

This section presents the results of our correlation analysis between AGN and IR luminosities in the constructed quasar samples. Additionally, we investigate the dependence of the CF on luminosities and present the results of the EP test. Finally, we assess the potential evolution of the CF with redshift.

\begin{figure}[!htp]

  \includegraphics[clip,width=1\columnwidth]{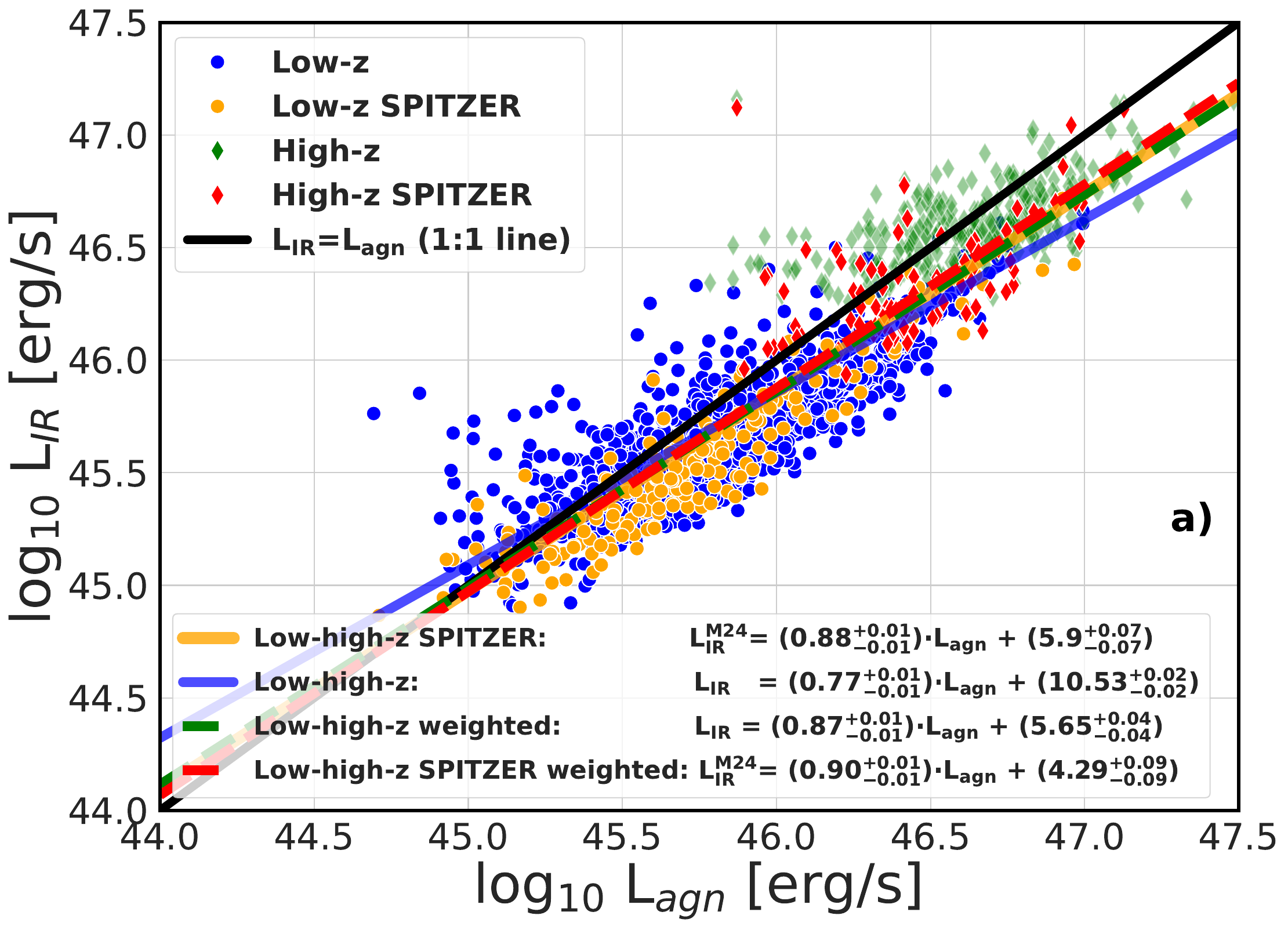}%

 \includegraphics[clip,width=1\columnwidth]{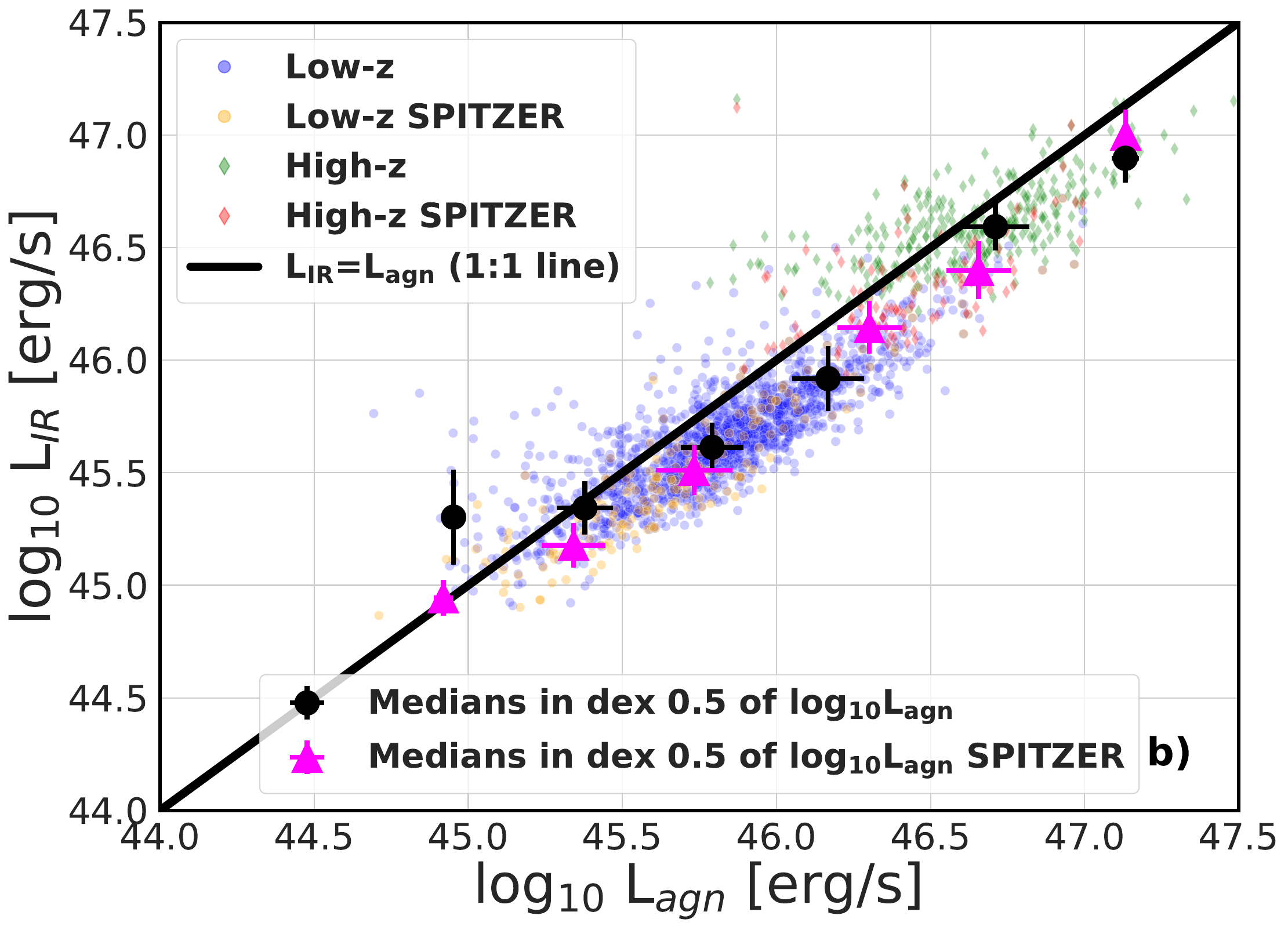}%

\caption{Relation between $\log L_{\rm IR}$ and $\log L_{\rm agn}$ with SNR$_{\rm W3\&W4}>5$. Blue and orange circles indicate Low-$z$ quasars, whereas green and red diamonds indicate High-$z$ sources. Orange circles and red diamonds indicate Low-$z$ and High-$z$ quasars with SPITZER M24 data. On the \textit{a) panel}, blue line denotes the best fitted Bayesian regression for the Low-high-$z$} data with SNR$_{\rm W3\&W4}>5$, orange line is the Bayesian regression for the Low-high-$z$ SPITZER data from both Low-$z$ and High-$z$, while green and red dashed lines represent the best Bayesian fit with the Low-$z$ sample weighted, and Low-$z$ with SPITZER weighted, respectively; black line gives 1:1 scaling relation between $L_{\rm IR}$ and $L_{\rm agn}$. On the \textit{b) panel}, big black circles represent the medians calculated for the 0.5\,dex in $\log L_{\rm agn}$, for both redshift samples (Low-high-$z$), and orange big triangles stand for the medians calculated for both redshift samples with the SPITZER data (Low-high-$z$ SPITZER). For both black and magenta points errorbars represent MAD. For exact values of fitted parameters see Table \ref{tab:regressions}.
\label{fig:SNR5}
\end{figure}

\subsection{Relations Between Luminosities}
\label{sec:LIR_Lagn}

Figure,\ref{fig:SNR5} illustrates the relationship between $L_{\rm IR}$ and $L_{\rm agn}$ for the analyzed samples, where we compare luminosities calculated with the WISE data to those derived based on SPITZER M24. Lines in the figure denote the results of the Bayesian regression analysis, for which the obtained model parameters are summarized in Table\,\ref{tab:regressions}. The data were restricted to SNR$_{\rm W3\&W4}>5$ (for the entire WISE sample, see Appendix\,\ref{sec:Relations}). Notably, we observe more WISE outliers despite the applied high-quality criterion, while the SPITZER sample exhibits greater condensation, particularly in the Low-$z$ range. Apparently, the Low-$z$ and High-$z$ samples tend to form a single correlation trend, thus we decided to fit a single power-law to both Low-$z$ and High-$z$ sources simultaneously - the ``Low-high-$z$ sample''. The best-fit regression lines can be found in Figure\,\ref{fig:SNR5}. The resulting power slope for the Low-high-$z$ data is $0.77\pm0.01$. For the Low-high-$z$ SPITZER data, where W4 was replaced with MIPS 24$\mu$m, the index is $0.88\pm0.01$. These values exhibit only slight differences, and visually, the fits closely align.
The major feature here is the lack of low luminosity objects in the top panel of the figure (compared to the whole sample without the SNR restriction, see Figure\,\ref{fig:Lir_Lbol_lamaniec_Spitzer} a). This can be explained by the fact that the objects with the lowest luminosities have unreliable measurements in terms of the SNR in the W3 and W4 filters. 

The SNR$_{\rm W3\&W4}>5$ Low-$z$ sample is almost five times larger than the High-$z$ sample. This can cause the regression fit to be largely determined by the Low-$z$ sources. To mitigate this effect, we conducted a simple test. From the Low-$z$ sample, 20\% of the whole SNR$_{\rm W3\&W4}>5$ datapoints were randomly selected (``Low-$z$ reduced''). A similar procedure was performed for the Low-$z$ SPITZER data, but only 50\% of the Low-$z$ SPITZER sample with SNR$_{\rm W3\&W4}>5$ was taken into account (``Low-$z$ SPITZER reduced''). The Low-$z$ reduced sample was then merged with the High-$z$ sample with SNR$_{\rm W3\&W4}>5$ (``Low-high-$z$ weighted''); likewise the Low-$z$ SPITZER reduced with the High-$z$ sample with SNR$_{\rm W3\&W4}>5$ (``Low-high-$z$ SPITZER weighted''). The Bayesian regression analysis was repeated for each of these combined samples as outlined in Section\,\ref{sec:methods}. This procedure was executed 50 times and the final parameters were taken as the median of the 50 fits. This test is described as the weighted fit and is represented by the green line in Figure\,\ref{fig:SNR5} a). The resulting slopes for the weighted data are $0.87 \pm 0.01$ (Low-high-$z$), and $0.90 \pm 0.01$ (Low-high-$z$ SPITZER). This means that the larger Low-$z$ sample dominates the fit, in particular the tail below the dex 45.5 in $\log L_{\rm agn}$. 

To address the issue of different sample sizes between the Low-$z$ and High-$z$ sources, a second test was conducted using binning for the entire dataset. The bins were created as follows: starting from $\log L_{\rm agn}=44.5$ the subsequent bins were created every 0.5\,dex of $\log L_{\rm agn}$. For each bin, the median values and the median absolute deviations (MAD) were calculated both in $\log L_{\rm agn}$ and $\log L_{\rm IR}$. 
This procedure was applied separately for the Low-high-$z$ and Low-high-$z$ SPITZER samples, and the results are given in the lower panel b) of Figure\,\ref{fig:SNR5}. The problematic low-luminosity segment of the Low-$z$ sample is now even more visible, exceeding the 1:1 scaling. A similar trend is observed in the SPITZER sample. However, due to a larger number of outliers toward higher $L_{\rm IR}$, the low luminosity bins in the WISE sample diverge more significantly from the overall trend, while in the SPITZER sample, the situation is far less severe. It's worth noting that the lowest luminosity bin contains however just a few sources in both WISE and SPITZER samples, and this is reflected by a large error bar in the WISE sample, even though in the corresponding SPITZER bin the variance is much smaller.

\begin{figure}[!htp]

  \includegraphics[clip,width=1\columnwidth]{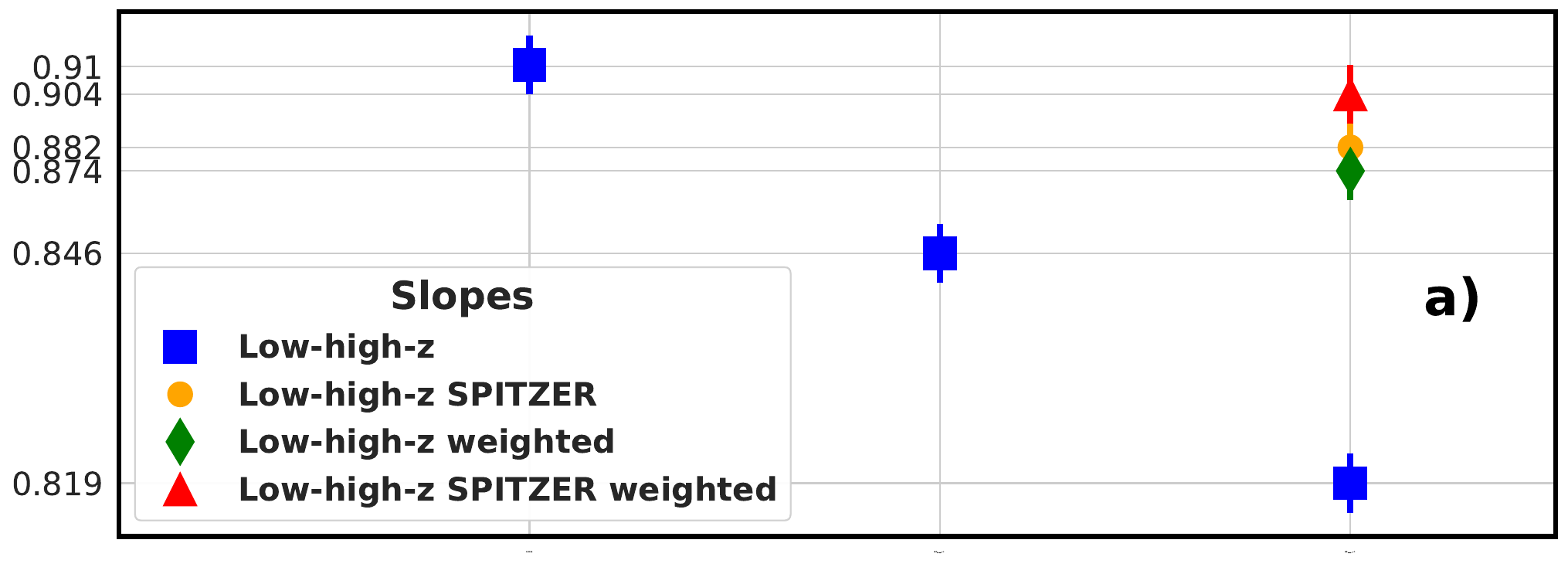}%
\vspace{-0.25cm}
\hspace*{0.11cm}\includegraphics[clip,width=0.986\columnwidth]{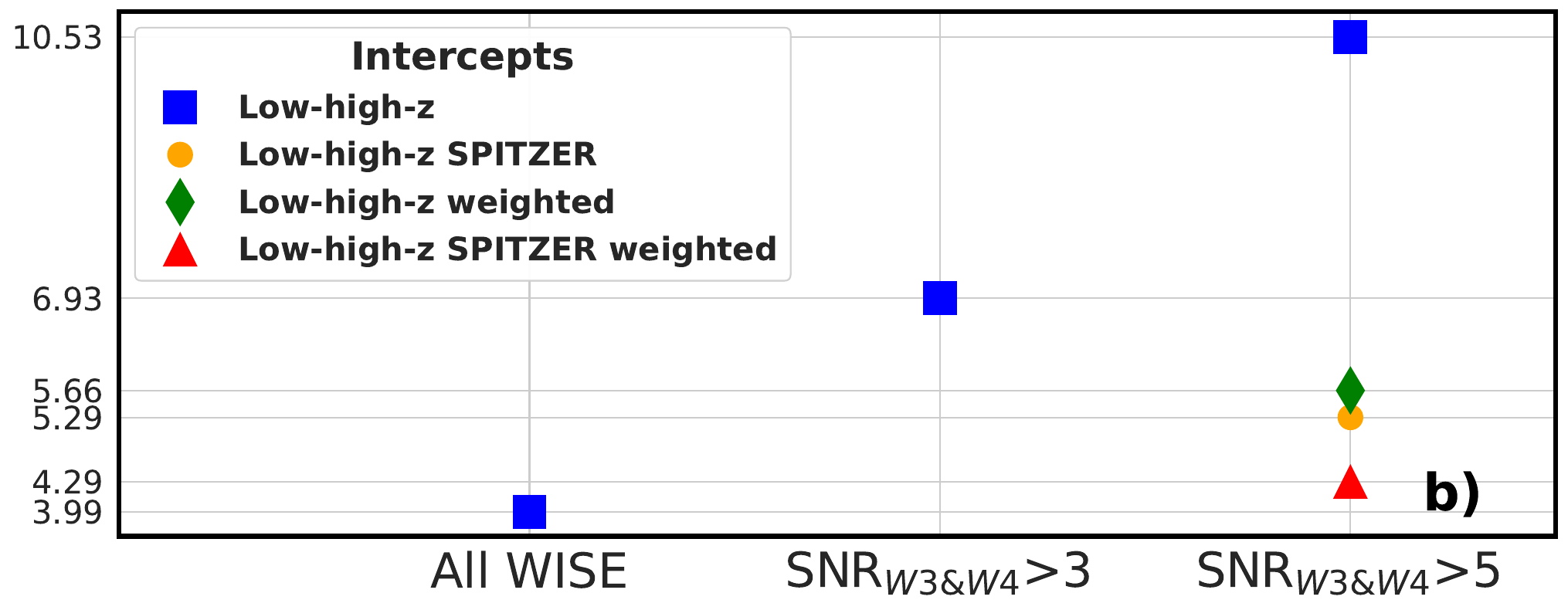}%

\caption{Comparison between the Bayesian regression analysis model parameters: a) slopes, b) intercepts, obtained for different analyzed samples of quasars (see Table\,\ref{tab:regressions} and Section\,\ref{sec:LIR_Lagn}).}

\label{fig:SLOPES}
\end{figure}

To have some insight into how the restrictions on the SNR influence the sample, one can refer to medians and their errors, as quantified in Table \ref{tab:medians}. It's evident that as we move towards higher SNR in the WISE filters, all the median luminosities increase. This is expected since more luminous sources have usually higher flux in a given redshift bin. The deviation in the Low-$z$ sample decreases with increasing SNR from 0.29 in $\log  L_{\rm agn}$ and 0.24 in $\log  L_{\rm IR}$, to 0.19 and 0.12, respectively, i.e. approaches the level observed in the High-$z$ population. The High-$z$ sample is however significantly smaller, consisting of 309 high-SNR objects compared to 7,999 without the SNR restriction, and so increasing the SNR does not result in this case in decreasing the variance. In Figure\,\ref{fig:SLOPES} we summarize the best-fit slopes (panel a) and intercepts (panel b) for the five best Bayesian fits for the WISE data, SNR$_{\rm W3\&W4}$>3, SNR$_{\rm W3\&W4}>5$, SPITZER and the SNR$_{\rm W3\&W4}>5$ samples with the weighted Low-$z$ sample. The general trend in the data with WISE W4 (blue squares) is that with the higher SNR the lower the slope value, so the deviation from the 1:1 relation between $L_{\rm IR}$ and $L_{\rm agn}$ becomes larger.

\begin{figure}[!htp]

  \includegraphics[clip,width=1\columnwidth]{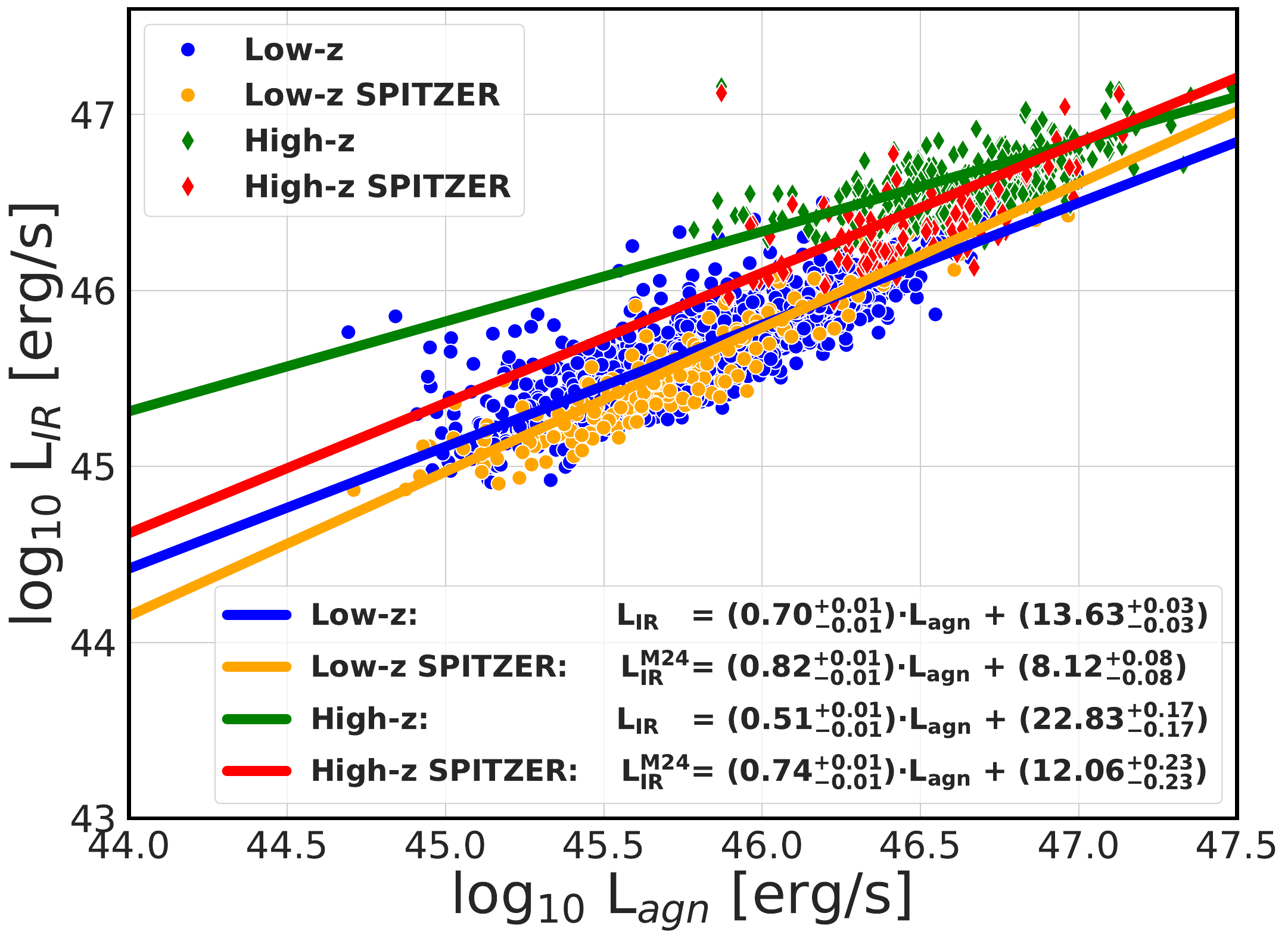}%

  \caption{Regression analysis results for $\log L_{\rm agn}$ and $\log L_{\rm IR}$ for Low-$z$, Low-$z$ SPITZER, and High-$z$, High-$z$ SPITZER sources separately. The model parameters of the Bayesian linear regression analysis are given in Table\,\ref{tab:regressions}.}
  \label{fig:Regression_Lir_Lagn}
\end{figure}

As mentioned earlier, the Low-$z$ and High-$z$ samples follow a similar correlation between $L_{\rm agn}$ and $L_{\rm IR}$. In the case of the best-quality data with SPITZER, the obtained best-fit regression lines in logarithmic space only show slight differences, as illustrated in Figure\,\ref{fig:Regression_Lir_Lagn}. For data without SPITZER the difference is more apparent, in particular, the Low-$z$ fitted slope is $0.70\pm0.01$, whereas the High-$z$ slope is $0.51\pm0.01$. A shallower High-$z$ correlation line diverges at the high luminosity tail of the sample. This is due to the majority of the members laying at bolometric luminosities below 46.5\,dex.

\begin{table*}[]
\caption{Model parameters $\Theta=(m,b)$ obtained from the bayesian linear regression analysis $\log L_{\rm IR}=m \times \log L_{\rm agn} + b$ as presented in Figures\,\ref{fig:SNR5} and \ref{fig:Regression_Lir_Lagn}. Columns describe 1) The subset name; 2) the 'm' parameter (slope); 3) the intercept 'b'.}

\begin{tabular}{|p{0.35\textwidth}|p{0.15\textwidth}|p{0.15\textwidth}|}
\hline
 Subset & m & b   \\ \hline \hline
 Low-high-$z$ SNR$_{\rm W3\&W4}>5$  & 0.77$^{+0.01}_{-0.01}$ & 5.90$^{+0.07}_{-0.07}$\\[4pt] \hline
 Low-high-$z$ SPITZER SNR$_{\rm W3}>5$  & 0.88$^{+0.01}_{-0.01}$ & 10.53$^{+0.02}_{-0.02}$\\[4pt] \hline
 Low-high-$z$ weighted SNR$_{\rm W3\&W4}>5$  & 0.87$^{+0.01}_{-0.01}$ & 5.65$^{+0.04}_{-0.04}$\\[4pt] \hline
 Low-high-$z$ SPITZER weighted SNR$_{\rm W3}>5$  & 0.90$^{+0.01}_{-0.01}$ & 4.29$^{+0.16}_{-0.16}$\\[4pt] \hline \hline
 Low-$z$ SNR$_{\rm W3\&W4}>5$  & 0.70$^{+0.01}_{-0.01}$ & 13.63$^{+0.03}_{-0.03}$\\[4pt] \hline
 Low-$z$ SPITZER SNR$_{\rm W3}>5$  & 0.82$^{+0.01}_{-0.01}$ & 8.12$^{+0.08}_{-0.08}$\\[4pt] \hline
 high-z SNR$_{\rm W3\&W4}>5$  & 0.51$^{+0.01}_{-0.01}$ & 22.83$^{+0.17}_{-0.17}$\\[4pt] \hline
 high-z SPITZER SNR$_{\rm W3}>5$  & 0.74$^{+0.01}_{-0.01}$ & 12.06$^{+0.23}_{-0.23}$\\[4pt] \hline

\end{tabular}
\label{tab:regressions}
\end{table*}

\subsection{Relations Between CF And Luminosities}

\begin{figure}[!htp]

  \includegraphics[clip,width=1\columnwidth]{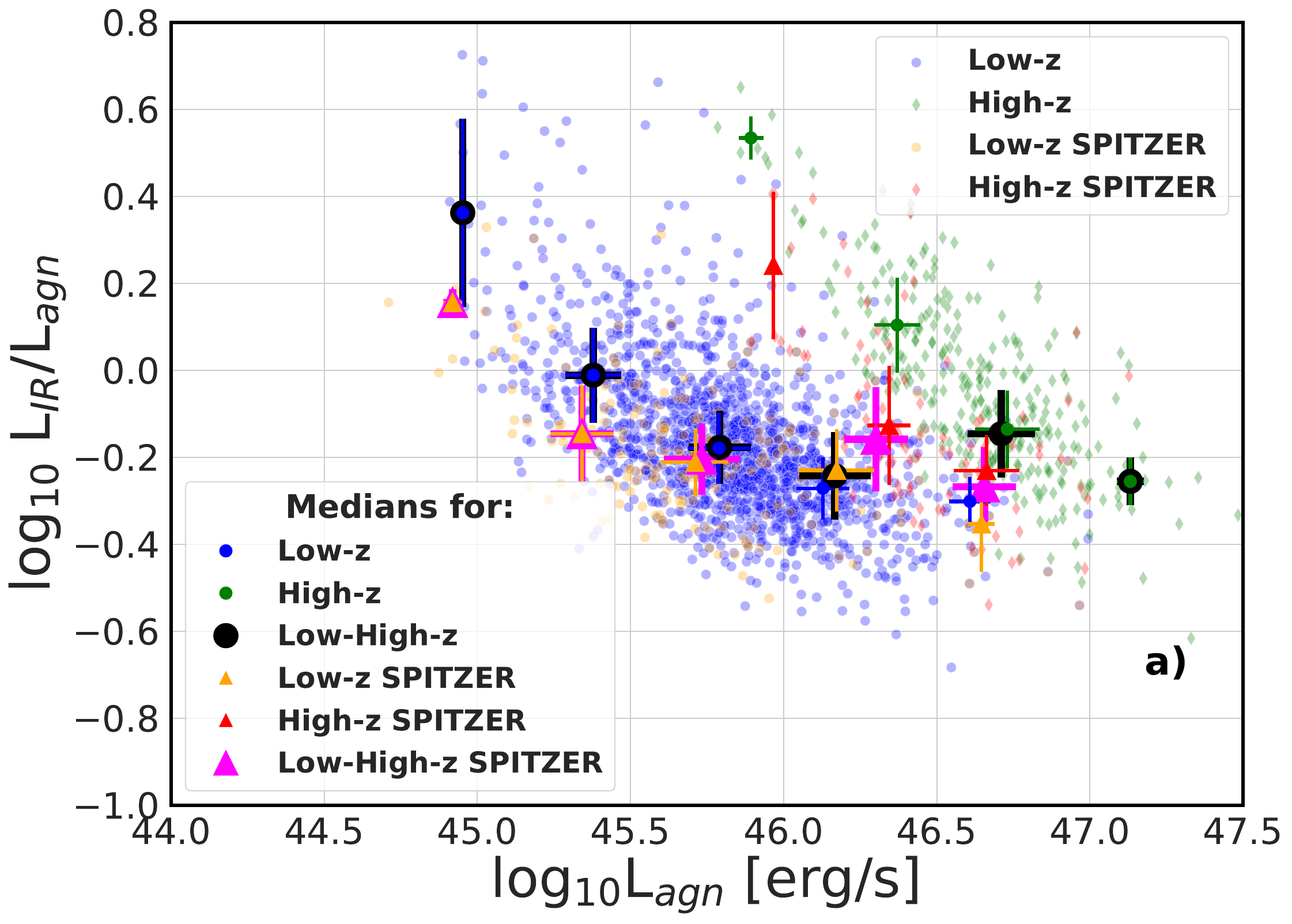}%

  \includegraphics[clip,width=1\columnwidth]{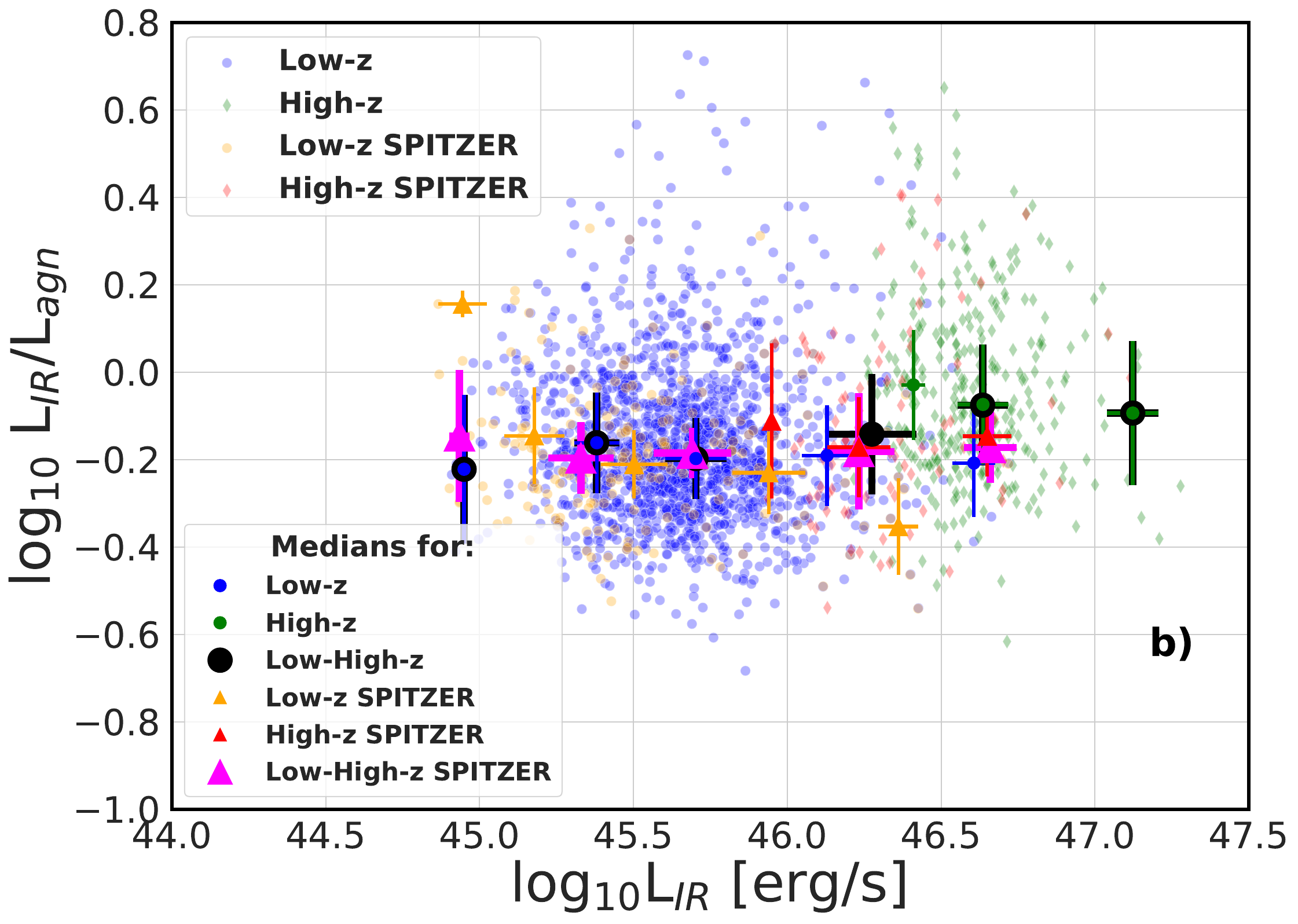}%

\caption{The relationships} between CF and $L_{\rm IR}$ (upper panel a) and $L_{\rm agn}$ (lower panel b), including only the sources with SNR$_{\rm W3\&W4}>5$. Blue circles indicate Low-$z$ quasars, whereas green and red diamonds indicate High-$z$ sources. Orange circles and red diamonds indicate Low-$z$ and High-$z$ quasars with the SPITZER M24 data. Additionally, the median statistics were calculated for the samples, as described in the legend in the lower left corner. Error bars represent the MAD errors.
\label{fig:SNR5ab}
\end{figure}

Figure\,\ref{fig:SNR5ab} shows the relations CF vs. $L_{\rm IR}$ and CF vs. $L_{\rm agn}$. Despite precise estimates of both luminosities, a substantial spread is still evident, particularly in the in $\log {\rm CF} - \log L_{\rm IR}$ relation. The corresponding medians, calculated in the same way as in Figure\,\ref{fig:SNR5}, are also displayed in Figure\,\ref{fig:SNR5ab}. In panel a), an anticorrelation between $\log  {\rm CF}$ and $\log L_{\rm agn}$ is apparent, consistent with the regression fit presented in Figure,\ref{fig:SNR5}. What is particularly evident here is the influence of the low-luminosity segment of the Low-$z$ sample, below $\sim \log L_{\rm agn}/$\,erg\,s$^{-1}$\,$=46$. All luminosities are in cgs units, unless stated otherwise. The Low-$z$ quasars (blue dots in the Figure) are in general dominating up to the bin at $\log L_{\rm agn} \sim 46.2$; the High-$z$ (green points) quasars dominate above 46.5\,dex in $\log L_{\rm agn}$. This is especially visible for the median calculated for both the Low-$z$ and High-$z$ samples together (Low-high-$z$; big black dots). A similar trend is present in the Low-$z$ and High-$z$ SPITZER sub-samples. On the other hand, no anticorrelation between CF and $L_{\rm IR}$ is present in panel b) of Figure\,\ref{fig:SNR5ab}. The medians here, calculated with the bins every 0.5\,dex $\log L_{\rm IR}$, form a horizontal sequence, with the only outliers being the low-luminosity tail of the Low-$z$ SPITZER sample. The differences between the Low-$z$ and High-$z$ samples in Figure\,\ref{fig:SNR5ab}a) are likely caused by the instrumental effects. Especially the low-luminosity bin in the High-$z$ sample (first green diamond), which has the highest CF, would simply be absent with the increased SNR threshold. Meanwhile, the first bin in High-$z$ SPITZER (red diamond) has the highest variance, with many outliers. The second point in High-$z$ has also higher values than the trend for both samples combined, which is not the case for the second bin in the corresponding SPITZER sample.

\begin{table*}[]
\caption{Medians of the basic primary quantities calculated with the all-points method. $L_{\rm agn}$ and $L_{\rm IR}$ are in units of erg\,s$^{-1}$, and $M_{\rm BH}$ are given in the $M_{\odot}$ units. Errors are the median absolute deviations. Columns: 1) Subset name, 2) $\log $ of the AGN luminosity (L$_{\rm agn}$) in  [erg/s] units, 3) $\log $ of the IR luminosity (L$_{\rm IR}$) in [erg/s] units, 4) $\log $ of the Covering Factor (CF), 5) $\log $ of the SMBH mass (M$_{\rm BH}$) in [$M_{\odot}$] units, 6) the accretion rate $\lambda_{\rm Edd}$ calculated from equation \ref{eq:3}.}

\begin{tabular}{|p{0.15\textwidth}|p{0.13\textwidth}|p{0.13\textwidth}|p{0.13\textwidth}|p{0.13\textwidth}|p{0.13\textwidth}|}
\hline
 Subset & $\log  L_{\rm agn}$ & $\log  L_{\rm IR}$ & $\log  CF$ & $\log  M_{\rm BH}$ & $\lambda_{\rm Edd}$  \\ \hline \hline
 Low-$z$ & 45.37$\pm$0.29 & 45.19$\pm$0.24  & $-$0.18$\pm$0.11 & 8.46$\pm$0.40 & $-$1.17$\pm$0.34\\ \hline
 High-$z$ & 46.15$\pm$0.18 & 46.15$\pm$0.11 & $-$0.01$\pm$0.13 & 9.12$\pm$0.36 & $-$0.83$\pm$0.34 \\ \hline
 Low-$z$ SNR>3 & 45.70$\pm$0.22 & 45.50$\pm$0.19 & $-$0.19$\pm$0.11  & 8.64$\pm$0.31 & $-$1.02$\pm$0.23 \\ \hline

 High-$z$ SNR>3 & 46.45$\pm$0.20 & 46.38$\pm$0.12 & $-$0.08$\pm$0.14 & 9.31$\pm$0.22 & $-$0.69$\pm$0.19 \\ \hline
 
 Low-$z$ SNR>5 & 45.83$\pm$0.19 & 45.65$\pm$0.16 & $-$0.18$\pm$0.10  & 8.80$\pm$0.30 & $-$0.91$\pm$0.22 \\ \hline

 High-$z$ SNR>5 & 46.65$\pm$0.19 & 46.59$\pm$0.12 & $-$0.05$\pm$0.13 & 9.41$\pm$0.21 & $-$0.61$\pm$0.19 \\ \hline
 
 Low-$z$ SPITZER & 45.52$\pm$0.28 & 45.32$\pm$0.26 & $-$0.19$\pm$0.11  & 8.60$\pm$0.27 & $-$1.09$\pm$0.24 \\ \hline
 High-$z$ SPITZER & 46.28$\pm$0.21 & 46.14$\pm$0.21 & $-$0.17$\pm$0.11 & 9.21$\pm$0.24 & $-$0.81$\pm$0.24 \\ \hline

\end{tabular}
\label{tab:medians}
\end{table*}

\subsection{Testing For The Luminosity Evolution}
\label{sec:EP_result}

The challenges in addressing the luminosity evolution of quasars in the analyzed samples arise from (i) different flux limits in various survey releases and (ii) the integration of source luminosities across broad frequency ranges. The SDSS DR16Q data were folded from DR7Q and DR12Q, and each of these releases had a different truncation and magnitude limit. In the end, we have chosen $r = 22$\,mag as the limit for $L_{\rm agn}$, which was the most restrictive limit among the SDSS DR16Q data, and W4\,$ = 6$\,mJy for $L_{\rm IR}$ (the most restricting limit of the IR photometry). The fluxes were then converted to monochromatic luminosities in erg\,s$^{-1}$ units. Next, we followed the steps of the EP test described in the previous Section\,\ref{sec:EP} for the joint Low-$z$ and High-$z$ population, working on (a) the Low-high-$z$ sample without SNR cuts, 
(b) the sample with SNR$_{\rm W3\&W4}>3$, (c) the Low-high-$z$ sample with SNR$_{\rm W3\&W4}>5$, (d) the Low-high-$z$ SPITZER sample. 
For each of these sub-samples, the EP test was performed separately for $L_{\rm IR}$ and $L_{\rm agn}$. The resulting test values of $\tau$ are: (a) $\tau_{\rm IR}=91.6$, $\tau_{\rm agn}=103.9$, 
(b) $\tau_{\rm IR}=54.4$, $\tau_{\rm agn}=53.0$, 
(c) $\tau_{\rm IR}=33.1$, $\tau_{\rm agn}=30.4$,
(d) $\tau_{\rm IR}=13.8$, $\tau_{\rm agn}=15.6$. 
In all those cases, the hypothesis of independence between $L_{\rm IR}$ and $z$, as well as between $L_{\rm agn}$ and $z$, should be therefore rejected, and so a strong redshift evolution of the luminosities in both bands is observed.

\subsection{Testing For The CF Evolution}
\label{sec:CF_evo}

\begin{figure}[!htp]
  \includegraphics[clip,width=1\columnwidth]{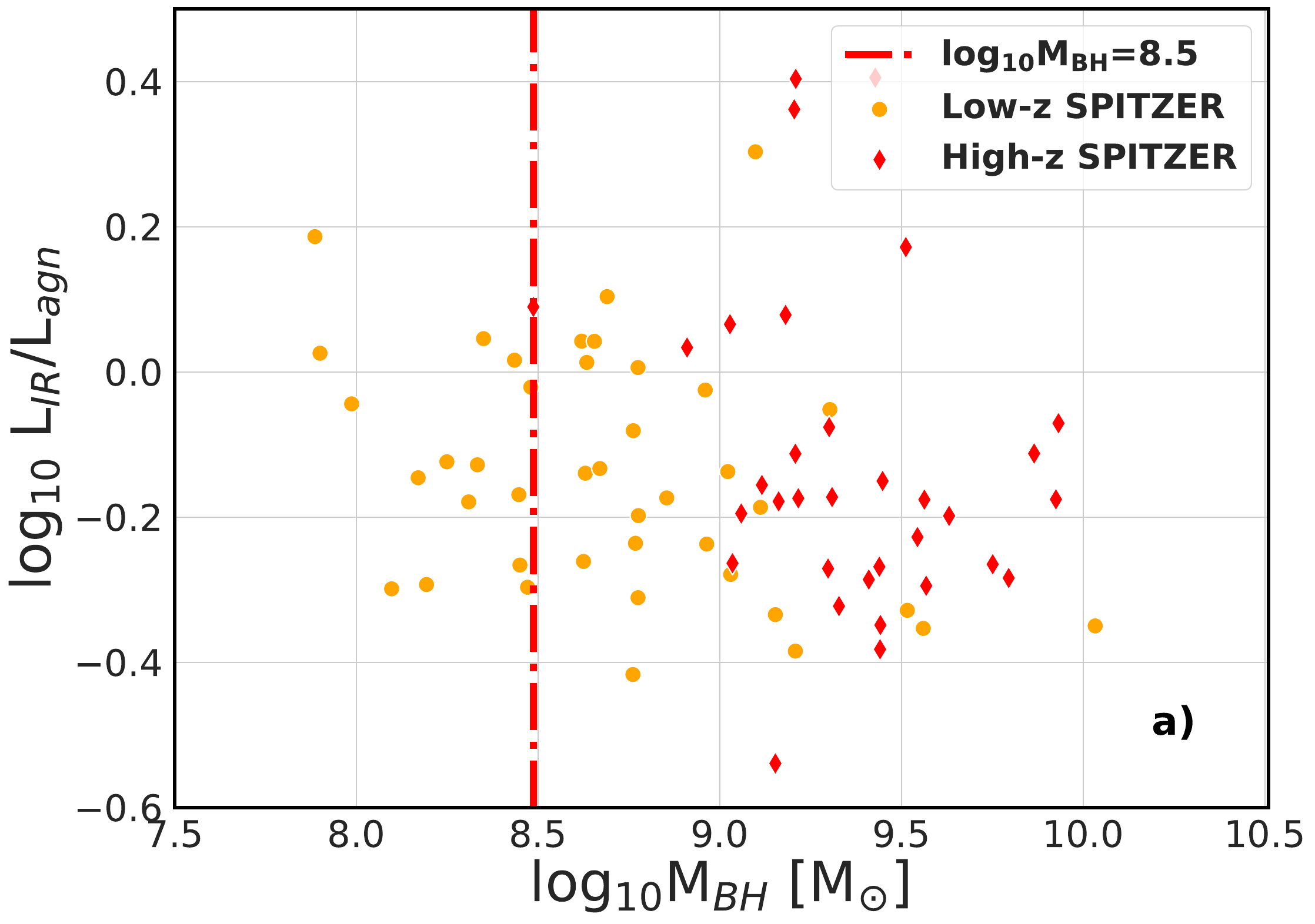}
  \includegraphics[clip,width=0.98\columnwidth]{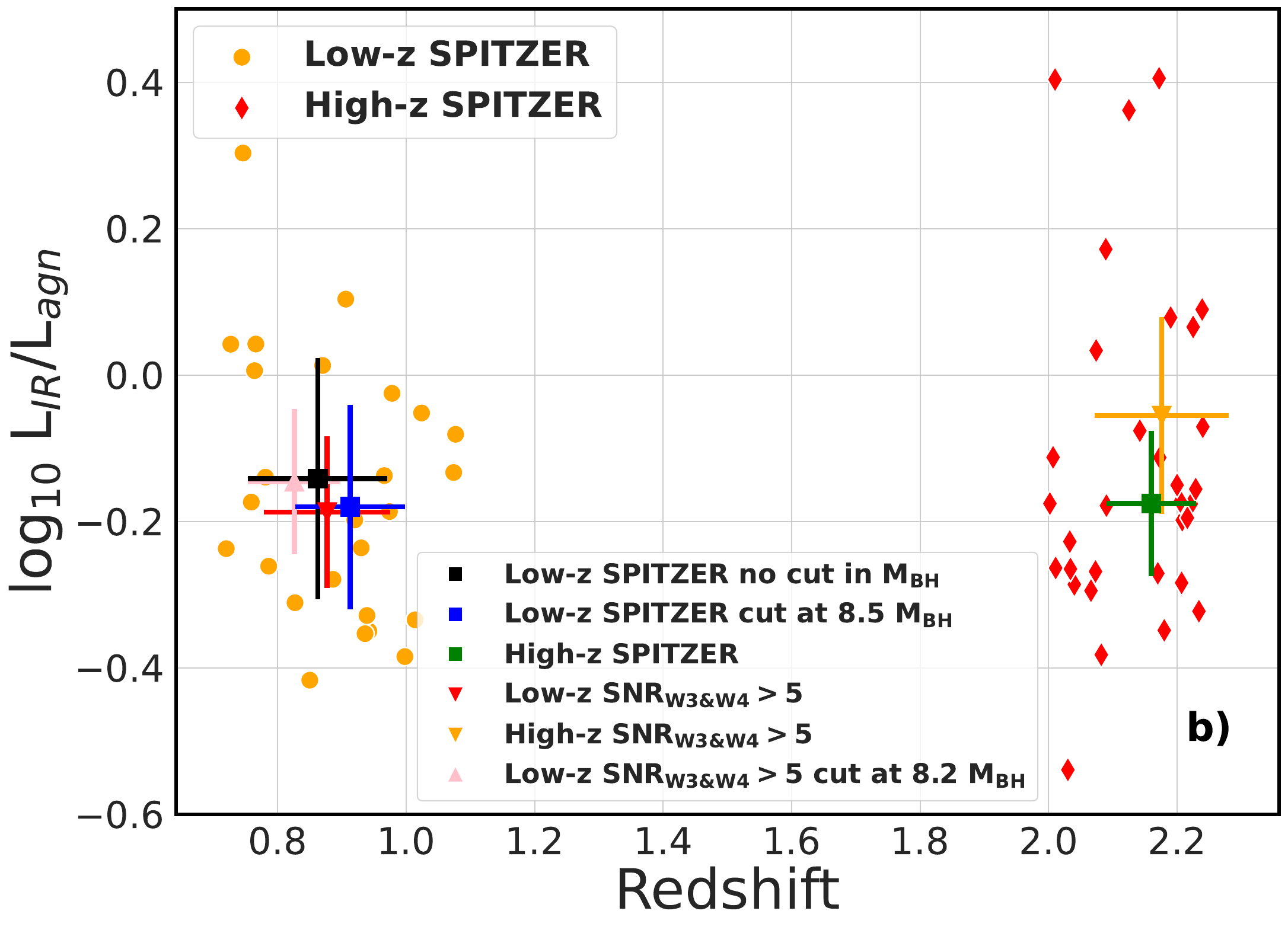}
  \caption{Top panel a) shows the relation between $\log $\,CF and $\log  M_{\rm BH}$ for quasars with the SPITZER M24 and VAC data. Red line represents the cut on $\log  M_{\rm BH}/M_{\odot} = 8.5$. Bottom panel b) shows the relation between $\log $\,CF and redshift for quasars with $\log  M_{\rm BH}/M_{\odot}>8.5$.  Orange and red points in both panels represent the Low-$z$ and High-$z$ quasars, respectively. Large symbols with MAD error bars stand for the medians calculated for each data sample: Low-$z$ with no cut in $M_{\rm BH}$ (red triangle), High-$z$ (orange triangle), Low-$z$ with the cut in $\log M_{\rm BH}/M_{\odot}>8.2$ (pink triangle), Low-$z$ SPITZER with no cut in $M_{\rm BH}$ (black square), High-$z$ SPITZER (green square), Low-$z$ SPITZER with the cut in $\log M_{\rm BH}/M_{\odot}>8.5$ (blue square). For luminosity and Eddington ratio selection please see Appendix \ref{sec:other_cuts}.}
  \label{fig:MBH_cut}
\end{figure}

The EP test results described earlier indicate the evolution of both $L_{\rm IR}$ and $L_{\rm agn}$ with redshift. Although relevant, this result still does not answer the question of a possible evolution of the CF. To address this issue, we selected quasars from our sample of the objects with comparable $M_{\rm BH}$ values. Quasars selected in this way should have corresponding physical properties regardless of their redshifts. Moreover, in addition to the spectroscopic-based virial $M_{\rm BH}$ estimates, we required also good-quality SPITZER M24 data for the targets. Finally, only the objects with the $M_{\rm BH}$ values above the threshold of either $\log M_{\rm BH}/M_{\odot} = 8.5$ (which is the lowest value for the High-$z$ population), were selected within the sub-sample with SNR$_{\rm W3\&W4}>5$. The final data sample for this test consists of 59 objects, including 26 Low-$z$ objects, and 33 High-$z$ sources. Figure\,\ref{fig:MBH_cut} panel a) shows the resulting distribution of the CF values with $M_{\rm BH}$. As one can note, the analyzed data set reveals no significant differences between the Low-$z$ and High-$z$ targets in the projected $\log$\,CF distribution. To quantify this statement, a one-dimensional two-sample Kolmogorov-Smirnov test was performed. The resulting p-value\,$= 0.80$ is high, meaning that the hypothesis that the two samples were drawn from the same distribution (the null hypothesis), should not be rejected at a very high confidence level. In the final step, we performed the 2-dimensional generalization of the two-sample Kolmogorov-Smirnov test \citep{2KS2D} for the $\log M_{\rm BH}-\log$\,CF distribution. The p-value in this case turns out relatively small, below 0.01, implying that the hypothesis that the two-dimensional sample is from the same distribution, can be rejected at the $>99\%$ confidence level. Additional tests, however, revealed that this was due to a shift in the mean $M_{\rm BH}$ values between our Low-$z$ and High-$z$ populations, consistent with a trend observed earlier and discussed in \cite{McLure2004}. 

Figure\,\ref{fig:MBH_cut} panel b) shows the distribution of the CF with respect to the redshift for the Low-$z$ SPITZER sample (after the cut in $M_{\rm BH}$) and the High-$z$ SPITZER sample. Additionally, the medians in both $\log$\,CF, and $z$ for each data sample were calculated, as represented in the Figure by large symbols with error bars. 
As follows, all the medians are consistent within the errors with the scenario of no significant evolution of the CF with redshift. As discussed in more detail in \ref{sec:other_cuts}, similar results hold also for quasars selected based on luminosities or Eddington ratios.

\subsection{Dispersion Analysis}
\label{sec:dispersion}

The intrinsic scatters $\sigma_i$ in the $\log L_{\rm IR}$ vs. $\log L_{\rm agn}$ relations in the primary datasets with SNR cuts were analyzed, as described in Section \ref{sec:MC_IS}. The results are summarized below:

\begin{itemize}

\item[i)] Low-$z$  
SNR$_{\rm W3\&W4}$\,$>5$: $\sigma_{i} = 0.157$  ;

\item[ii)] High-$z$ 
SNR$_{\rm W3\&W4}$\,$>5$: $\sigma_{i} = 0.165$ ;

\item[iii)] Low-high-$z$ 
SNR$_{\rm W3\&W4}$\,$>5$: $\sigma_{i} = 0.208$ .

\end{itemize}

For the SPITZER data, on the other hand, we have obtained:

\begin{itemize}
    
\item[iv)] Low-$z$ SPITZER 
SNR$_{\rm W3}$\,$>5$: $\sigma_{i} = 0.150$ ;

\item[v)] High-$z$ SPITZER 
SNR$_{\rm W3}$\,$>5$: $\sigma_{i} = 0.222$ ;

\item[vi)] Low-high-$z$ SPITZER 
SNR$_{\rm W3}$\,$>5$: $\sigma_{i} = 0.148$ .

\end{itemize}

The $\sigma_{i}$ values are very similar in both Low-$z$ and Low-$z$ SPITZER. The difference is visible for the High-$z$ SPITZER sample, where the variance is large as the relative number of outliers is increased, especially at the low-luminosity end of the High-$z$ sample (Figure\, \ref{fig:Regression_Lir_Lagn}).

\begin{figure}[!htp]

  \includegraphics[clip,width=1\columnwidth]{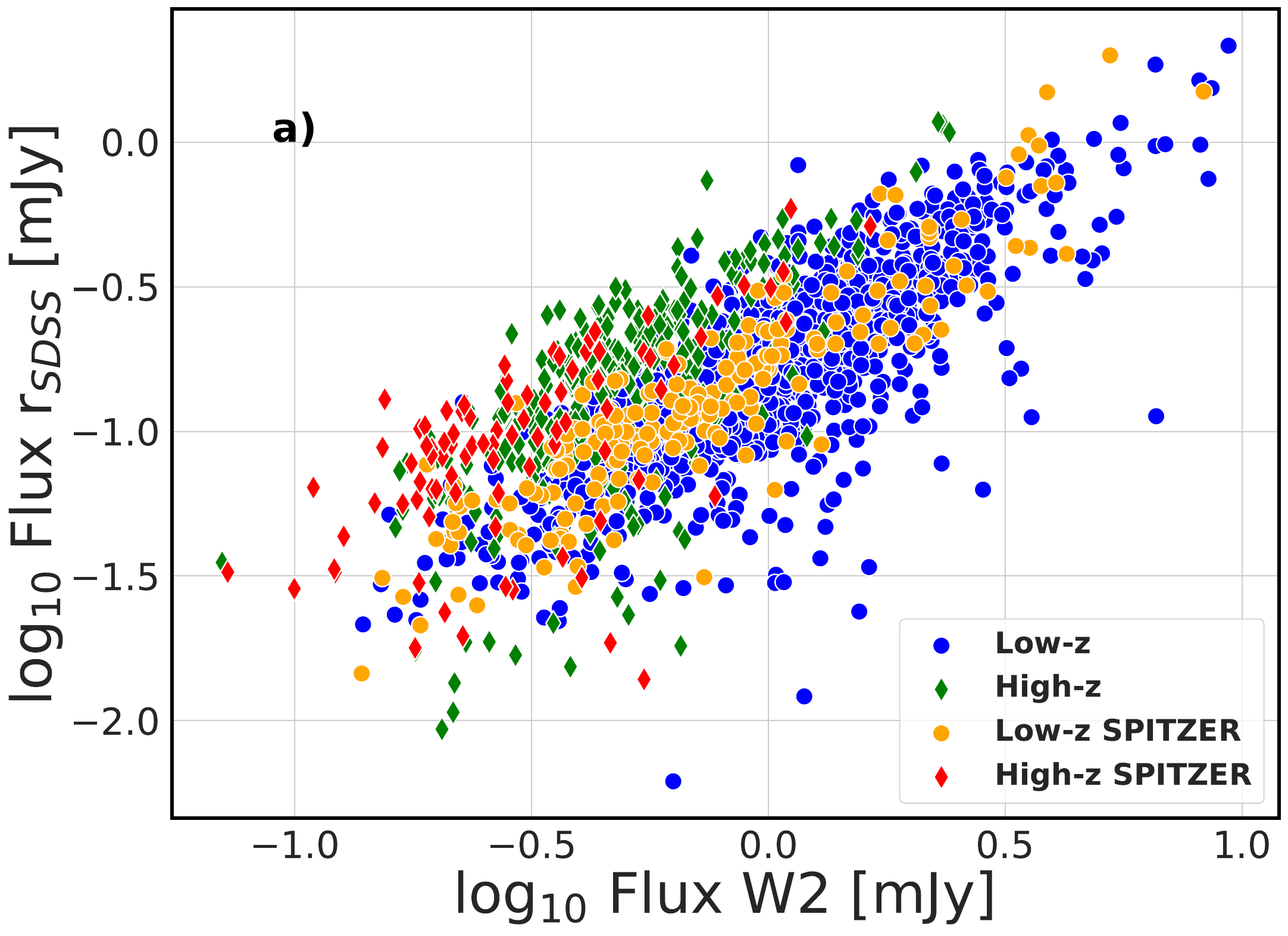}%

  \includegraphics[clip,width=1\columnwidth]{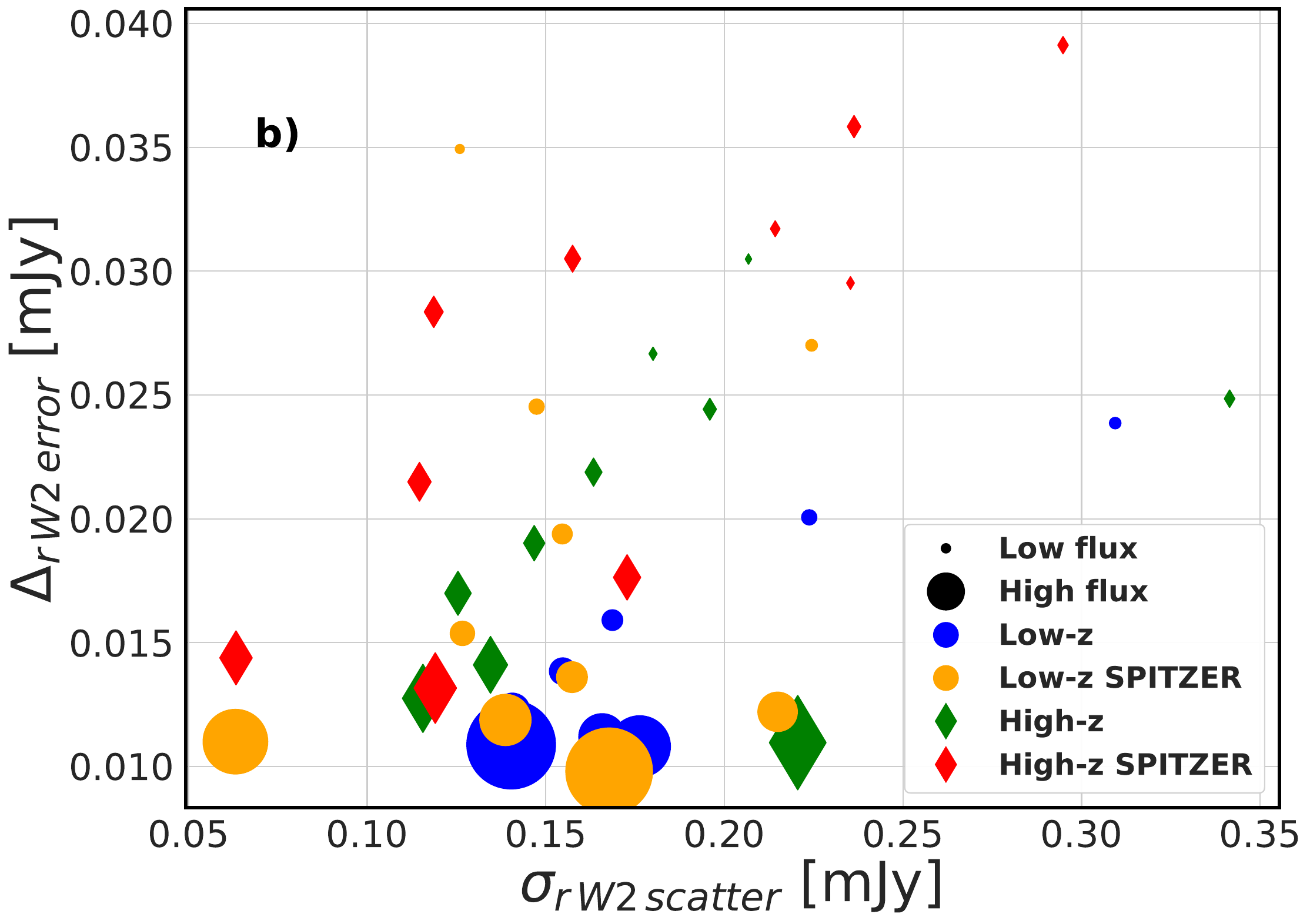}%

  \caption{\textit{Top panel:} Scatter plot for the extinction corrected $\log f_r$ and $\log f_{\rm W2}$, for Low-$z$, High-$z$, Low-$z$ SPITZER and High-$z$ SPITZER. \textit{Lower panel:} Standard deviation for binned fluxes in $r$ and W2 bands, and the error of $\log$ r-W2 relations $\sigma_{r\, W2\, scatter}$. The bins were calculated as equally distanced in $\log$ r space. Sizes of the symbols correspond to $\log f_r$ flux values in each bin.}
  \label{fig:Variance}
\end{figure}

The second test aimed to assess variances in both fitted regression and observational errors. This test was adapted from a similar approach presented in the literature \citep{Risaliti_Lusso2015}, with necessary modifications. To ensure a cosmology-free comparison, fluxes were used rather than luminosities. Specifically, we utilized the SDSS $r$ (the most limiting filter in $L_{\rm agn}$, that is present in both Low-$z$ and High-$z$ with and without SPITZER) and WISE W2 filters (the most limiting filter in $L_{\rm IR}$). Figure\,
\ref{fig:Variance}a) shows the relation between  the logarithms of the $r$-band and W2 extinction-corrected fluxes, $\log f_r$ and $\log f_{\rm W2}$, respectively.

The data were spread into 10 bins in $\log f_r$, with equal distances in logarithmic space. For each bin, standard deviations $\sigma$ were calculated considering observational (flux measurement) uncertainties, namely $\Delta_{r - {\rm W2} - {\rm error}}=\sqrt{\Delta(\log f_r)^2+\Delta(\log f_{\rm W2})^2}$. Moreover, for given flux distributions as given in Figure\,\ref{fig:Variance}a), we calculated also $\sigma_{r\, W2\, scatter}=\sqrt{\sigma\,(\log \, f_r)^2+\sigma\,(\log \, f_{W2})^2}$. We repeated this process for all four datasets (Low-$z$, High-$z$, Low-$z$ SPITZER, High-$z$ SPITZER). The results of the test are presented in Figure\,\ref{fig:Variance}b), where the size of the markers is coded to represents the mean $r$-band flux of each bin, with larger points indicating higher flux values.

From this test, one can see that: 1) The trend with $\sigma_{r\, W2\, scatter}$ is only visible for lower luminous bins (with higher flux, the $\sigma_{r\, W2\, scatter}$ is lower), with higher luminosity bins having no significant trend, 2) On average the dispersion of relation is higher than observational errors for all datasets. The dispersion of relation $\sigma_{r\, W2\, scatter}$ is comparable with values of $\sigma_i$. In our opinion two proposed tests give similar results.

\begin{figure}[!htp]
  \resizebox{\hsize}{!}{\includegraphics{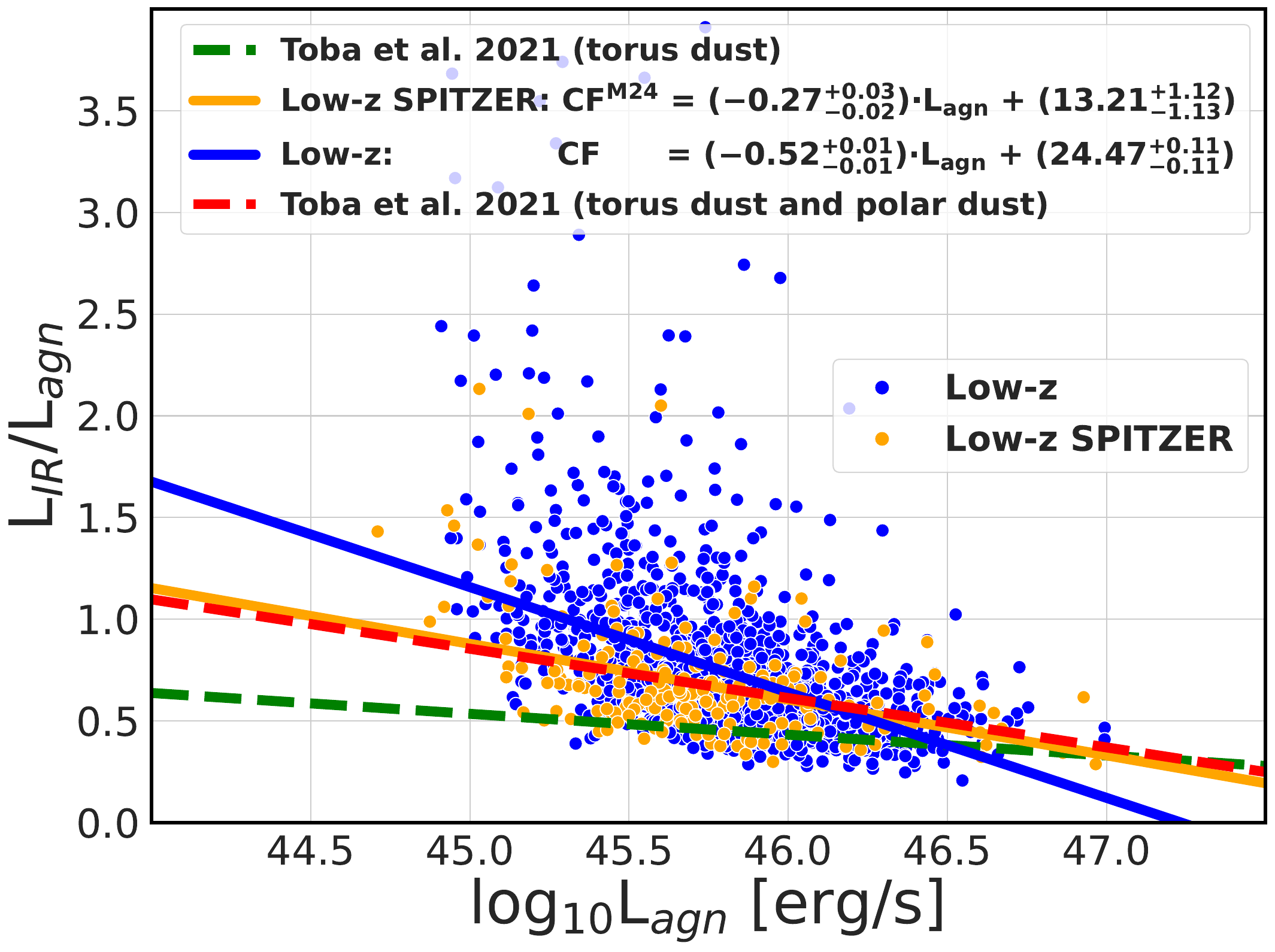}}
  \caption{Regression analysis results for the covering factor and bolometric luminosity for the Low-$z$ quasars with SNR$_{\rm W3\&W4}>5$. Blue and orange lines are the same as} in Figure\,\ref{fig:Regression_Lir_Lagn}.
  Green line corresponds to the regression from \citet{toba2021} fitted to the model with torus dust only, while red line is the sum of regressions fitted to the torus and polar dust models separately.
  \label{fig:Toba}
\end{figure}

\begin{figure}[!htp]

  \includegraphics[clip,width=1\columnwidth]{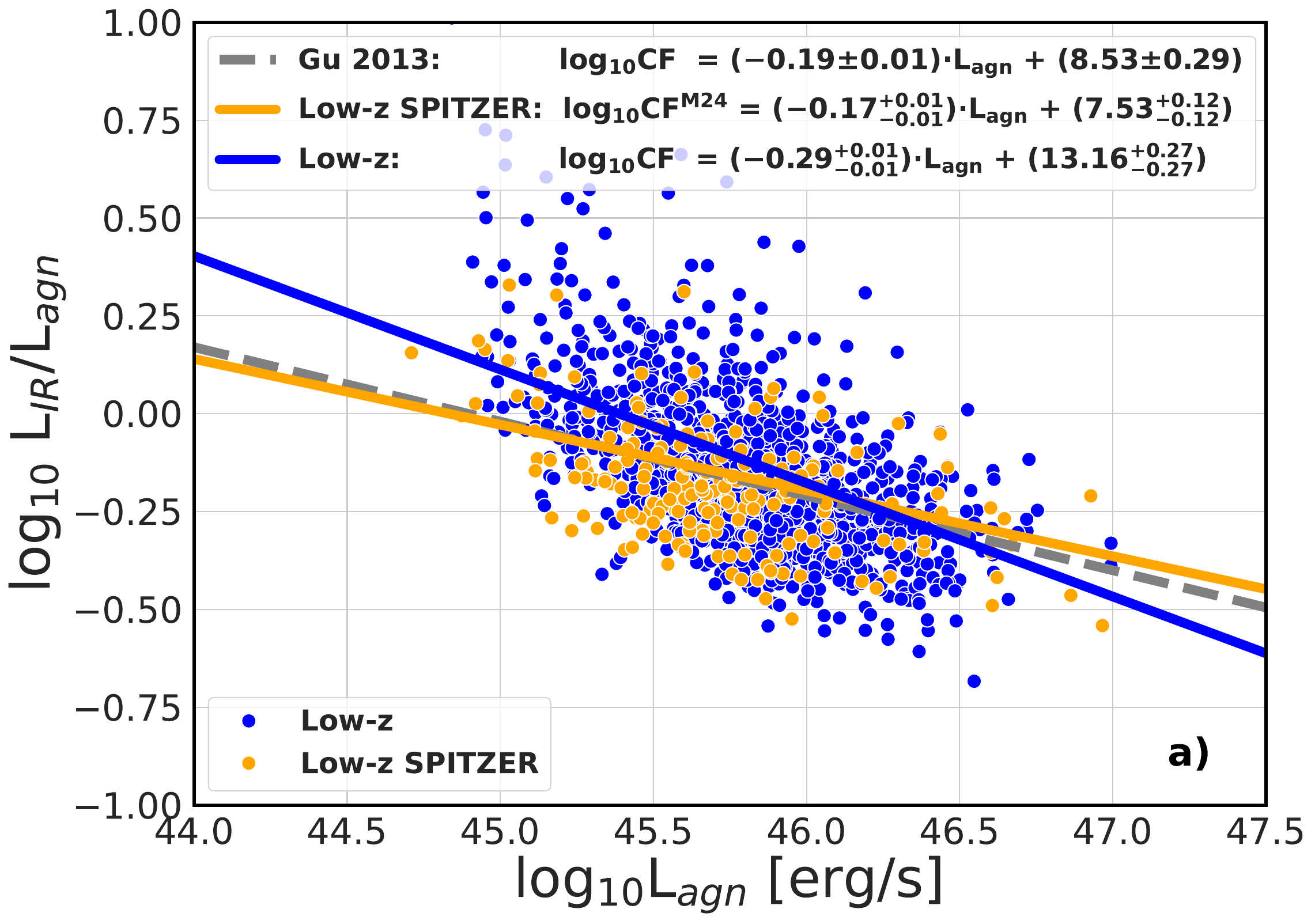}%

  \includegraphics[clip,width=1\columnwidth]{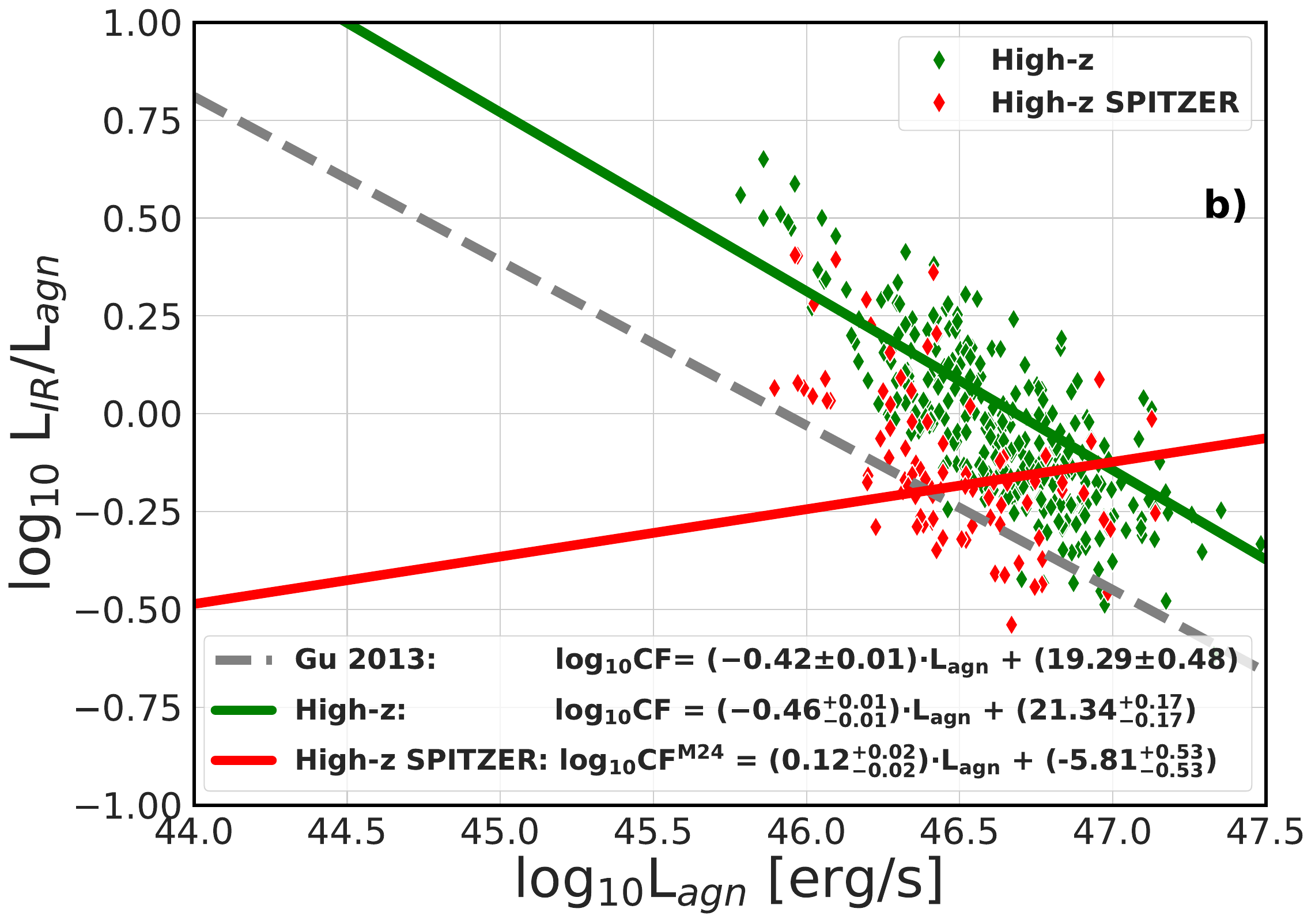}%

  \caption{Relation between $\log$\,CF and $\log L_{\rm agn}$ for the Low-$z$ and High-$z$ sources with SNR$_{\rm W3\&W4}>5$ (upper and lower panels, respectively). Regression lines in both panels fitted to our various sub-samples, with and without the SPITZER data, along with the \cite{2013ApJ...773..176G} regression lines for comparison, are described in the legends of the two panels.}
  \label{fig:Gu}
\end{figure}

\section{Discussion}
\label{sec:Discussion}

Our detailed analysis has confirmed that calculating quasar infrared and bolometric luminosities by integrating directly all available photometric data points is more accurate than when using a power-law approximation. (Figures\,\ref{fig:Lir_Lbol_both},\,\ref{fig:LirvsLir_LbolvsLagn}). This finding leads to an improved correlation between $L_{\rm agn}$ and $L_{\rm IR}$, as presented in Figure\,\ref{fig:Lir_Lbol_both}, especially in the High-$z$ sample, 
for which the power-law method underestimates the $L_{\rm agn}$ values and over-estimates $L_{\rm IR}$. The integral method also reduces the $L_{\rm IR}$ dispersion in the low-luminosity segment of the High-$z$ distribution, 
and improves the consistency between the Low-$z$ and High-$z$ samples in the overlapping luminosity range. 

The bias which remains is caused by the accuracy of the IR photometry. This was tested for the least accurate filter W4, in particular by comparing it with the SPITZER MIPS 24\,$\mu$m, whenever possible. In many cases, SPITZER photometry gave systematically lower fluxes than those measured in WISE W4. 
Furthermore, the SPITZER data have a smaller overall spread (Figure\,\ref{fig:Lir_Lbol_lamaniec_Spitzer}, top panel). Imposing data quality cut on the W3 and W4 filters by setting SNR\,$>5$, improves the situation in this respect, by tightening the luminosity-luminosity correlations (see Figures\,\ref{fig:SNR5} and \ref{fig:Appendix_M24_W4}, \ref{fig:W4vsM24}). Although the much reduced number of sources with the available M24 data limits the scope of the analysis, the medians of the primary parameters in the SPITZER-detected quasars are more representative for the entire sample than medians calculated with the high-quality WISE data (SNR$_{\rm W3\&W4}>5$), as given in Table\,\ref{tab:medians}.


%

We also checked variance vs. error for $L_{\rm agn}$ and $L_{\rm IR}$ following the method by \cite{Risaliti_Lusso2015}. It seems that the lowest variance is in the medium error bin. This means that the lowest spread is present near the middle of the distribution. Simply the medium luminosities have also the highest density of members in the bin.

Regarding the analyzed samples including either WISE or SPITZER data, the major differences follow from WISE W4 vs SPITZER MIPS 24\,$\mu$m photometry, as discussed in detail in Appendix\,\ref{sec:SED_comparison}. We believe that those differences are caused by instrumental effects rather than source variability. In particular, there is a systematic difference in the distribution of data points on the $\log L_{\rm agn} -\log L_{\rm IR}$ diagram between the ``W4'' and ``MIPS24'' samples, with higher-luminosity sources being underrepresented in the SPITZER data.

Our main point of interest was to investigate the distribution of the covering factor of the dusty torus within the quasar population. For that purpose, we adopted a simple definition of the CF parameter as the $L_{\rm IR}$ to $L_{\rm agn}$ ratio. 
When we consider the entire sample of sources binned in $\log {\rm CF}$ and $\log L_{\rm agn}$, the bins above 45.5 in $\log L_{\rm agn}$ form a horizontal sequence with $\log CF \simeq -0.2$ (see Figure\,\ref{fig:SNR5ab} panel a). The difference between the WISE and SPITZER data emerges here only at the lowest luminosity bins; for those, the median CF values are in general higher, and noticeably divergent (although still within the errors). On the other hand, in the $\log {\rm CF}-\log L_{\rm IR}$ binned distribution, we see constant value $\log CF \simeq -0.2$ throughout the entire luminosity range.

When considering separately the Low-$z$ and High-$z$ samples, in both cases we observe an anti-correlation between $\log {\rm CF}$ and $\log L_{\rm agn}$. The low-luminosity tail of the Low-$z$ sample is however sensitive to the adopted instrument at 24\,$\mu$m: the slope coefficient of the line fitted to the W4 data is almost twice smaller than that obtained with the SPITZER data. In Figure\,\ref{fig:Toba}, we compare our regression results in this respect with the results by \cite{toba2021}, who performed the broad-band quasar SED fitting including the dusty torus and polar dust. In their analysis, the CF of the torus exclusively turns out weakly dependent on $L_{\rm agn}$, but when the polar dust is added as the SED model component, the anti-correlation becomes steeper and, in fact, basically the same as observed in our Low-$z$ SPITZER sub-sample. 
This may suggest that the integrated IR quasar luminosities, as considered in our work, are dominated by the torus in the high-luminosity regime, and are affected by a significant contamination from the polar dust in the low-luminosity Low-$z$ regime.

Somewhat puzzling, in this context, is that the CF in the High-$z$ sample alone also decreases with $L_{\rm agn}$. This is well visible in Figures\,\ref{fig:SNR5ab} and \ref{fig:Gu}. An interesting finding here is that, with the SPITZER data, we obtain different slopes of the regression lines for both the Low-$z$ and High-$z$ sub-samples ($-0.17 \pm 0.01$ and $0.12 \pm 0.01$, respectively; see Figure\,\ref{fig:Gu}). The Low-$z$ relation can be accounted for, as argued above, by the polar dust contamination, with the $\log {\rm CF}$ dropping from 0 to --0.25 within the $\log  L_{\rm agn}$ range from 45 to 46, while in the High-$z$ sources, same drop occurs on $\log L_{\rm agn}$ from 46 to 47. 
It is important to note that, unlike Low-$z$, the High-$z$ fits for luminosity ratios using different methods give inconsistent results. It is very well visible in Fig. \ref{fig:Gu} and \ref{fig:Bayes_comparison}, where W4 vs. M24 data are used or not-weighted vs. error-weighted regression is used. This can suggest that the data quality and distribution do not allow for good constrain on the slope parameter. Thus it is uncertain what physical behavior is present in the high-z sample CF.


\cite{Ricci2017} and \cite{ricci2022} pointed out that the column density of material in the torus or material above the torus anticorrelates with the Eddington ratio. In their radiation-regulated unification scenario in low-accretion rate AGN ($\lambda_{\rm Edd} \geq -2$) the solid angle is larger suggesting thicker tori. The study focused on constraining X-ray absorption in the local low-luminosity AGN population which is well-represented in the Swift's Burst Alert Telescope (BAT) catalog. This finding is generally compatible with studies by other authors based on SED fitting. For example \cite{toba2021} found an anticorrelation of the covering factor with AGN luminosity and pointed out that polar dust emission adds up to the torus emission leading to elevated CF values. On the other hand study by \cite{yamada2023} for a specific sample of local U/LIRGs (Ultra Luminous Infra Red Galaxy) shows that luminosities of the torus and polar dust components are correlated with each other and with the Eddington ratio. However the ratio of $L_{\rm torus} / L_{\rm agn}$ and $L_{\rm polar} / L_{\rm agn}$ are not correlated with the Eddington ratio.
Those studies were done for local AGNs. How they scale up to High-$z$ sources and higher luminosity in that matter is an open question. \cite{ricci2022} 
noted that, due to a higher amount of ISM/host dust/obscuring material and mergers, the bulk of the dust-related absorption in High-$z$ sources may take place outside of SMBH sphere of influence, and as such should rather depend on $L_{\rm agn}$ rather than $\lambda_{Edd}$.

As we demonstrated when investigating the difference between W4 and MIPS24 flux measurements, the IR data accuracy is a particularly important factor, which may affect the main conclusions of the analysis. As discussed in detail in Appendix\,\ref{sec:results_spitzer_check}, the low-luminosity tails of both the Low-$z$ and High-$z$ samples, diverge from the overall $L_{\rm agn}-L_{\rm IR}$ correlation.


In the gathered sample of quasars, we do see significant bolometric and infrared luminosity evolution with redshift \citep[in this context see, e.g.,][]{Lusso2013}. These evolution manifest as a systematic difference between the Low-$z$ and High-$z$ luminosity values. The luminosity evolution is also formally confirmed by the Efron-Petrosian test.
However, if we limit the SPITZER sample to the higher SMBH masses (over $10^{8.5}M_{\odot}$) in both Low-$z$ and High-$z$ ranges, then we do not see any CF\,$\equiv L_{\rm IR}/L_{\rm agn}$ evolution with redshift (see Figure\,\ref{fig:MBH_cut}), and this observation is confirmed with the Kolmogorov-Smirnov test. The scenario with not-evolving CF means that the dusty torus properties are universal across a wide redshift range, at least in the high SMBH mass regime. This result stands, within the errors, with the selection of quasars with 1) similar $L_{\rm agn}$, 2) similar $\lambda_{\rm Edd}$, as described in Appendix \ref{sec:other_cuts}.

The CF distribution was compared with two previous works \citep{2013ApJ...773..176G, toba2021}. In Figure\,\ref{fig:Toba} we show the relationship between the CF and $L_{\rm agn}$ for the Low-$z$ subset exclusively, with the regression line fitted to our data, and along with the \cite{toba2021} regression lines. \citeauthor{toba2021} used the Bayesian approach, based on the likelihood function from the work by \cite{Kelly2007}, which uses error weights. The two different regression lines stand for different models used by \cite{toba2021} to calculate the luminosities: the torus dust (green line) is a simple basic model, while the torus and polar dust model (red line) also takes into account the dust placed in polar areas of an AGN.

In Figure\,\ref{fig:Gu} we compare our results regarding the $\log\,{\rm CF}-\log\,L_{\rm agn}$ anti-correlation with those of \cite{2013ApJ...773..176G}. It should be noted that \cite{2013ApJ...773..176G} used the power-law method of integrating the luminosities. The differences between the estimations were described in Section \ref{sec:integration_comparison}. The overall trend is similar for the Low-$z$ quasars, with the slope in both regressions having only a minor difference, namely $-0.29\pm0.01$ in our work and $-0.19\pm0.01$ in \cite{2013ApJ...773..176G}). 
For the High-$z$ objects, the slopes emerging from our work and \citeauthor{2013ApJ...773..176G}, are again similar ($-0.46\pm0.01$ and $-0.42\pm0.01$, respectively). The difference becomes however more pronounced when only the High-$z$ SPITZER subsample is taken into  account (slope +0.12). In Figure \ref{fig:Regression_Lir_Lagn} the SPITZER subsample is in a better agreement with the \cite{2013ApJ...773..176G} correlations, with the Low-$z$ slope of $0.82\pm0.01$, and the High-$z$ slope of $0.74\pm0.01$. For a more in-depth comparison with the work by \cite{2013ApJ...773..176G}, see Appendix\,\ref{Gu:data}.

The CF value translates to the torus viewing angle in the type\,1 AGN population. Possible effects of the viewing angle evolution are crucial for the quasar-based cosmology \citep{Prince2021}. Our ``no CF evolution'' result is in agreement with the studies of the torus viewing angle evolution \citep{Prince2022}, although observational biases and systematics induced by applied methods must be further addressed. This result is, on the other hand, in disagreement with \cite{2013ApJ...773..176G}, who argued for an evolution of the CF with redshift. However, as we pointed out in this study, our improvement in deriving $L_{\rm IR}$ and $L_{\rm agn}$ leads to systematic differences in luminosity estimates in comparison to the \cite{2013ApJ...773..176G} method, and as a result also in the CF estimates.

To further improve the accuracy of our simple methodology we must further address several pending issues. As we have shown, using the W4 data, especially without a restricting quality cut, introduces a strong bias. SPITZER MIPS is a better choice in terms of data accuracy. Altogether, error-weighting is a good practice. We further describe those points in the Appendices of this article. Additionally, the low-luminosity tail of the Low-$z$ sample can be affected by an integration limit in $L_{\rm agn}$, due to hotter standard accretion disks around lower mass SMBH, and by an unknown intrinsic extinction. Both effects may cause an underestimation of the bolometric quasar luminosity. Also, further investigation of the data limitation in both the IR as well as the optical/UV ranges may be needed. Those will be subjects of further studies.






\section{Conclusions}
\label{sec:Conclusions}

In the gathered sample of quasars with the available IR--to--UV data, we see significant luminosity evolutions with redshift, but no significant scaling of the covering factor of the dusty torus --- defined as the IR to bolometric luminosity ratio --- with luminosities. That is, the Low-$z$ and High-$z$ sources follow a similar correlation between the IR and bolometric luminosities. Neither do we observe any covering factor evolution with redshift, at least in the regime of high SMBH masses, $M_{\rm BH} > 10^{8.5}M_{\odot}$, and when using more accurate SPITZER data. On the other hand, the low-luminosity tail of the Low-$z$ sample breaks the trend set by the high-mass segment. This is in line with the work by \cite{toba2021}, who proved strong polar dust contamination in the covering factor in the lower luminosity range of the Low-$z$ quasar population. Surprisingly, the High-$z$ sample alone shows also a similar kind of dependence but in a higher-luminosity regime, although it is rather caused by limited data quality. To verify if this effect is physical, we need to further test our methodology and data accuracy, while keeping in mind how the WISE W4 fluxes differ from the SPITZER MIPS 24\,$\mu$m data. 
This is because any potential scaling of the covering factor with luminosity should be robustly identified and calibrated for accurate quasar-based cosmology.


 \begin{acknowledgements}
 
We thank the anonymous referee for their remarks. We thank Bo\.zena Czerny for helpful discussions. This research was supported by the Polish National Science Center grants 2018/30/M/ST9/00757 and by Polish Ministry of Science and Higher Education grant DIR/WK/2018/12. {\L}.S. was supported by the Polish NSC grant 2016/22/E/ST9/00061. This research was funded by the Priority Research Area Digiworld under the program Excellence Initiative - Research University at the Jagiellonian University in Kraków.
 \end{acknowledgements}

\bibliographystyle{aa-note} 
\bibliography{ref} 

\begin{appendix}

\section{WISE W4 and MIPS M24 Datasets Comparison}
\label{sec:results_spitzer_check}

\subsection{Data Quality Comparison}
Here we address a calibration issue known as the ``redleak'', using the correction method described by \cite{redleak}. This problem arises from the dependence of the filter correction factors on the slope of an SED. While most of the WISE calibration was conducted with the use of stars, which appear ``blue'' in the IR band, galaxies due to the cold dust emission exhibit different SED slopes and appear ``red''. Consequently, applying a standard correction to extragalactic sources may underestimate W3 and overestimate W4 by about 9\% \citep{wright2010}. However, in our case, accounting for the redleak correction had a minimal impact.

\begin{figure}[!htp]

  \includegraphics[clip,width=1\columnwidth]{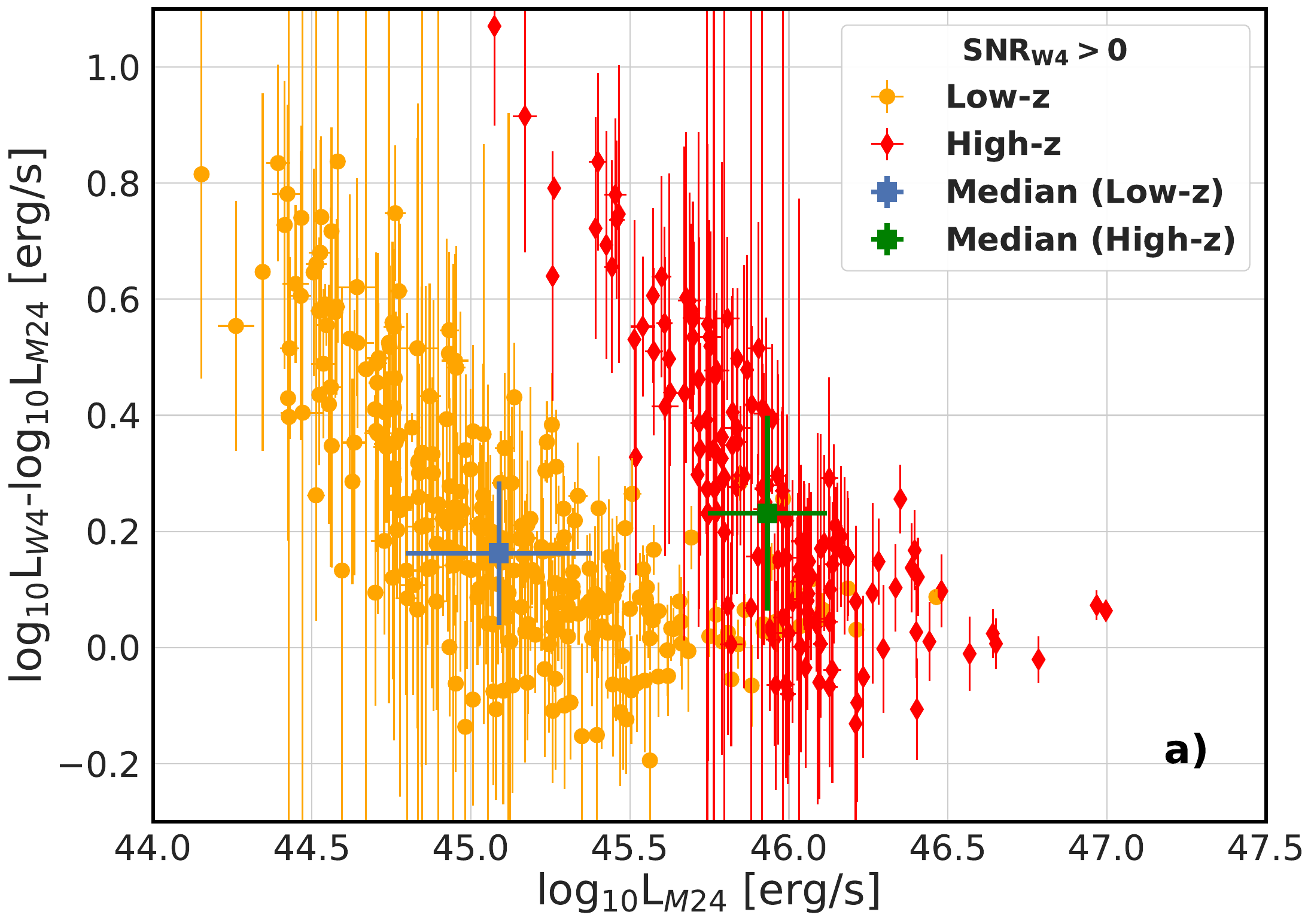}
  \includegraphics[clip,width=1\columnwidth]{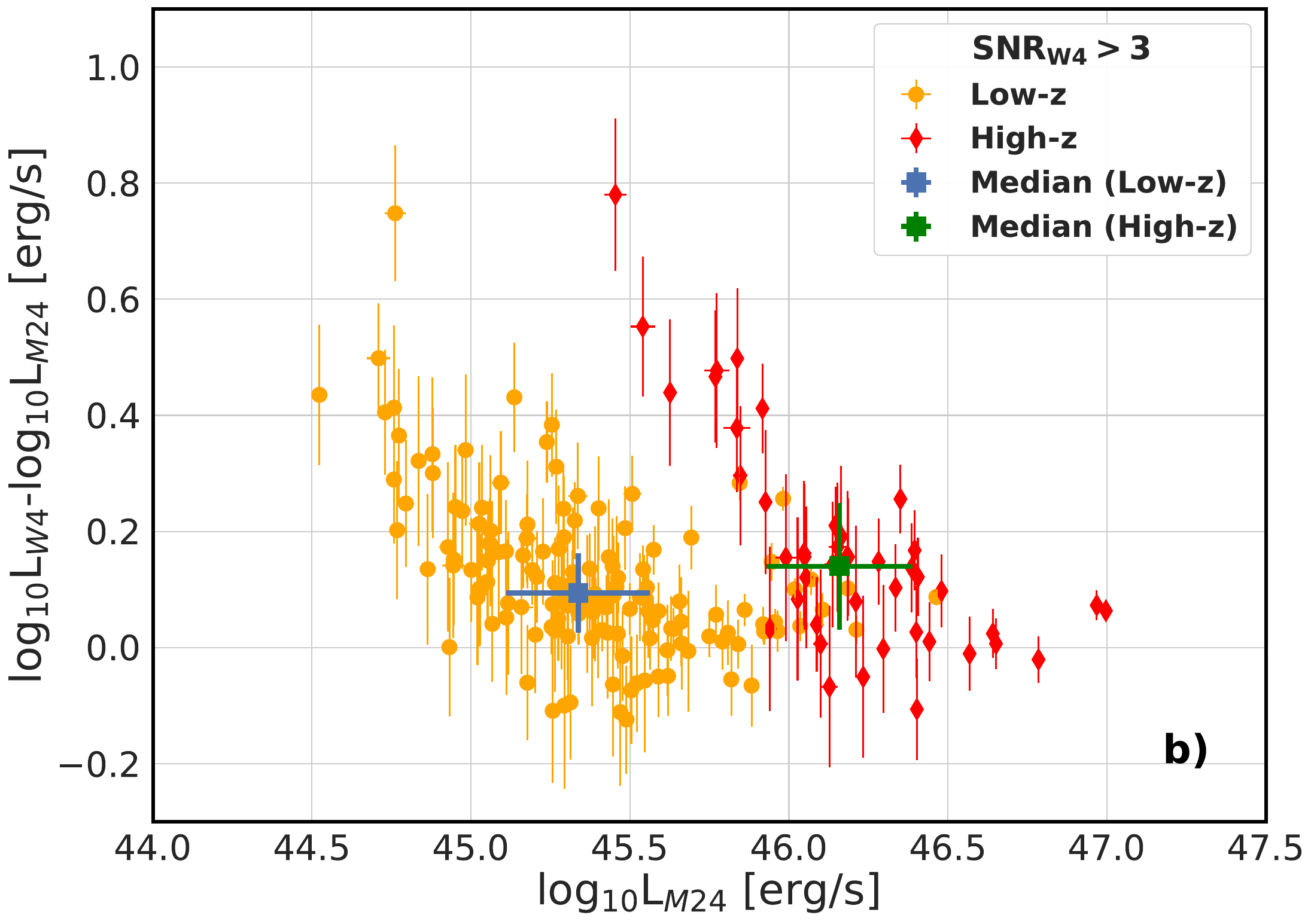}
    \includegraphics[clip,width=1\columnwidth]{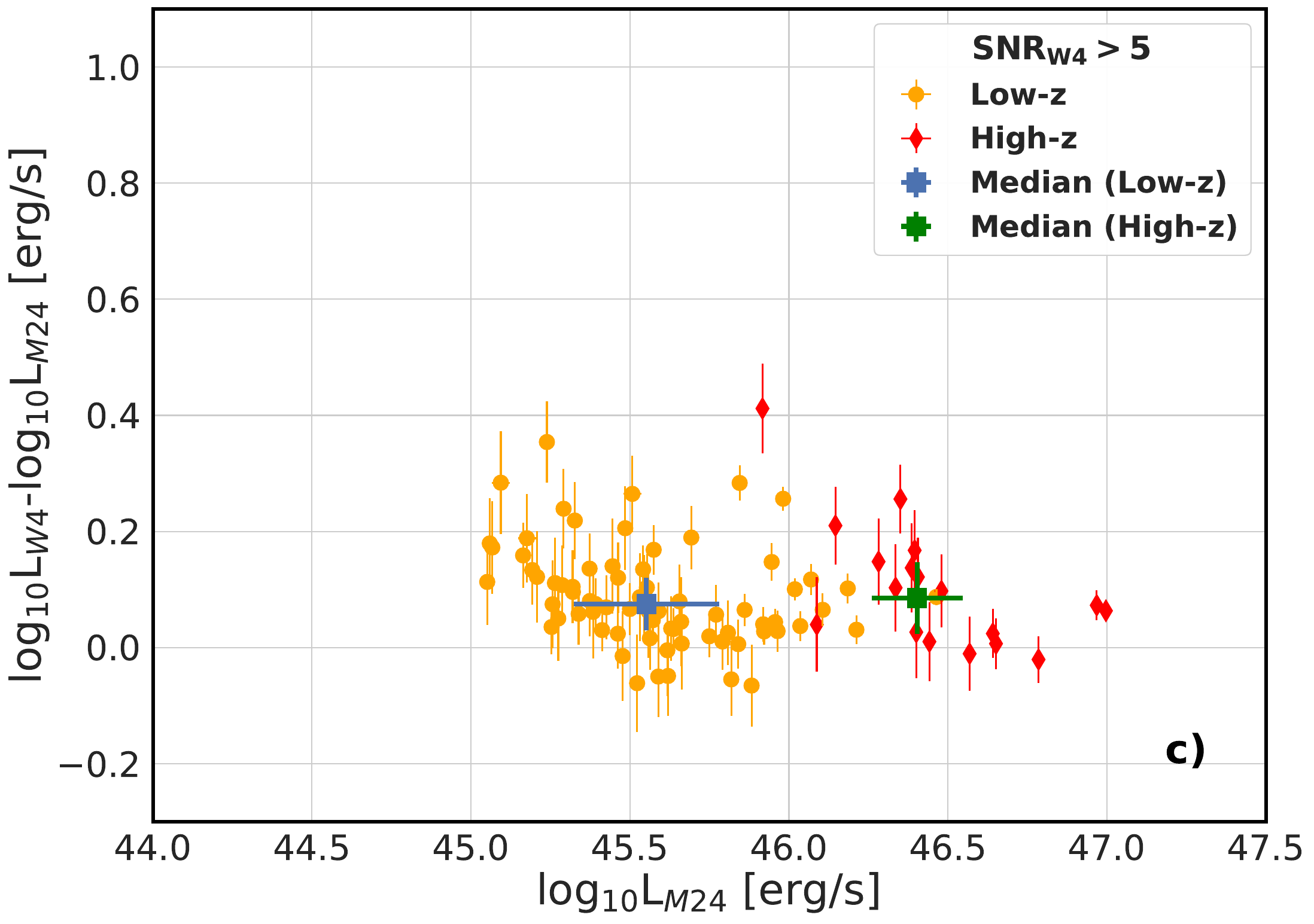}    

      \caption{Comparison of data quality for the M24 and W4 with different restrictions on SNR$_{\rm W4}$: panels from top to bottom presenting $\log\,L_{\rm W4}-\log\,L_{M24}$ vs $\log\,L_{M24}$ with SNR$_{\rm W4}>0$, $>3$, and $>5$, respectively. Orange circles mark the Low-$z$ quasars, while red diamonds the High-$z$ sources.}
  \label{fig:Appendix_M24_W4}
\end{figure}

\begin{figure}[!htp]

  \includegraphics[clip,width=1\columnwidth]{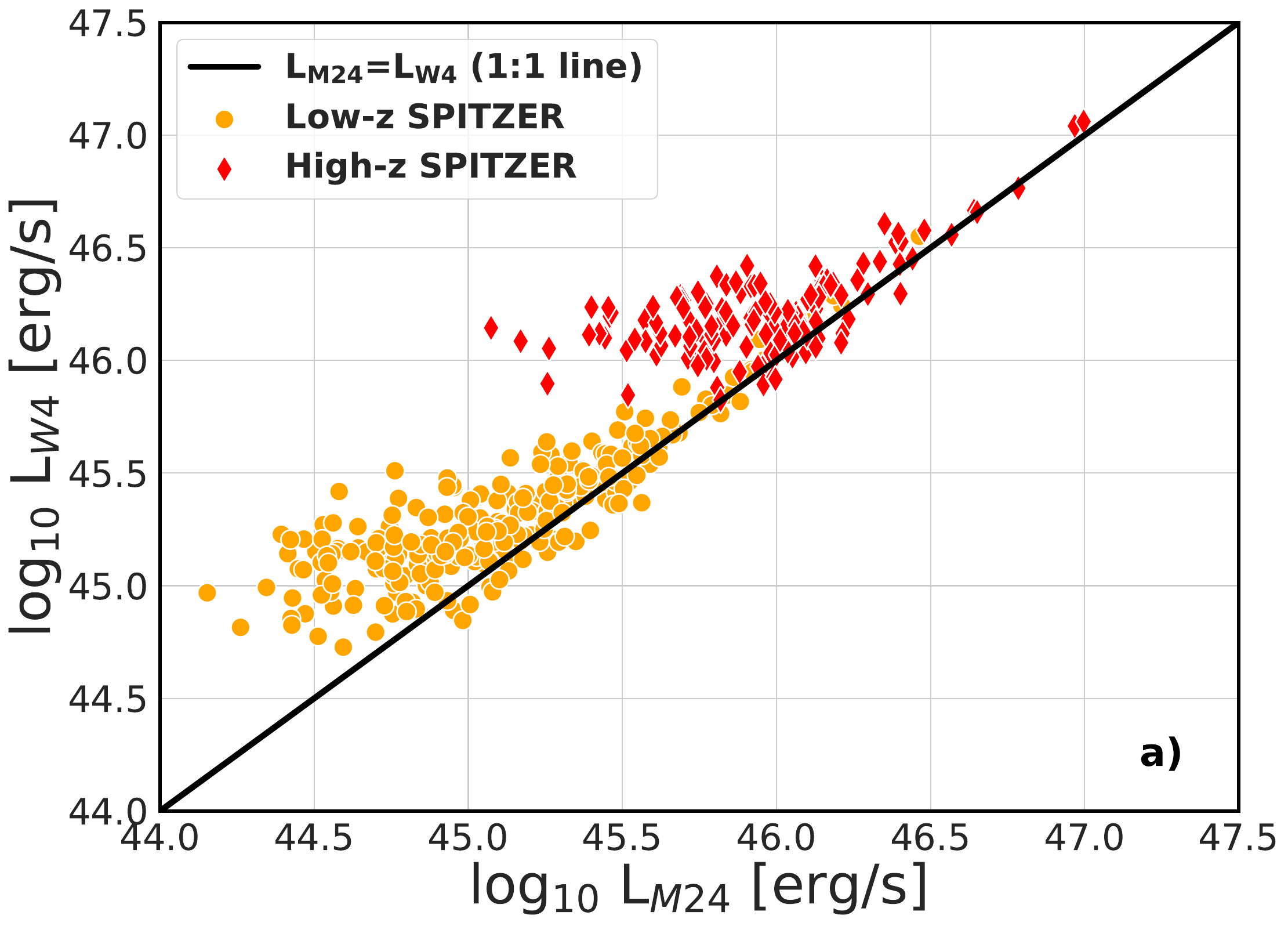}

\vspace{0.01cm}
  \includegraphics[clip,width=1\columnwidth]{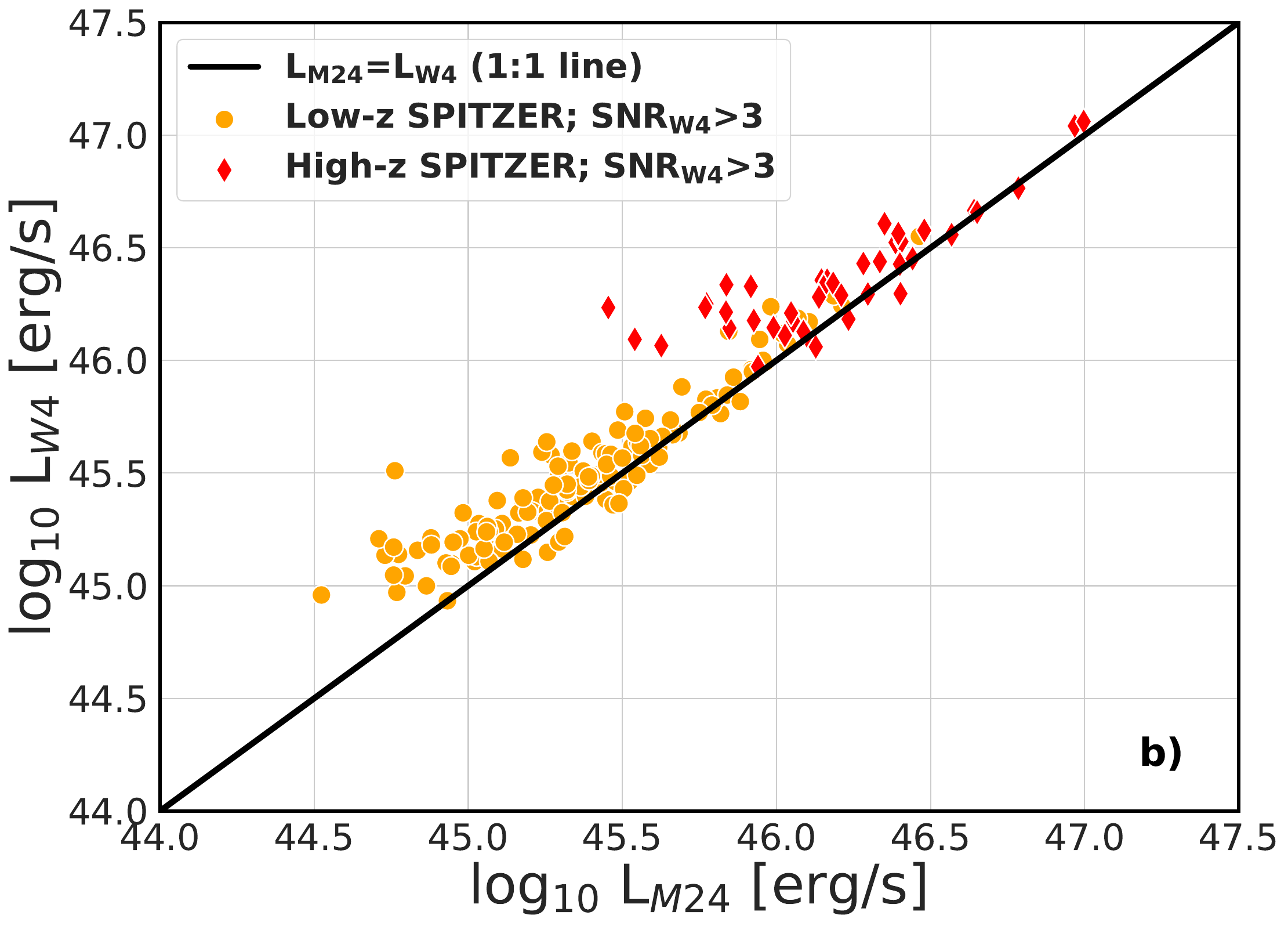}%

  \includegraphics[clip,width=1\columnwidth]{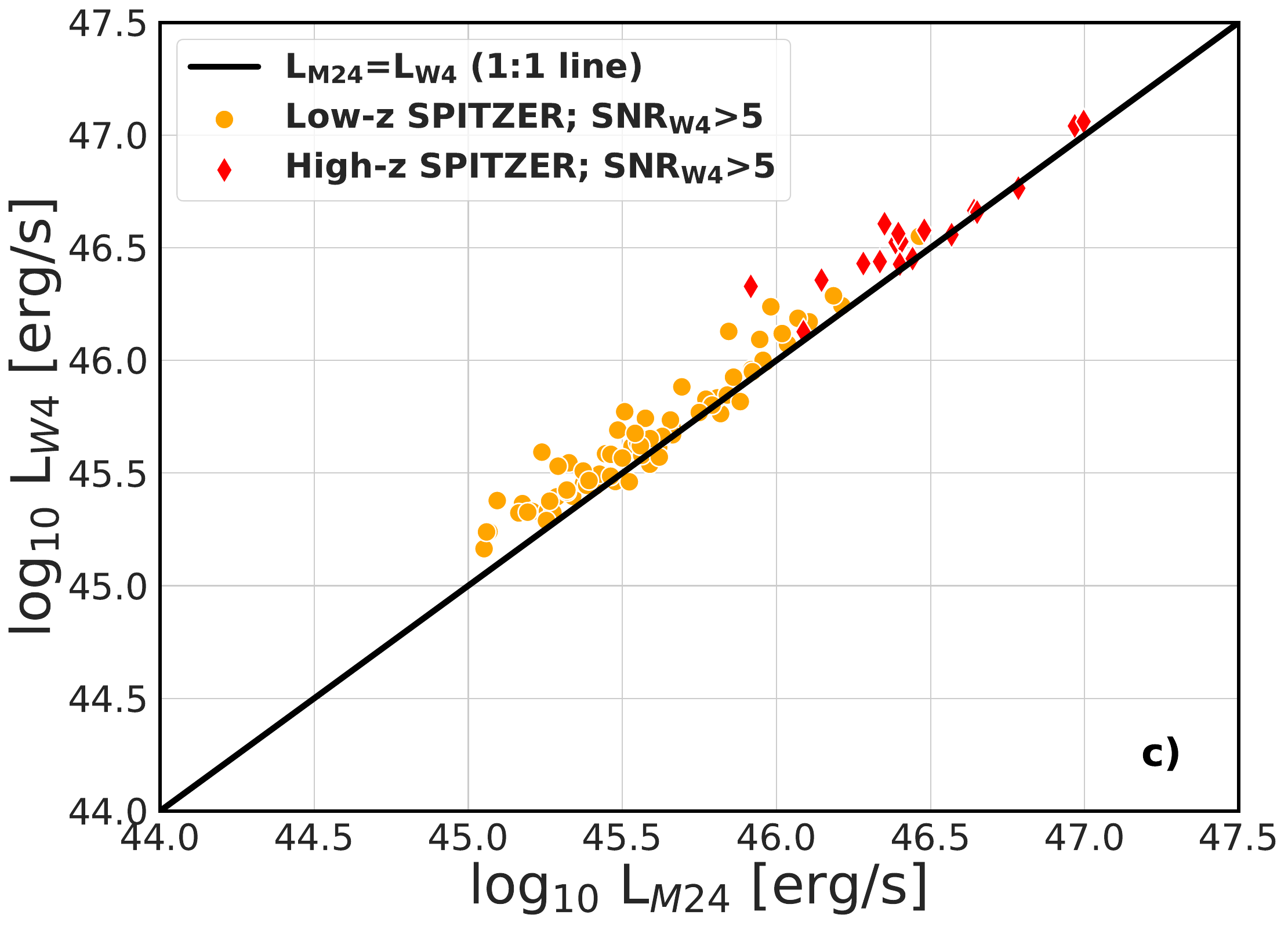}
 
\caption{Comparison between monochromatic WISE W4 and SPITZER M24 luminosities for both Low-$z$ and High-$z$ sources. The black line in the panel shows the 1:1 relation. Orange circles and red diamonds indicate Low-$z$ and High-$z$ quasars with SPITZER M24 data. Panels from top to bottom correspond to SNR$_{\rm W4}>0$, $>3$, and $>5$, respectively.}

\label{fig:W4vsM24}
\end{figure}

To validate the accuracy of the WISE W4 observations, SPITZER M24 data were examined. We conducted a cross-match using data from \cite{Spitzer_catalog2021} with a cross-match radius of 3\,arcsec. The data were available only for 290 objects in the Low-$z$ sample, and 157 objects in the High-$z$ sample. The SPITZER data are shown in Figures\,\ref{fig:Lir_Lbol_lamaniec_Spitzer} and \ref{fig:SNR3}. The relation $L_{\rm IR}$ vs. $L_{\rm agn}$ shown in Figure\,\ref{fig:Lir_Lbol_lamaniec_Spitzer} top panel, is characterized by a much reduced dispersion, but follows a similar, roughly linear trend. Figure\,\ref{fig:Lir_Lbol_lamaniec_Spitzer} middle panel shows a slightly different relation for both redshift samples, with SPITZER data generally having a lower variance in $\log$\,CF. Notably the $\log$\,CF values of Low-$z$ and High-$z$ sources are much closer to the M24-based $L_{\rm IR}$, as can be seen in Table\,\ref{tab:medians}. The change in the amount of SPITZER data points between Figure\,\ref{fig:Lir_Lbol_lamaniec_Spitzer} and Figure\,\ref{fig:SNR3} is due to the SNR$_{\rm W3}>3$ criterion. The resultant data sets of the SPITZER include 239 Low-$z$ sources and 122 High-$z$ targets. In our analysis, we carefully examined the difference in luminosity estimates based on the W4 and M24. Even with SNR$_{\rm W3\&W4}>5$, there is still some discrepancy between each monochromatic luminosities, with W4 tending to overestimate the luminosity values compared to M24 (see Figure\,\ref{fig:W4vsM24}, lower panel).

Figure\,\ref{fig:W4vsM24} top panel shows the comparison between the monochromatic luminosities of W4 and M24, both calculated with the all-points method. Notably, there is a clear deviation from the 1:1 line at lower luminosities, in both redshift ranges. This discrepancy between W4 and M24 diminishes as luminosities increase. The major deviation from the 1:1 line signals significant issues with the photometry measurements in one of the bands. This issue remains even in the case of the best-quality W4 measurements with the SNR$_{\rm W4}>3$, as shown in the bottom panel of Figure\,\ref{fig:W4vsM24}. We conducted additional test on the quality of both W4 and M24 by directly comparing photometry observations from maps with the determination of measured flux, and comparing it with the values from the archive. In general, the W4 tends to extract fluxes from larger areas on the sky when compared with higher angular-resolution M24 ($2.55''$ vs $12''$).

\begin{figure*}[!htp]
  \includegraphics[clip,width=1\linewidth]{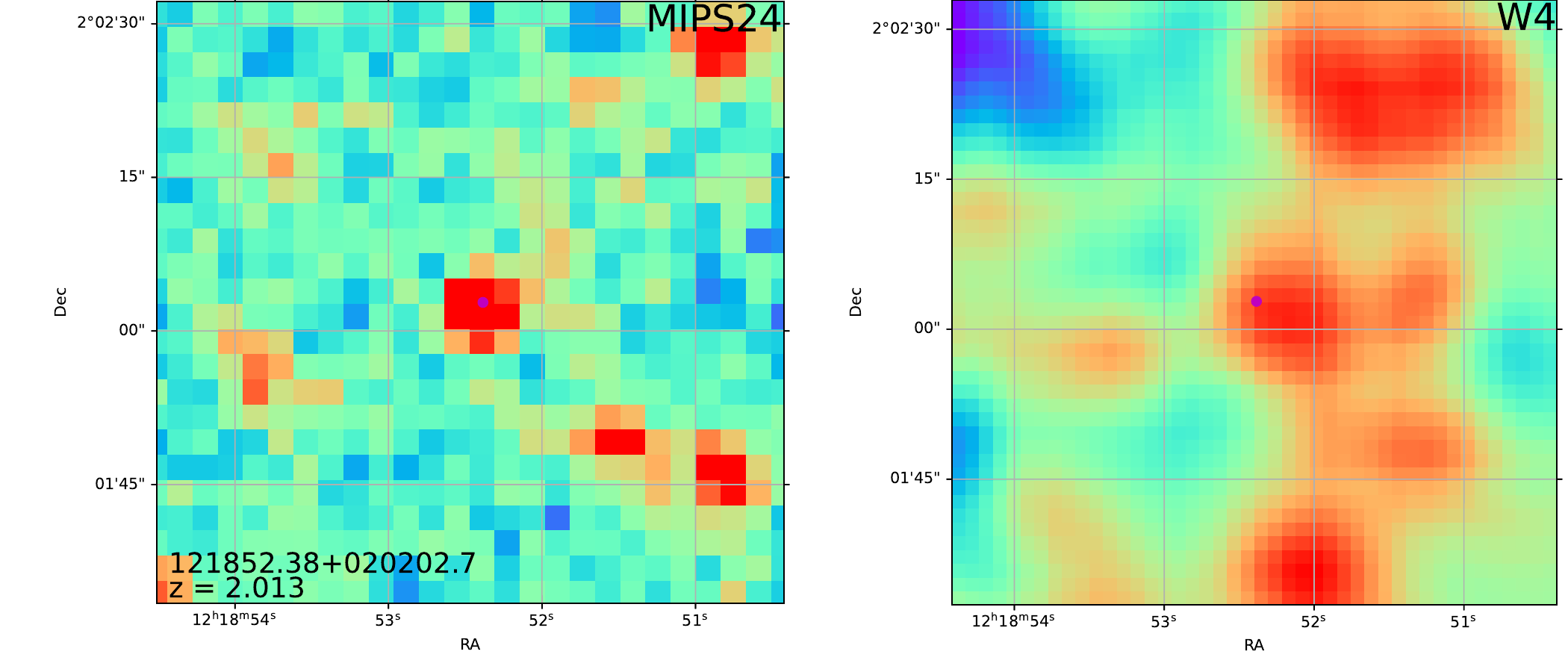} \newline
  \includegraphics[clip,width=1\linewidth]{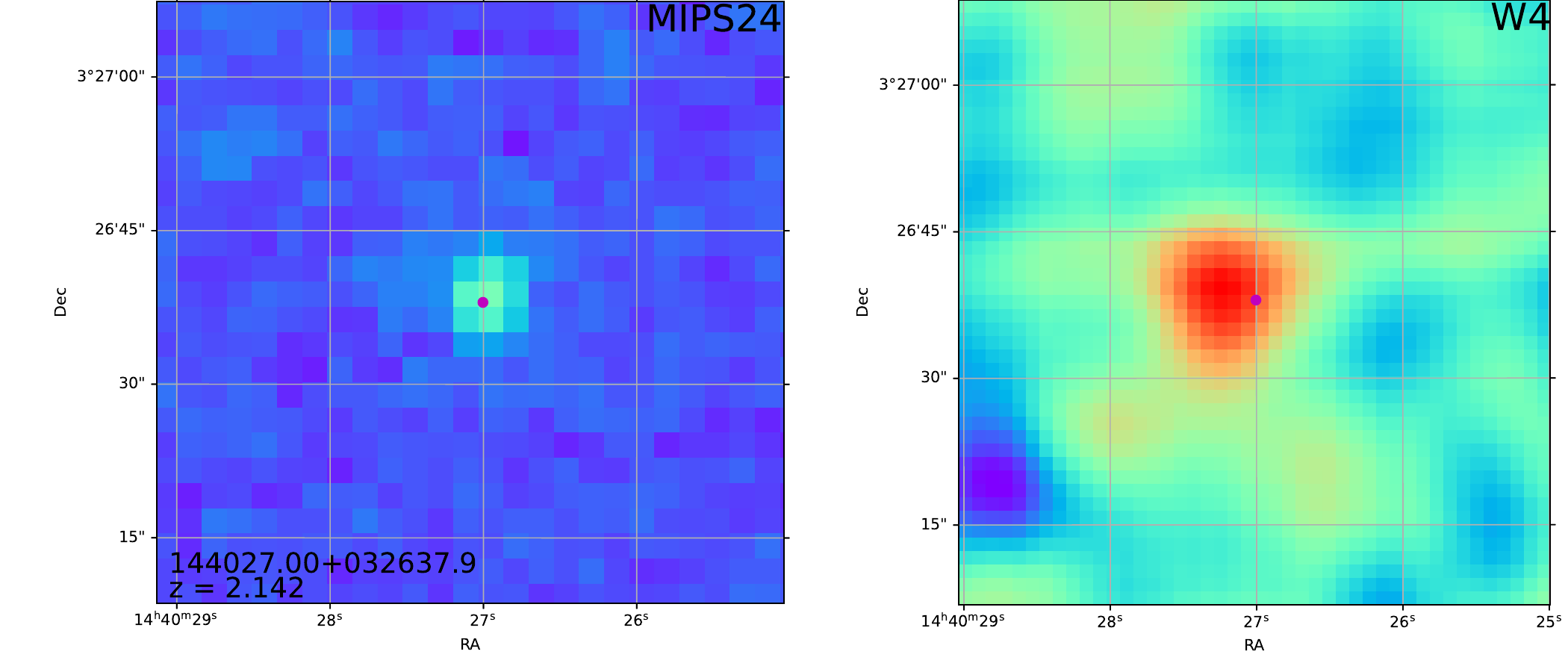}
  \caption{Sample images of the SPITZER MIPS 24\,$\mu$m and WISE W4 frames. The source position is marked with the magenta point at the center of each frame. Colors represent flux levels, arbitrarily in each frame. Violet/dark blue is assigned for low flux, while red shows the highest flux level.}
  \label{fig:Appendix_images}
\end{figure*}

SPITZER and WISE instruments differ in spatial resolution and point spread functions (PSFs). As we show in Figure\,\ref{fig:Appendix_images}, in some cases W4 blends neighboring sources.
Additionally, the wide field character of WISE increases the probability of a light pollution caused by a bright star in the field of view, or non-sidereal objects such as asteroids. Lower-luminosity sources are, obviously affected in this respect to a larger extents, as shown directly in Figure\,\ref{fig:Appendix_M24_W4} and \ref{fig:W4vsM24}.

\subsection{Spectral Energy Distribution}
\label{sec:SED_comparison}
\begin{figure}[!htp]

  \includegraphics[clip,width=1.\columnwidth]{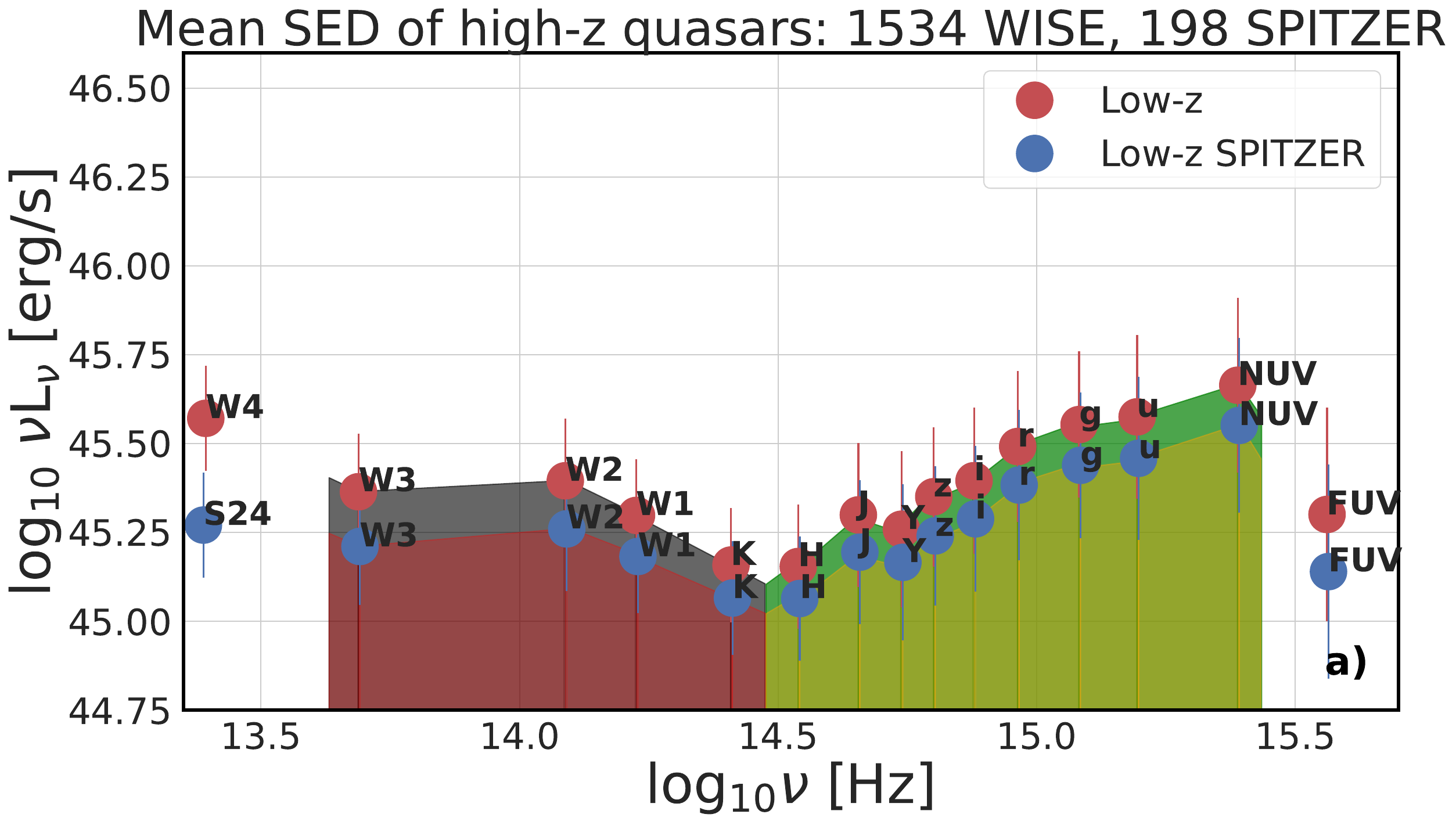}%

  \includegraphics[clip,width=1.\columnwidth]{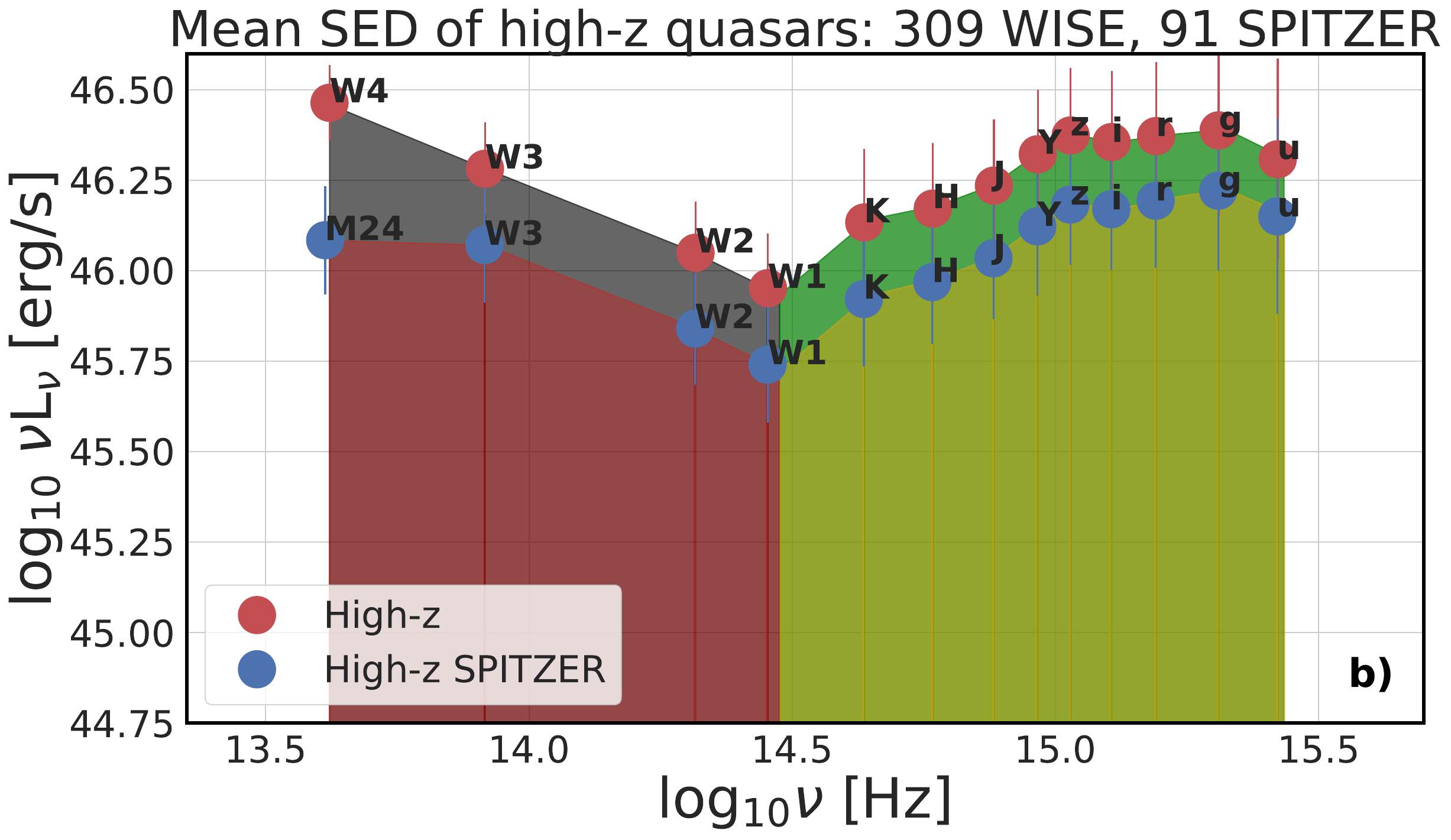}%

\caption{The mean rest-frame SEDs for the Low-$z$ and High-$z$ quasars (top and bottom panels, respectively). Red points represent data obtained solely from WISE IR (Low-$z$ top panel, and High-$z$ lower panel) and blue points have SPITZER M24 (Low-$z$ SPITZER top panel and High-$z$ SPITZER lower panel). Low-$z$ and High-$z$ shown data have SNR$_{\rm W3\&W4}>5$, while SPITZER datasets have SNR$_{\rm W3}>5$. The black and green colors indicate the areas used for the calculations of $L_{\rm IR}$ and $L_{\rm agn}$, respectively for both Low-$z$ and High-$z$. The red and yellow colors indicate the areas used for the calculations of $L_{\rm IR}$ and $L_{\rm agn}$, respectively for both Low-$z$ SPITZER and High-$z$ SPITZER. Each photometric observation is signed with the filter name. The errorbars are defined as the 1st and 3rd quartiles.}
\label{fig:SED_comp}
\end{figure}

To quantify the differences between WISE and SPITZER photometry even further, we compared the mean SEDs for Low-$z$, High-$z$ (SNR$_{W3\&W4}$>5) and Low-$z$ SPITZER, High-$z$ SPITZER (SNR$_{W3}$>5). The comparison is shown in Figure\,\ref{fig:SED_comp}. As described in the data selection Section\,\ref{sec:Data}, the main difference between data with and without SPITZER is the replacement of the WISE W4 filter with MIPS 24$\mu$m. It's worth noting that for both redshift bins SPITZER-based datasets have lower $L_{\rm IR}$ and $L_{\rm agn}$ luminosities (approximately 0.1 dex luminosity difference for Low-$z$ and 0.2 dex for High-$z$ samples). The difference in luminosity is even larger between W4 and M24 filters and is visible for both redshift bins. 
Generally, despite the systematic differences in luminosity, the average SEDs for W4 and M24 data exhibit very similar shapes. Similar SED shapes support the idea that average CF should be comparable and the apparent difference should be caused by the systematic difference between WISE and SPITZER photometry. It's worth noting that the SPITZER sample mostly contains sources of lower values of $L_{\rm IR}$ distribution of the WISE sample (with SNR$_{W3}$>5).

\section{Different Likelihood for Bayesian Fitting}
\label{sec:likelihood}

\begin{figure*}[!htp]

  \includegraphics[clip,width=1\columnwidth]{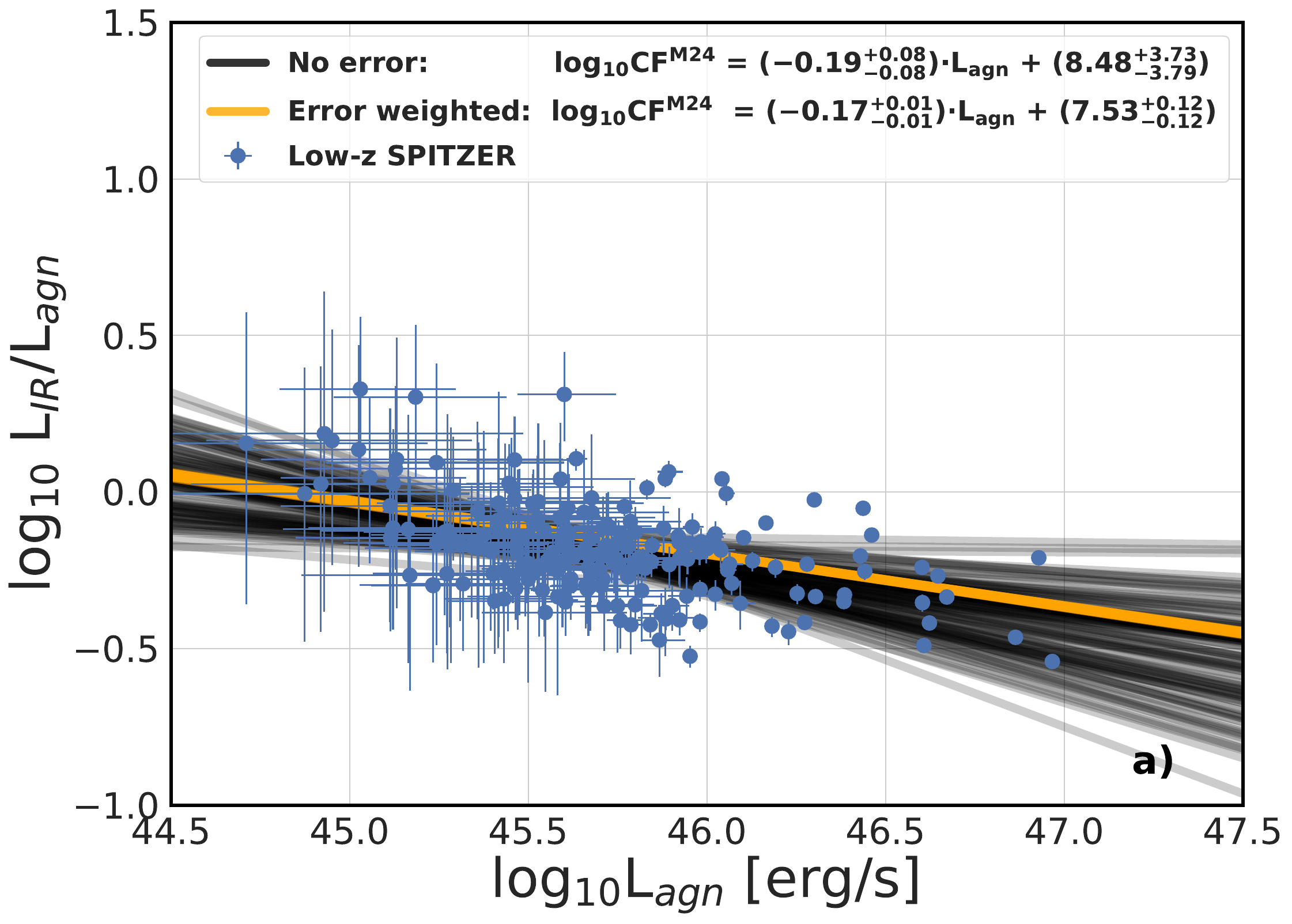}
  \includegraphics[clip,width=1\columnwidth]{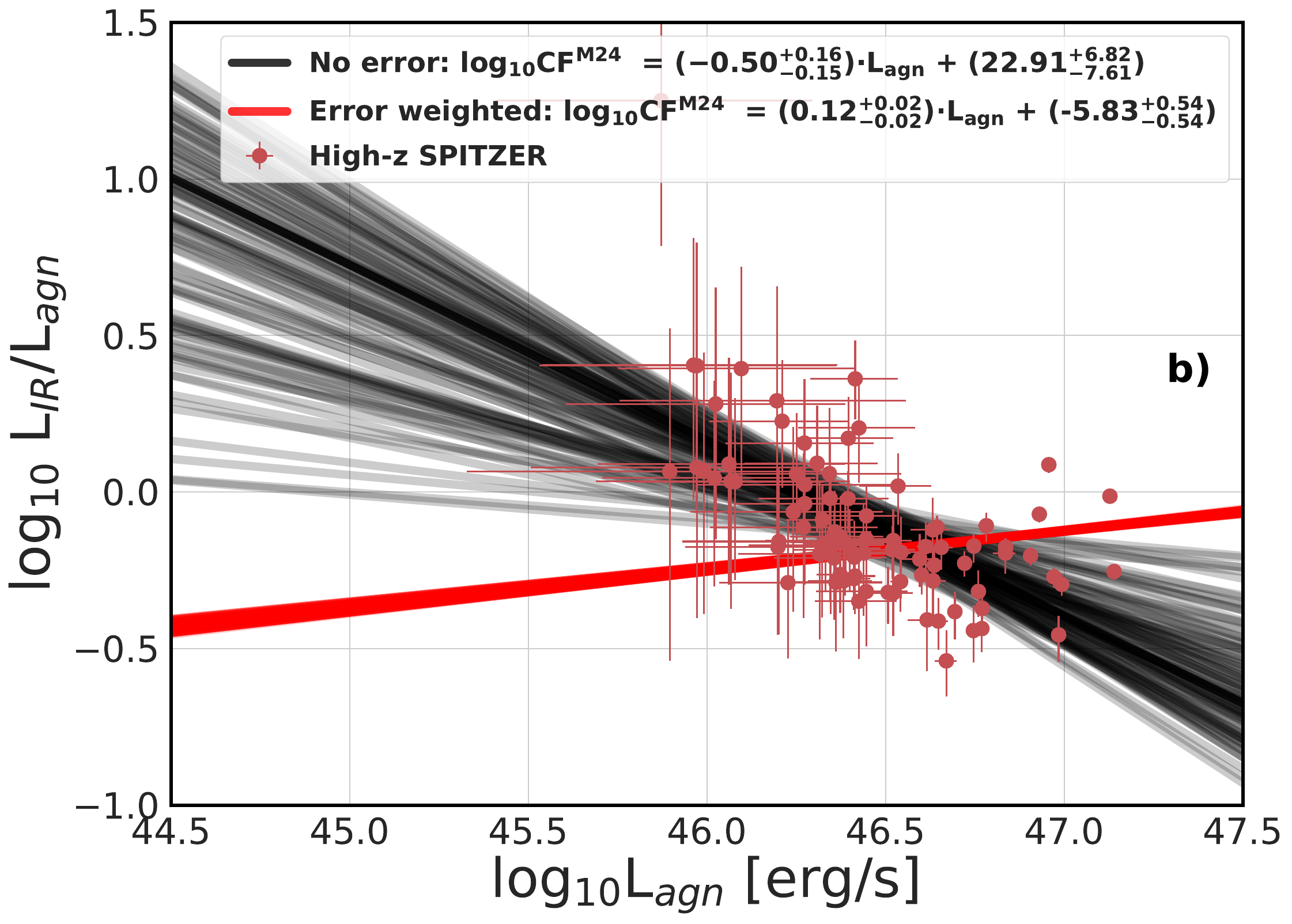}%

\caption{Comparison of two different likelihoods in Bayesian fitting methods for SPITZER data between $\log$\,CF vs $\log L_{\rm agn}$ for a) Low-$z$ sources, and b) High-$z$ sources. Orange and red lines represent the same Bayesian fitting as in Figure\,\ref{fig:Gu}, while black lines represent the fitting based on the likelihood without error weights.}

\label{fig:Bayes_comparison}
\end{figure*}

One of the tested likelihoods was similar to the main one used in the analysis, but without the error weights:
\begin{equation}\label{eq:like_no_error}
    P(D|\Theta,M) = -\sum_{n}\left[(y_n-mx_{n}-b)^2 \times \ln\left((y_n-mx_{n}-b)^2\right) \right]
\end{equation}
The fitting is shown in Figure\,\ref{fig:Bayes_comparison}. Black lines represent Bayesian fitting with the likelihood function \ref{eq:like_no_error}. The high spread is visible on both a) and b) panels. Although the fitted lines for Low-$z$ sources are similar to weighted likelihood, for the High-$z$ targets the difference is significant (slopes 0.12 and --0.50 for likelihood with and without error weighted accordingly). The cause for this discrepancy is most likely the errors of low luminous objects which is present in both Low-$z$ and High-$z$ samples, but affect mostly the High-$z$ fit.

\section{Luminosity Scalings}
\label{sec:Relations}

\subsection{The Entire WISE Sample}
\begin{figure}[!htp]

  \includegraphics[clip,width=1.0\columnwidth]{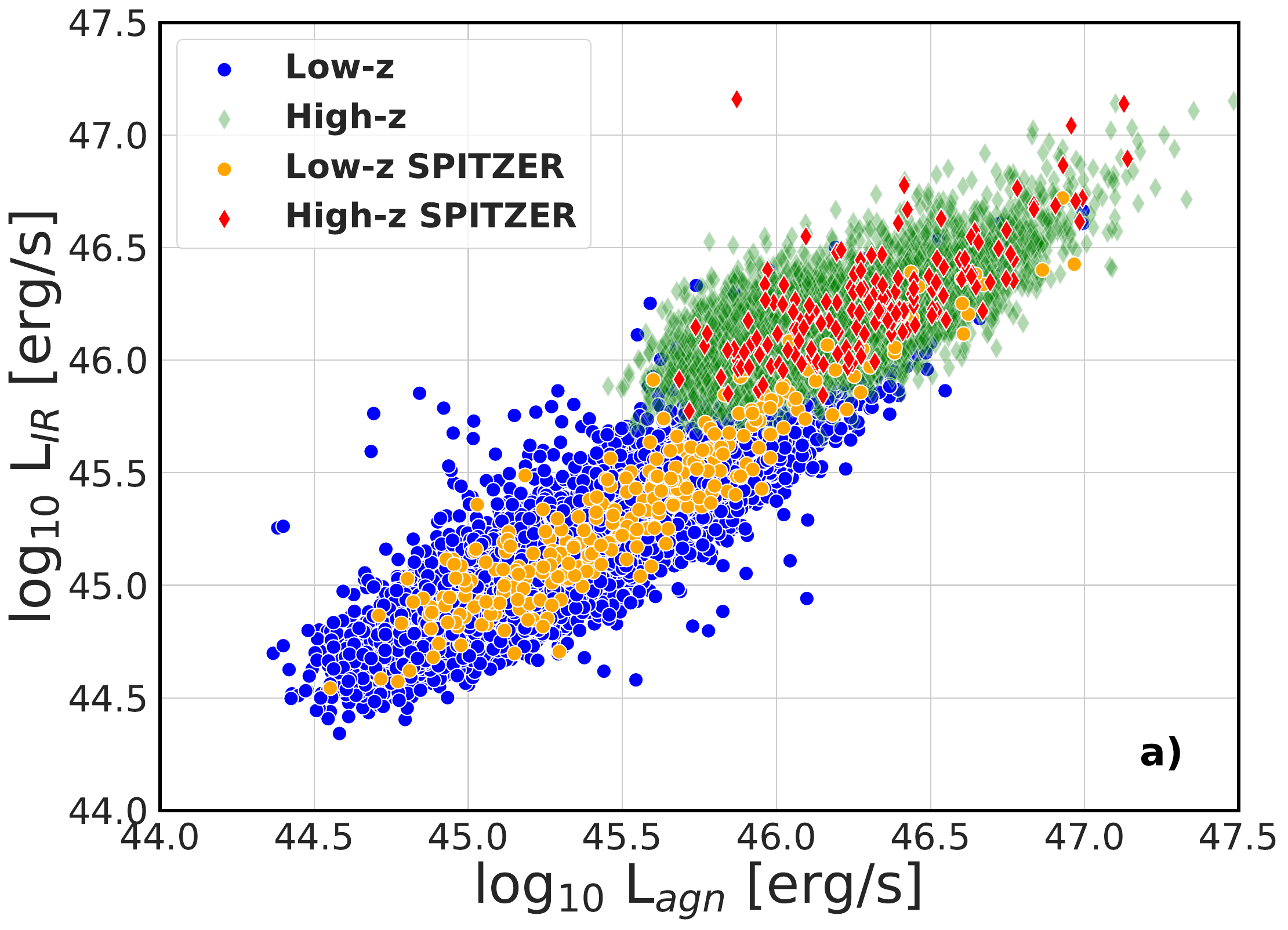}%

  \includegraphics[clip,width=1.0\columnwidth]{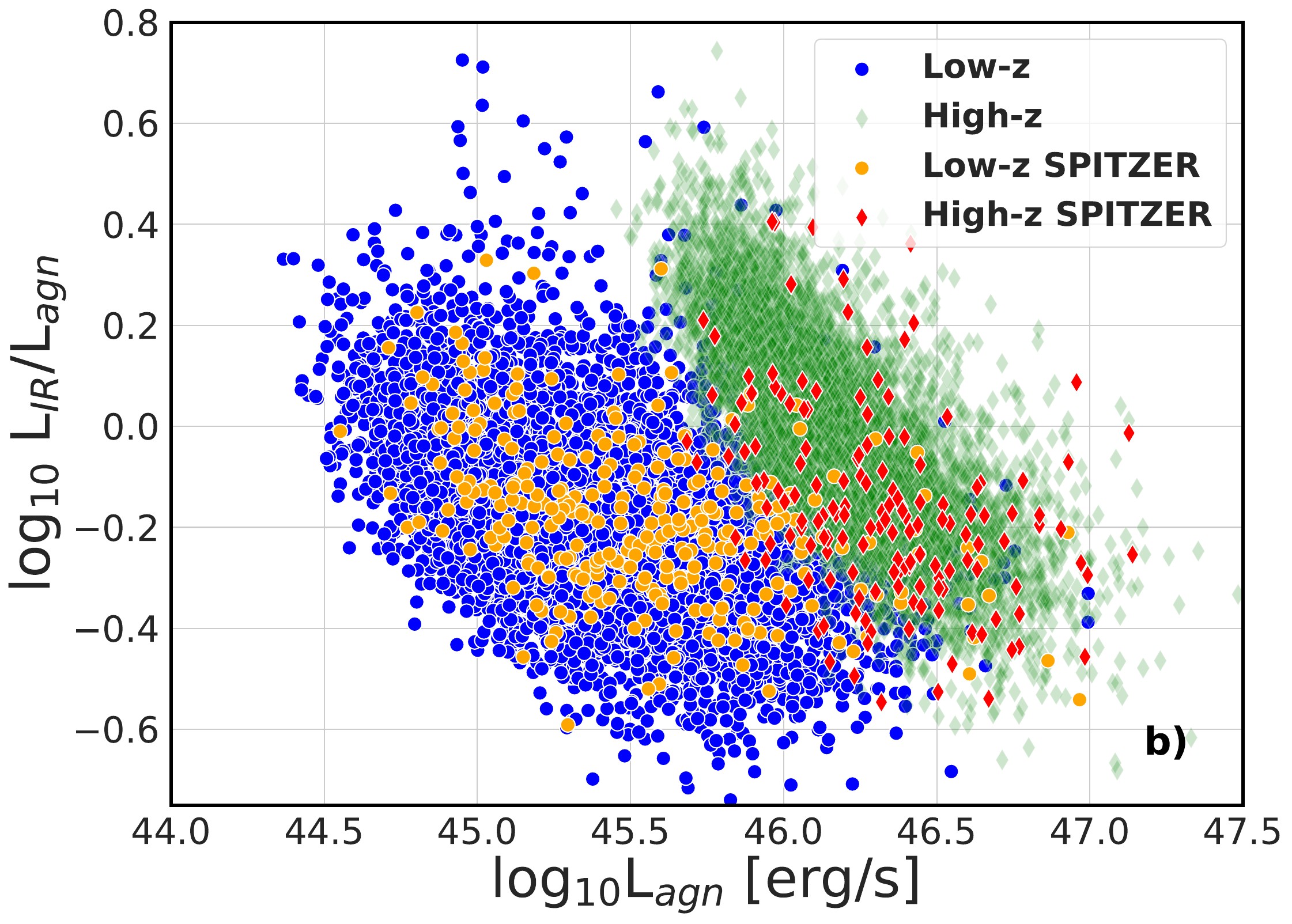}%

  \includegraphics[clip,width=1.0\columnwidth]{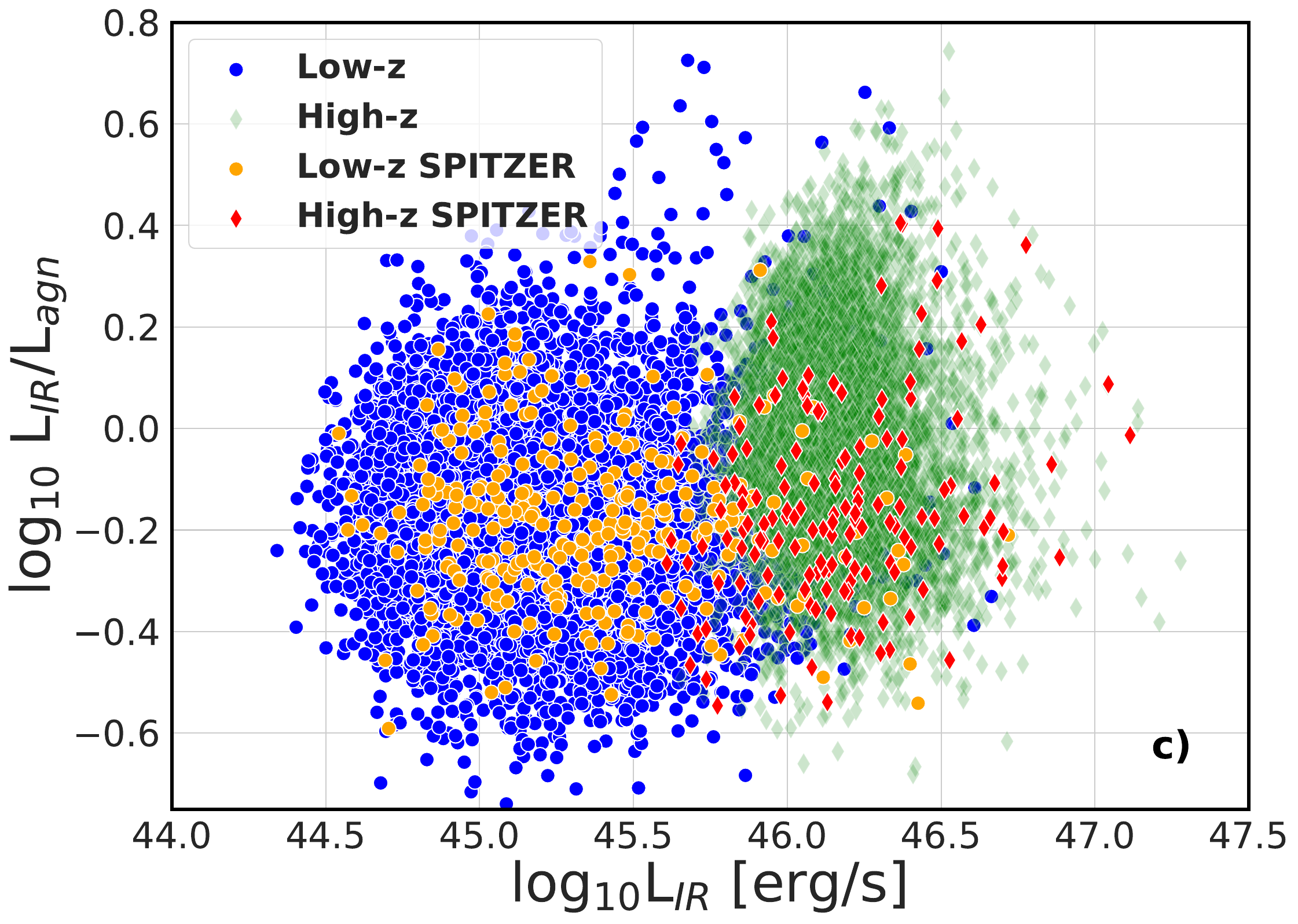}

\caption{Luminosity scalings within the complete WISE datasets for Low-$z$ and High-$z$ quasars}. Top panel shows $\log L_{\rm IR}$ vs $\log L_{\rm agn}$, middle panel illustrates $\log$\,CF vs $\log L_{\rm agn}$, and lower panel displays $\log$\,CF vs $\log L_{\rm IR}$. Blue circles indicate Low-$z$ quasars, whereas green and red diamonds indicate High-$z$ sources. Yellow circles and red diamonds indicate Low-$z$ and High-$z$ quasars with the SPITZER M24 data.
\label{fig:Lir_Lbol_lamaniec_Spitzer}

\end{figure}

Figure\,\ref{fig:Lir_Lbol_lamaniec_Spitzer} illustrates the relations between the CF, $L_{\rm IR}$, and $L_{\rm agn}$, for the complete WISE sample. Additional SPITZER data, which are also shown in the Figures, are discussed further in Appendix\,\ref{sec:results_spitzer_check}. Figure\,\ref{fig:Lir_Lbol_lamaniec_Spitzer} top panel shows, in particular, the relation between $\log L_{\rm IR}$ and $\log L_{\rm agn}$. The Pearson correlation coefficient values for this relationship are 0.9 for the Low-$z$ sample and 0.70 for the High-$z$ sample, as determined by the best-fitted least squares linear regression. Notably, both the Low-$z$ and High-$z$ quasars exhibit a similar trend.

Figure\,\ref{fig:Lir_Lbol_lamaniec_Spitzer} middle panel shows the relation between $\log {\rm CF}$ and $\log L_{\rm agn}$. Both the Low-$z$ and High-$z$ quasars exhibit similar scaling, but the Low-$z$ quasars have a greater spread in $L_{\rm agn}$. On the other hand, the High-$z$ quasars have higher values of $\log$\,CF, and the general scaling relation is more robust. An anticorrelation between $\log$\,CF and $\log L_{\rm agn}$ is present, albeit with a high variance. However, no such anticorrelation is evident in the lower panel of Figure\,\ref{fig:Lir_Lbol_lamaniec_Spitzer}, presenting the relation between $\log$\,CF and $\log L_{\rm IR}$.

\subsection{SNR $>3$}

\begin{figure}[!htp]

  \includegraphics[clip,width=1\columnwidth]{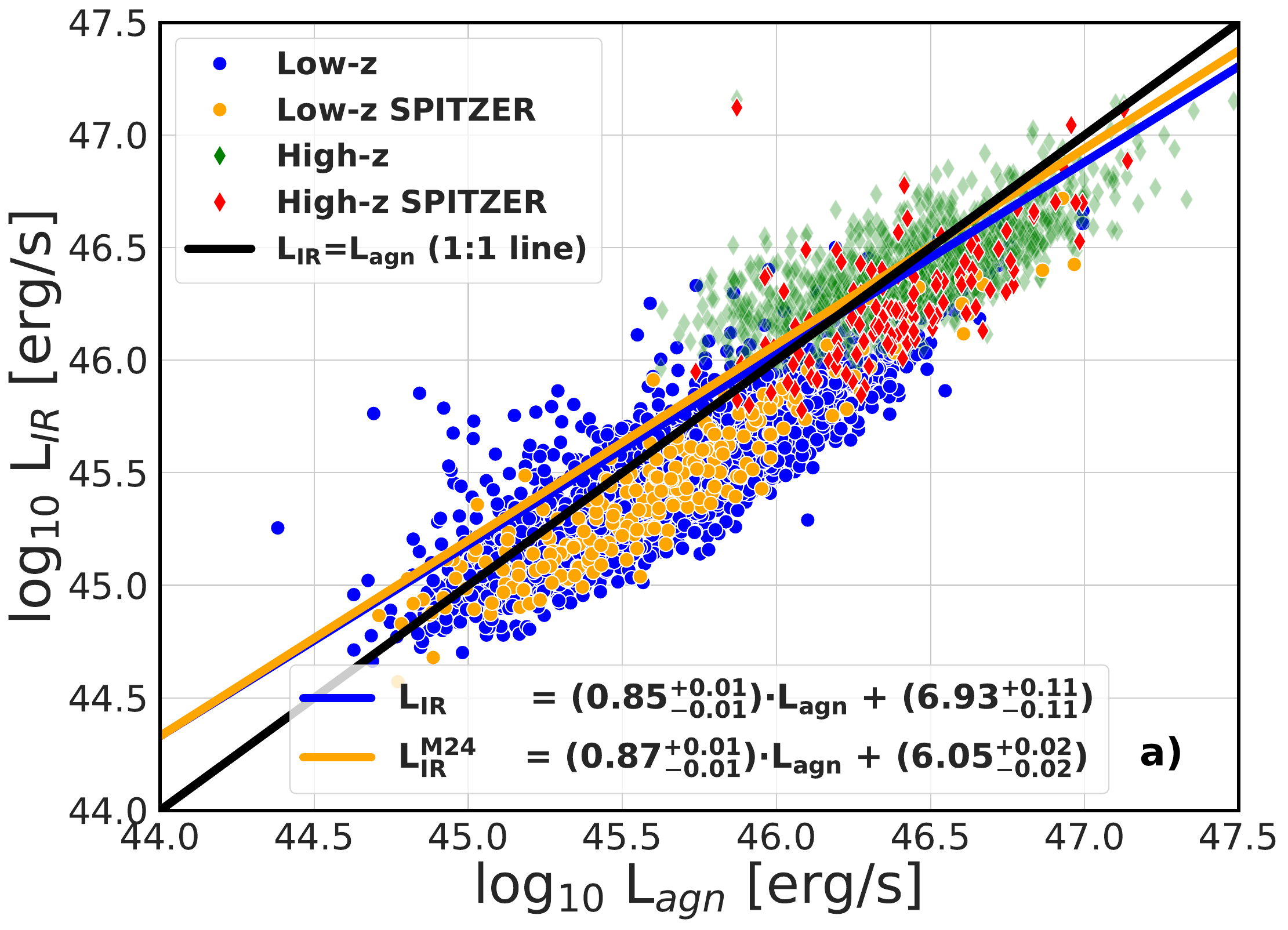}%

  \includegraphics[clip,width=1\columnwidth]{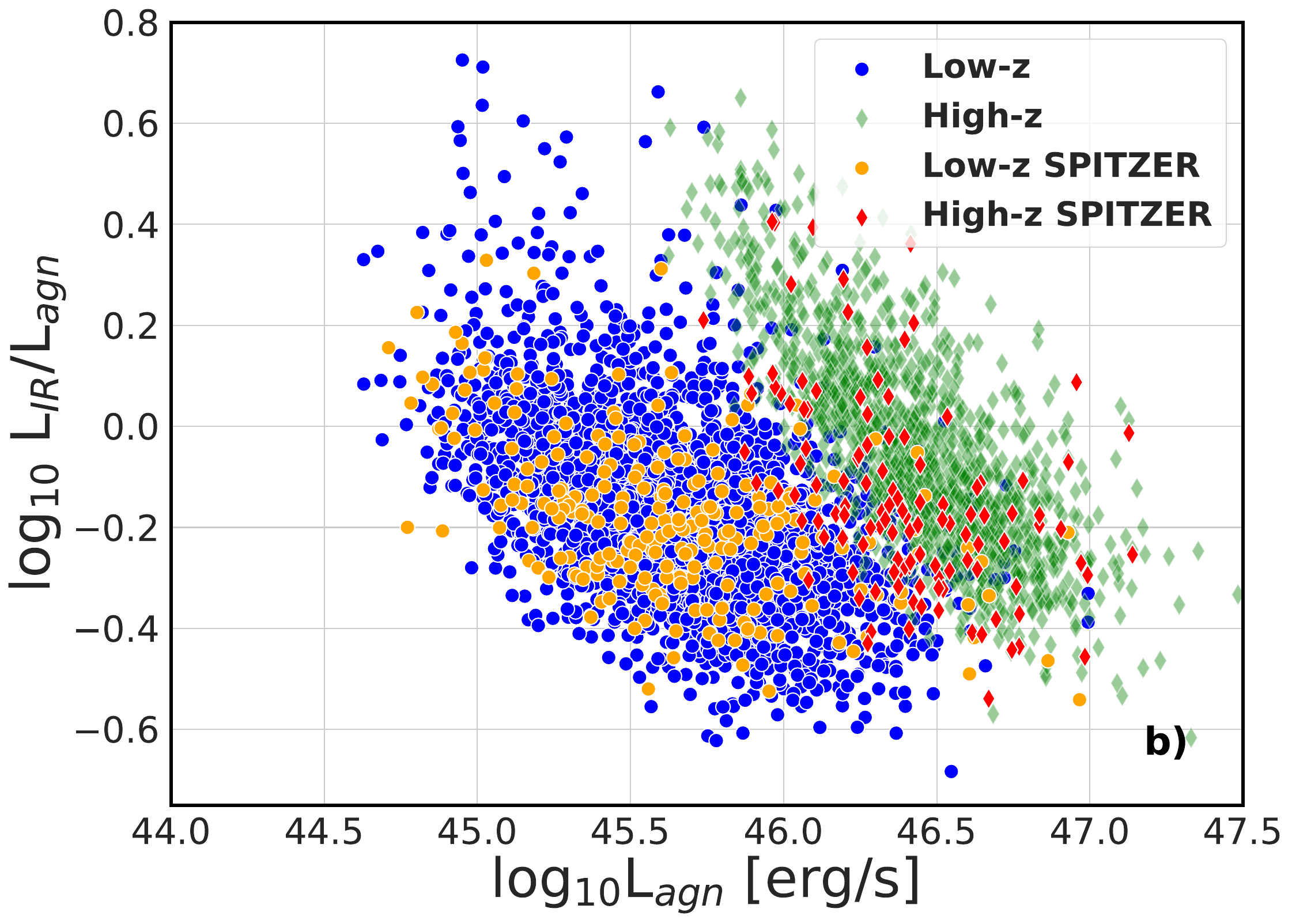}%

  \includegraphics[clip,width=1\columnwidth]{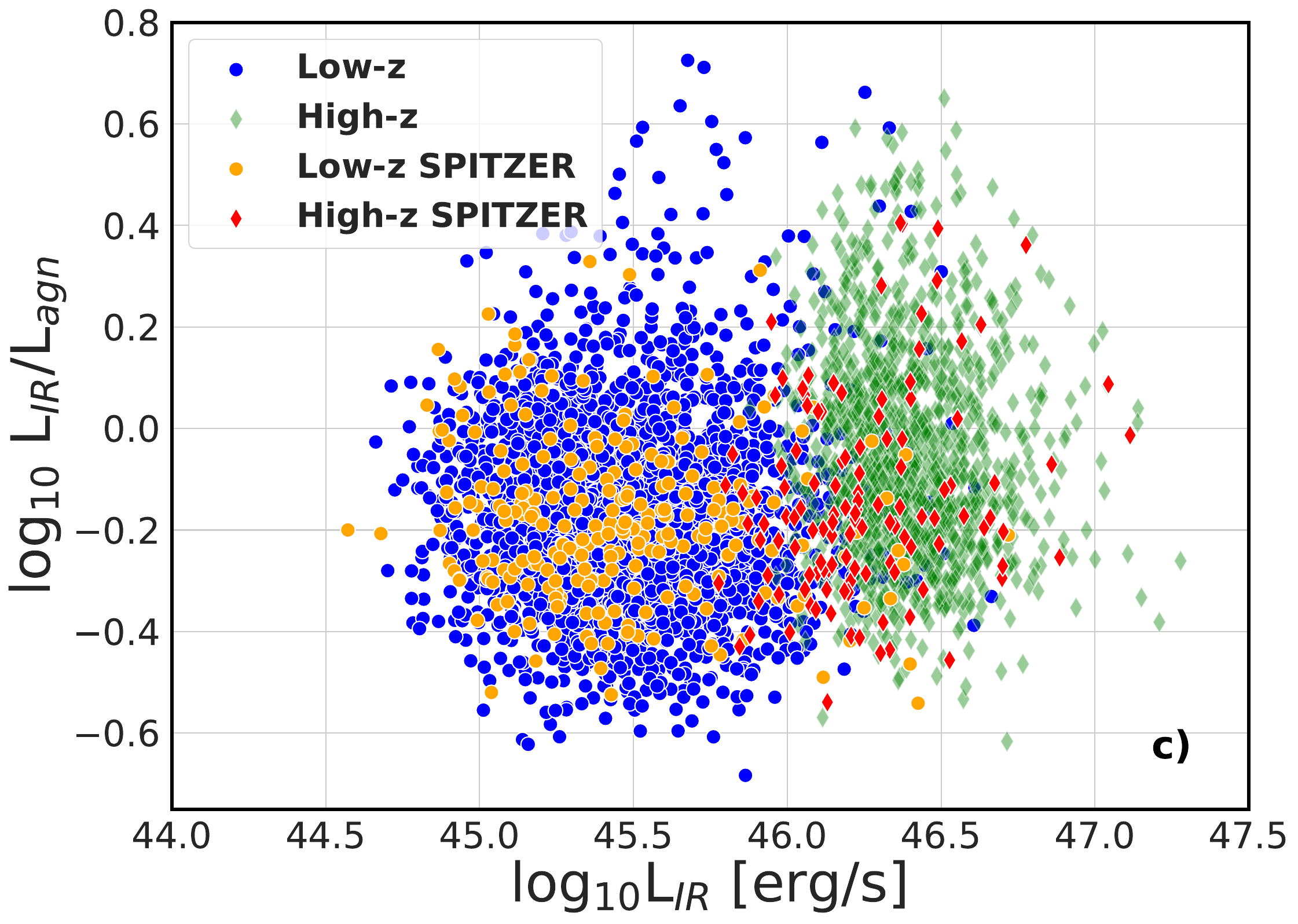}%

\caption{Luminosity scalings for the WISE samples of the Low-$z$ and High-$z$ quasars with SNR$_{\rm W3\&W4}>3$. Top panel shows $\log L_{\rm IR}$ vs $\log L_{\rm agn}$, middle panel $\log$\,CF vs $\log L_{\rm agn}$, and lower panel $\log$\,CF vs $\log L_{\rm IR}$. Blue circles indicate Low-$z$ quasars, whereas green and red diamonds indicate High-$z$ sources. Yellow circles and red diamonds indicate Low-$z$ and High-$z$ quasars with the SPITZER M24 data. On the top panel a), blue line denotes the Bayesian regression for the data with SNR$_{\rm W3\&W4}>3$, orange line is the Bayesian regression for the SPITZER data from both Low-$z$ and High-$z$ sub-samples, and the black line is the 1:1 relation between the $L_{\rm IR}$ and $L_{\rm agn}$.}
\label{fig:SNR3}
\end{figure}

The Low-$z$ sample contains 3,229 objects, while the High-$z$ sample includes 1,279 objects, all of which meet the SNR${\rm W3\&W4}>3$ criterion. Figure\,\ref{fig:SNR3} shows the relations between the $L_{\rm IR}$, $L_{\rm agn}$ and CF parameters in the direct analogy to those given in Figure\,\ref{fig:Lir_Lbol_lamaniec_Spitzer}, but with the SNR$_{\rm W3\&W4}>3$ cut.
One notable difference is the absence of low-luminosity objects in the top panel of the figure, when compared to Figure\,\ref{fig:Lir_Lbol_lamaniec_Spitzer} top panel.

\section{Comparison with Gu 2013}
\label{Gu:data}

The ``power-law method'' luminosity estimation method proposed by \cite{2013ApJ...773..176G} involved fitting two power-law functions to the IR and optical-UV segments of the SED for each source accordingly. Here we followed this method for a direct comparison with our all-points method, therefore interpolating luminosities at specific wavelengths of 0.01\,$\mu$m ($L_{0.11\,\mu \textrm{m}}$), 1\,$\mu$m ($L_{1\,\mu  \textrm{m}}$), and 7\,$\mu$m ($L_{7\,\mu \textrm{m}}$) from the same filters.
The calculation of $L_{\rm IR}$ involved integrating the area under the power-law fit between $L_{7\,\mu \textrm{m}}$ and $L_{1\,\mu \textrm{m}}$, while $L_{\rm agn}$ was calculated as the integral of the fitted power law between the luminosities $L_{1\,\mu \textrm{m}}$ and $L_{0.11\,\mu \textrm{m}}$. 

\subsection{Comparison of Luminosity Estimates}
\label{sec:integration_comparison}

In this section, we compare the results obtained using the all-points method for luminosity estimation with those obtained using the power-law method. As previously described, in the power-law method both $L_{\rm IR}$ and $L_{\rm agn}$ are calculated as the areas under the power-law fitted curves between the two points of the interpolated monochromatic luminosities. The power-law method (red and blue areas in the Figure) agrees relatively well with photometry for Low-$z$ quasars, but deviates from monochromatic luminosities for High-$z$ quasars. A notable limitation of the power-law method is the underestimation of $L_{\rm agn}$ and the overestimation of $L_{\rm IR}$ for High-$z$ quasars. Consequently, this method may lead to a significant overestimation of the CF.

\begin{figure}[!htp]

  \includegraphics[clip,width=1.\columnwidth]{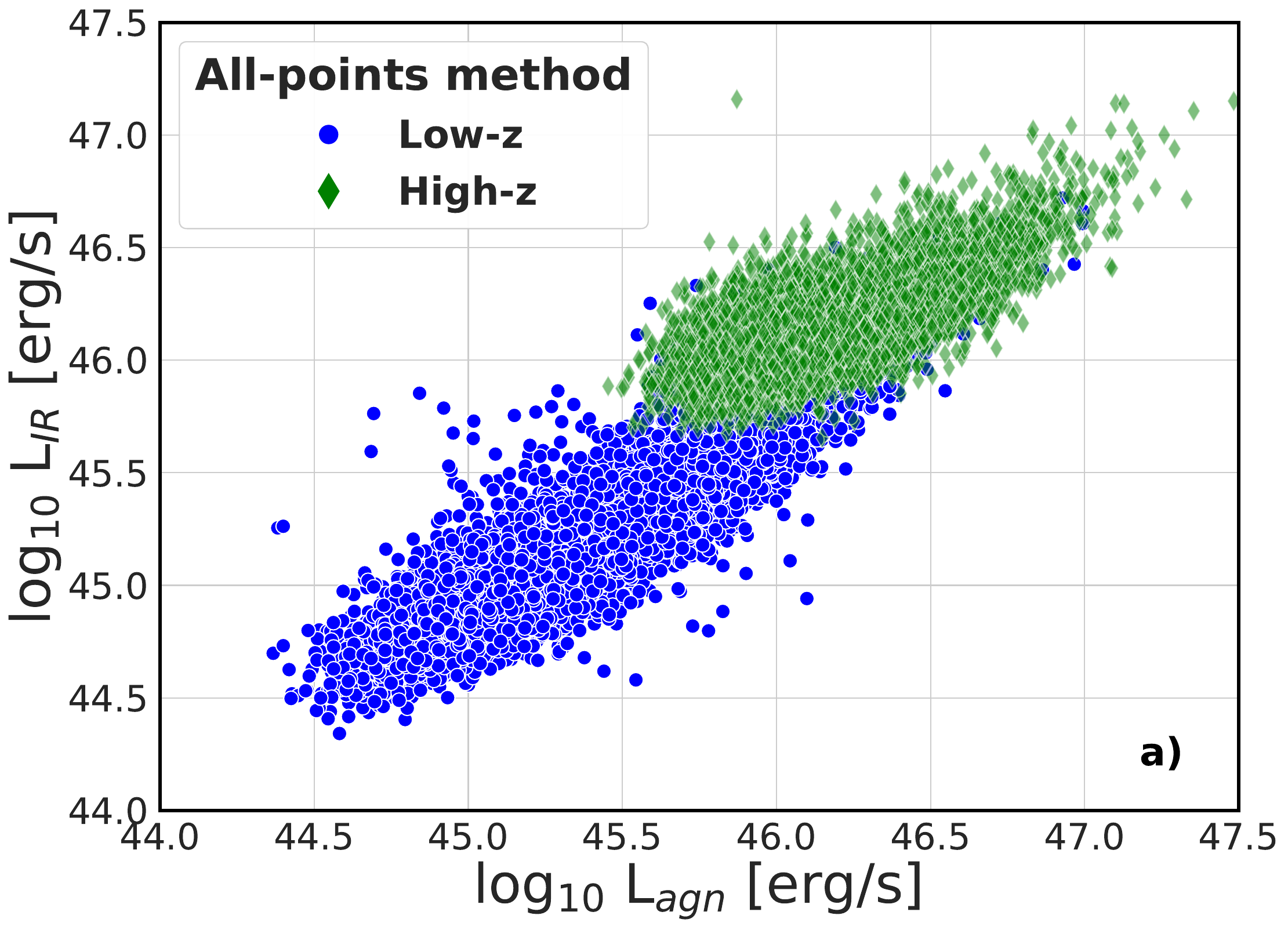}%

  \includegraphics[clip,width=1.\columnwidth]{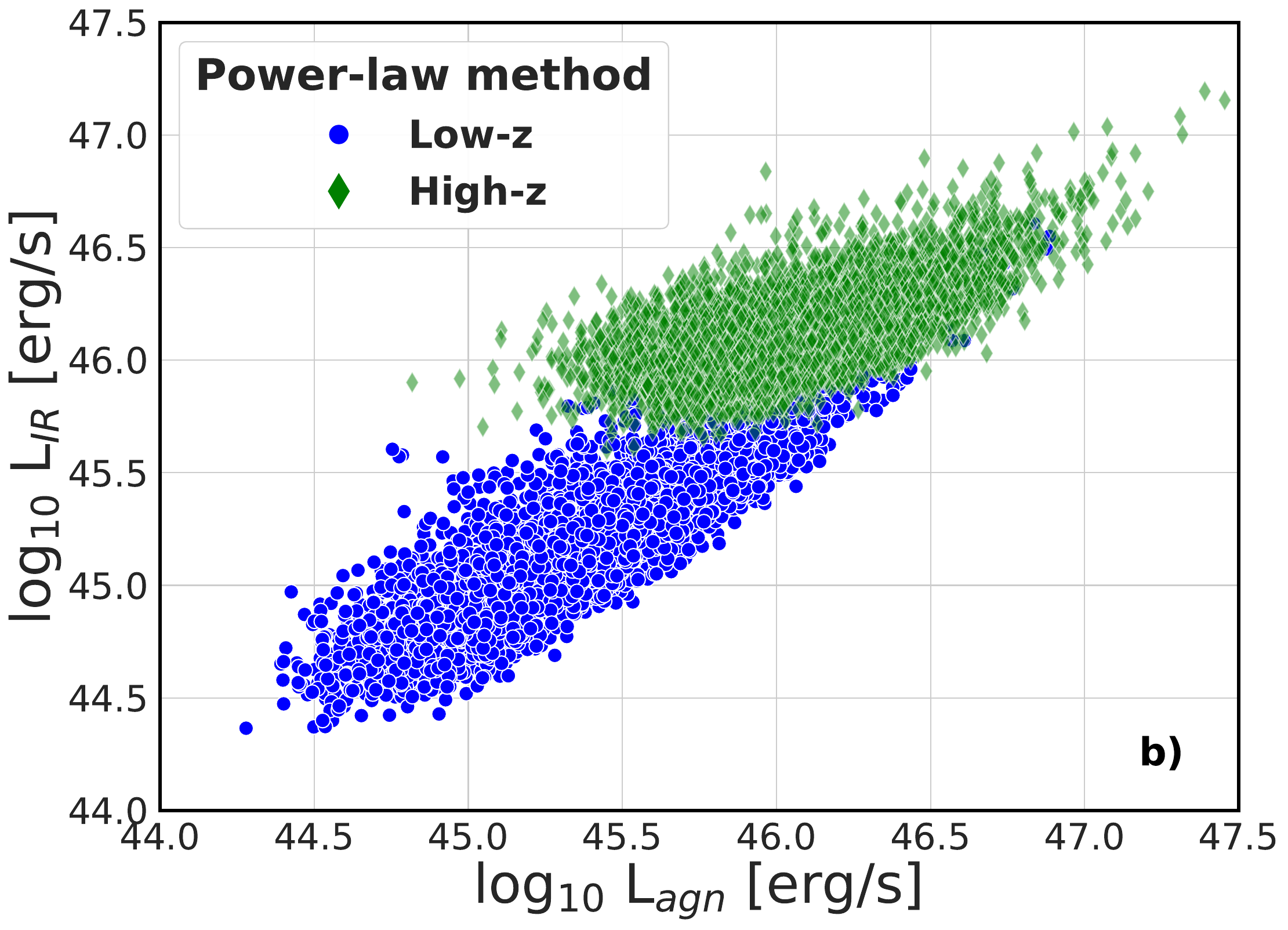}%

\caption{The relation between $\log L_{\rm agn}$ and $\log L_{\rm IR}$ calculated with the all-points method and the power-law method (upper and lower panels, respectively), for the Low-$z$ and High-$z$ quasars (blue circles and green diamonds, respectively).}
\label{fig:Lir_Lbol_both}
\end{figure}

Figure~\ref{fig:Lir_Lbol_both} presents the comparison of $L_{\rm IR}$ versus $L_{\rm agn}$ calculated with each method (upper and lower panels for the all-points and power-law methods, respectively). As shown, the luminosities of the High-$z$ quasars following from the power-law method have a greater dispersion compared to the all-points method in the $\log L_{\rm IR}$ versus $\log L_{\rm agn}$ representation. However, the number of outliers in the Low-$z$ sample increases with the all-points method. The High-$z$ quasars feature a major difference in the $\log L_{\rm IR}$ vs. $\log L_{\rm agn}$ relation, having a slight ``tail'' at lower values of $\log L_{\rm agn}$.

\begin{figure}[!htp]

  \includegraphics[clip,width=1.\columnwidth]{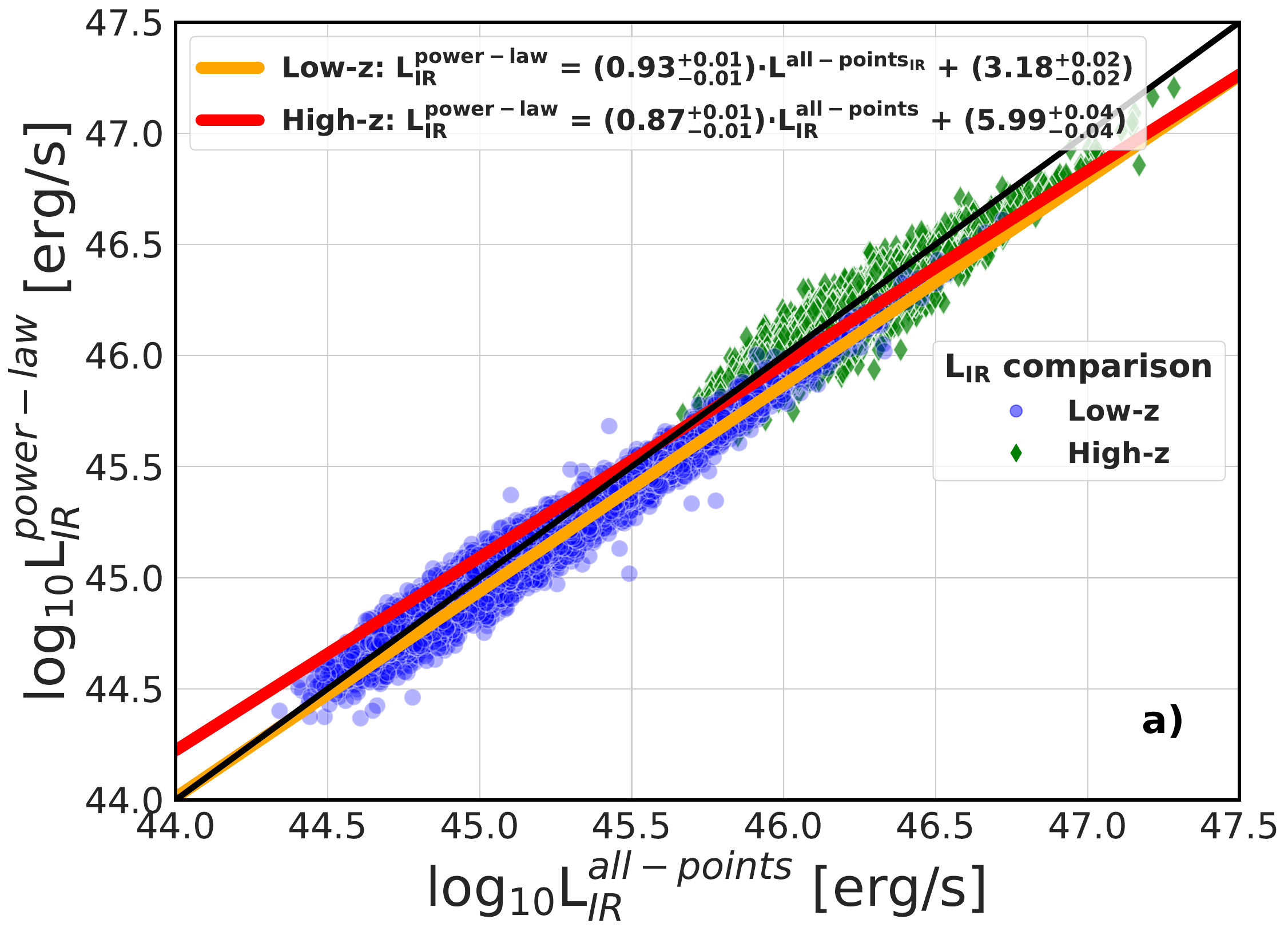}%

  \includegraphics[clip,width=1.\columnwidth]{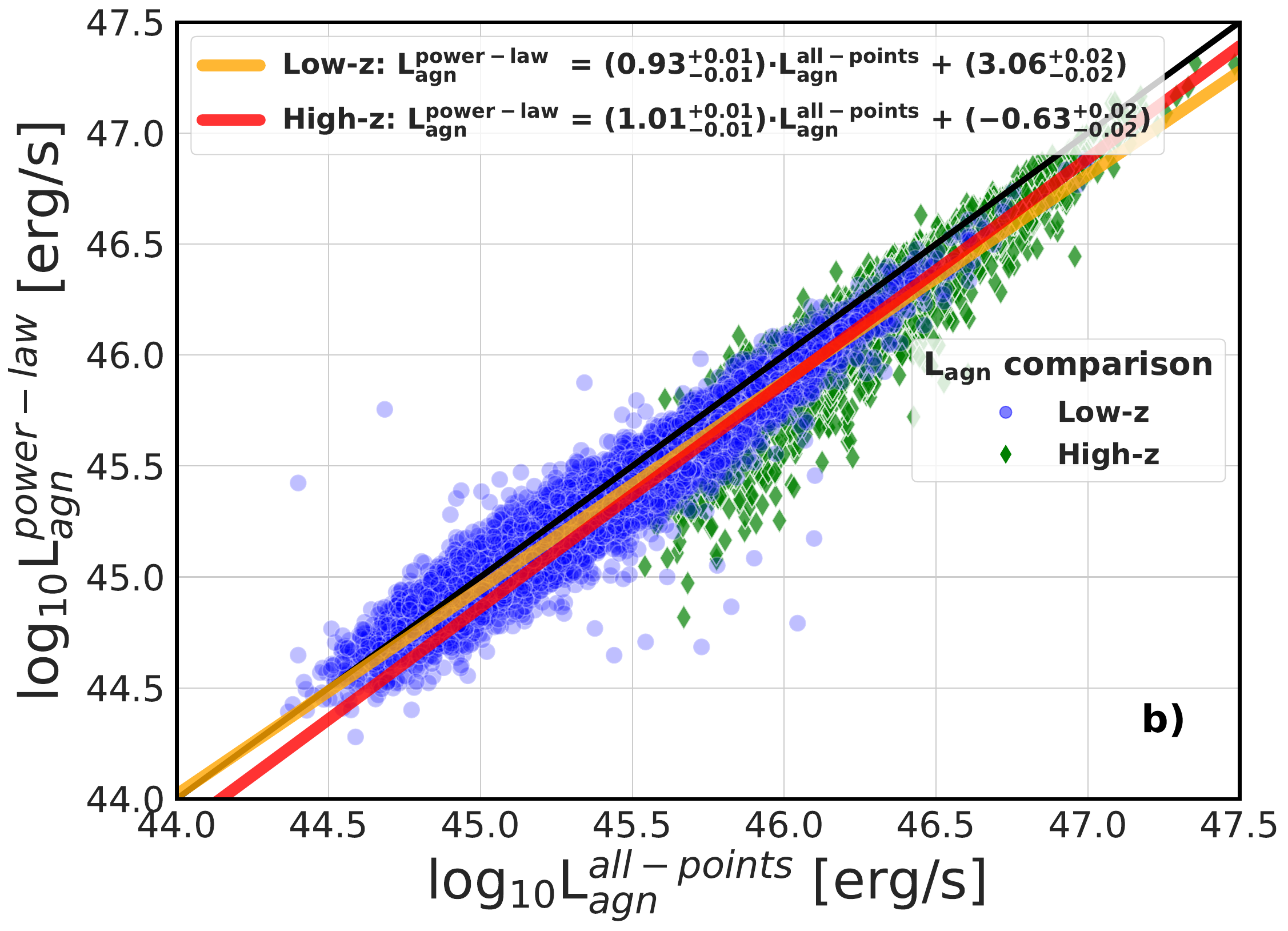}%

\caption{Comparison between luminosities based on the power-law (Y axis) and all-points (X axis) methods of integration. The upper panel a) shows the comparison between $L_{\rm IR}$, and the lower panel b) panel between $L_{\rm agn}$. The black line stands for the 1:1 relation. In both panels, blue circles and green diamonds denote the Low-$z$ and High-$z$ quasars, respectively. Red and orange lines represent the best Bayesian regression fit, with the best-fit relation given in each panel.}
\label{fig:LirvsLir_LbolvsLagn}
\end{figure}

Figure~\ref{fig:LirvsLir_LbolvsLagn} shows the direct comparison between the all-points and power-law methods for $L_{\rm IR}$ (upper panel) and $L_{\rm agn}$ (lower panel). For the majority of the Low-$z$ objects, the difference becomes significant above 45.5 for both $\log L_{\rm IR}$ and $\log L_{\rm agn}$, with the power-law method underestimating the values compared to the all-points method. For the High-$z$ sample, the power-law method tends to slightly overestimate IR luminosities up to $\log L_{\rm IR} =46.5$, above which the trend reverses. In comparison, the $L_{\rm agn}$ for High-$z$ quasars is overall significantly underestimated with the power-law method.

\subsection{Recreation of Gu 2013}
To recreate the results of the \cite{2013ApJ...773..176G} work, the same data were analyzed. While there may be minor discrepancies in the data due to potential updates in the database, we aimed for a direct comparison, and therefore we utilized the same SDSS data releases as in the \cite{2013ApJ...773..176G}, and the two exact redshift ranges were defined: the Low-$z$ range within $0.7 \leq z \leq 1.1$ and the High-$z$ range within $2.0 \leq z \leq 2.4$. Optical data for the Low-$z$ objects, SDSS DR7Q and DR9Q, were taken from \citet{SDSS7DR} and \citet{SDSS9DR}, and for the High-$z$ quasars only the SDSS DR9Q was used. The DR7Q contains 16,695 Low-$z$ quasars, while the DR9Q contains 10,553 Low-$z$ and 25,934 High-$z$ quasars. The SDSS data were then cross-matched within a 2\,arcsec radius with the WISE All-Sky Data Release. This returned 11,943 Low-$z$ quasars from DR7Q, 4,149 Low-$z$, and 16,228 High-$z$ quasars from the DR9Q. The following data cross-matching was done with the UKIDSS DR8 (also within 2\,arcsec radius), with the 4,855 objects in DR7Q Low-$z$, 2,517 in Low-$z$, and 2,630 in High-$z$ of DR9Q. Finally, only the Low-$z$ sample was cross-matched with the GALEX GR6/GR7 data release with a 3\,arcsec radius, resulting in 3,136 DR7Q and 1,134 DR9Q quasars. The final Low-$z$ sample contains 4,218 quasars (after removal of duplicates), while the final High-$z$ sample contains 2,630 quasars. In comparison, \cite{2013ApJ...773..176G} claimed 3,940 and 2,056 objects in the Low-$z$ and High-$z$ samples, respectively. 

Despite differences in sample size, our results concerning integrated luminosities and the CF are consistent within the error margins. The median integrated luminosities for the Low-$z$ sample are $\log L_{\rm IR} = 45.91 \pm 0.44 $, $\log L_{\rm agn} = 46.08 \pm 0.42$, while for the High-$z$ sample $\log L_{\rm IR} = 46.13 \pm 0.11$, and $\log L_{\rm agn} = 46.01 \pm 0.17$. The median values of the CF for the Low-$z$ and High-$z$ data samples are: $CF_{\textrm{low}-z}=0.70\pm0.11$ and $CF_{\textrm{high}-z}=1.23\pm 0.37$. These error estimates are the median absolute deviations.

The OLS regressions were fitted to the data, with the following results:
\begin{equation}
\log L_{\rm IR} = (0.96\pm0.01) \log L_{\rm agn} + (1.48\pm0.12)
\nonumber
\end{equation}
for the High-$z$ quasars, and
\begin{equation}
\log L_{\rm IR} = (0.42\pm0.01) \log L_{\rm agn} + (22.84\pm0.33)
\nonumber
\end{equation}
for the Low-$z$ quasars;
\begin{equation}
\log L_{\rm IR}/L_{\rm agn} = (-0.04\pm0.01) \log L_{\rm agn} + (1.48\pm0.12)
\nonumber
\end{equation}
for the High-$z$ sample, and
\begin{equation}
\log L_{\rm IR} / L_{\rm agn} = (-0.58\pm0.008) \log L_{\rm agn} + (22.84\pm0.33)
\nonumber
\end{equation}
for the Low-$z$ sample.

All in all, the exact recreation of the data sample from \cite{2013ApJ...773..176G} turned out quite problematic. We carefully checked that, for example, for one of the quasars for which the SED were shown (SDSS J001600.60-003859.2), the current version of the SDSS DR12Q catalog gives negative flux measurements, so it is unclear how the target was selected in \cite{2013ApJ...773..176G}.

\section{Other physical properties cuts}
\label{sec:other_cuts}

\begin{figure*}[!htp]

  \includegraphics[clip,width=1\columnwidth]{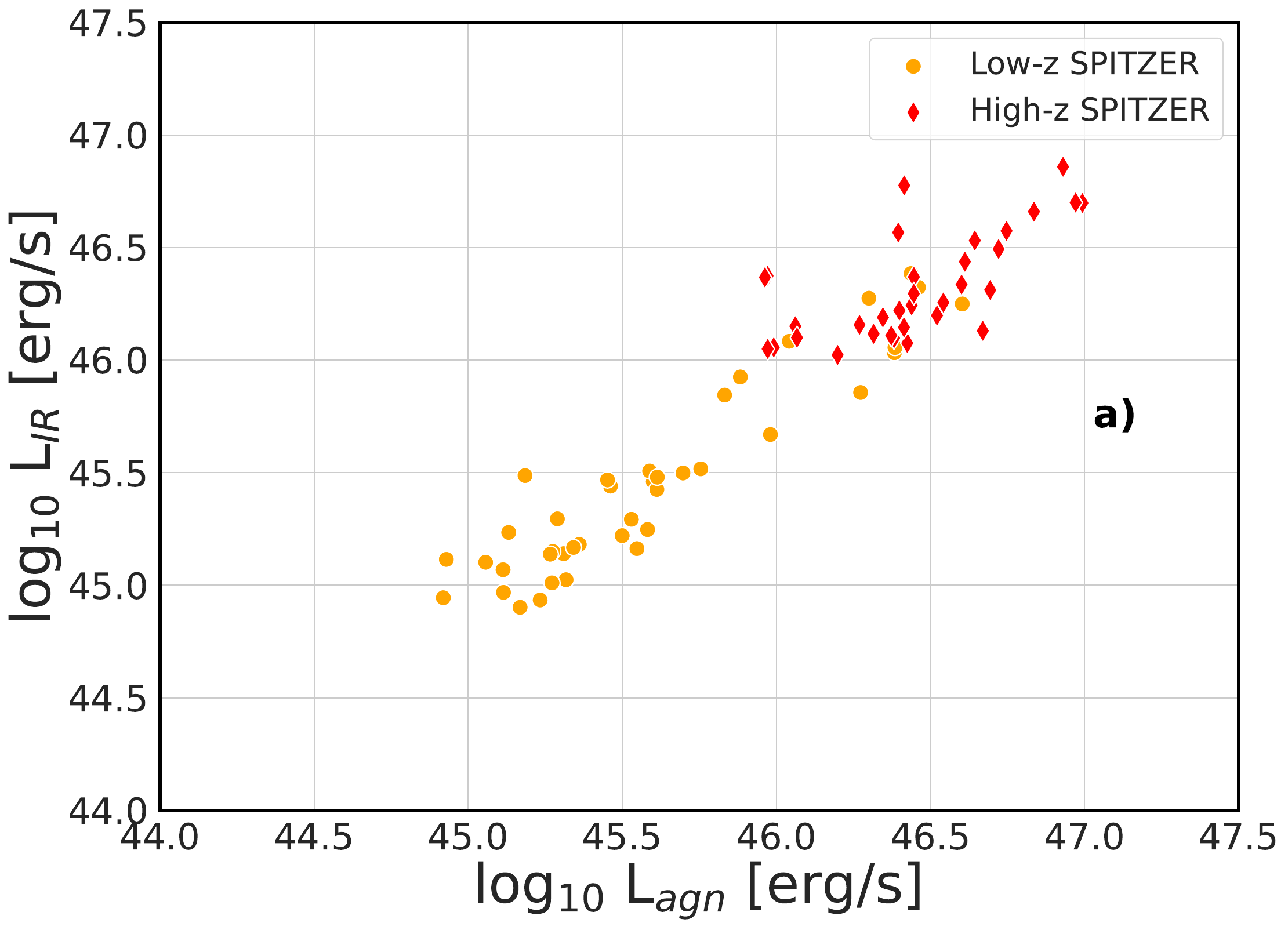}
  \includegraphics[clip,width=1\columnwidth]{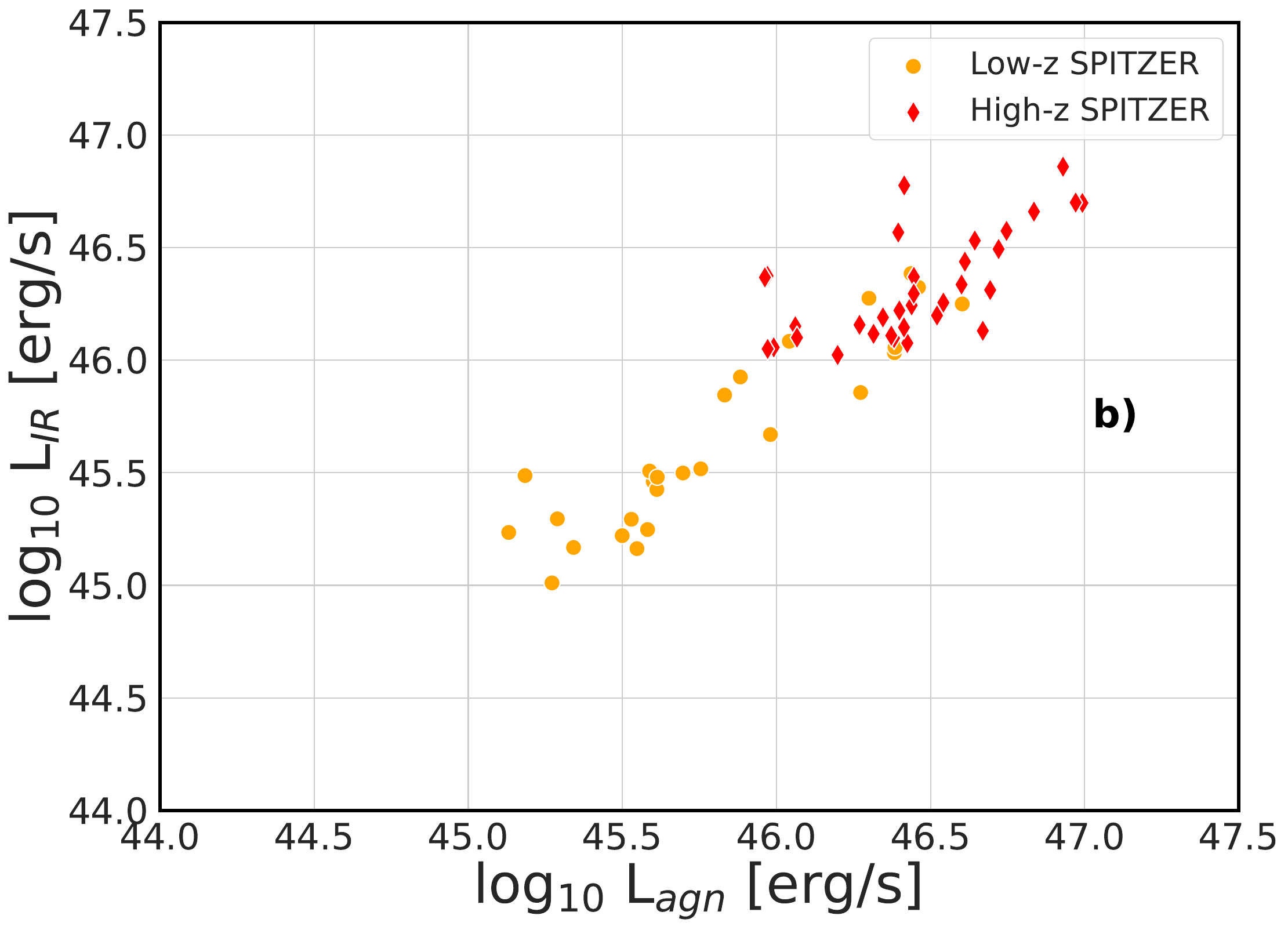}%

\caption{$L_{\rm IR}$ vs $L_{\rm agn}$ relation for SPITZER data. Left panel shows the objects without the $\log M_{\rm BH} = 8.5$ cut, right panel shows the objects above the $\log M_{\rm BH} = 8.5$ threshold.}

\label{fig:test_MBH}
\end{figure*}

In the main article, we primarily focused on selecting objects with similar SMBH masses, as shown in Figure\,\ref{fig:MBH_cut}. However, this approach has a limitation because objects with similar $M_{BH}$ can exhibit very different luminosities, depending on the accretion rate. To double-check different possible cuts we also performed: a similar luminosity selection and a similar Eddington ratio cut. We also checked how the  $M_{SMBH}$ cut selected the Low-$z$ SPITZER sources in $L_{\rm agn}$ and $L_{\rm IR}$. In Figure\,\ref{fig:test_MBH} we present objects before and after the cut at $\log M_{BH}/M_{\odot}>10.5$. As shown, the $M_{BH}$ cut excluded approximately half of the Low-$z$ SPITZER objects with luminosities below $\log L_{\rm IR\, \& \, agn} < 46.0$.

\begin{figure*}[!htp]

  \includegraphics[clip,width=1\columnwidth]{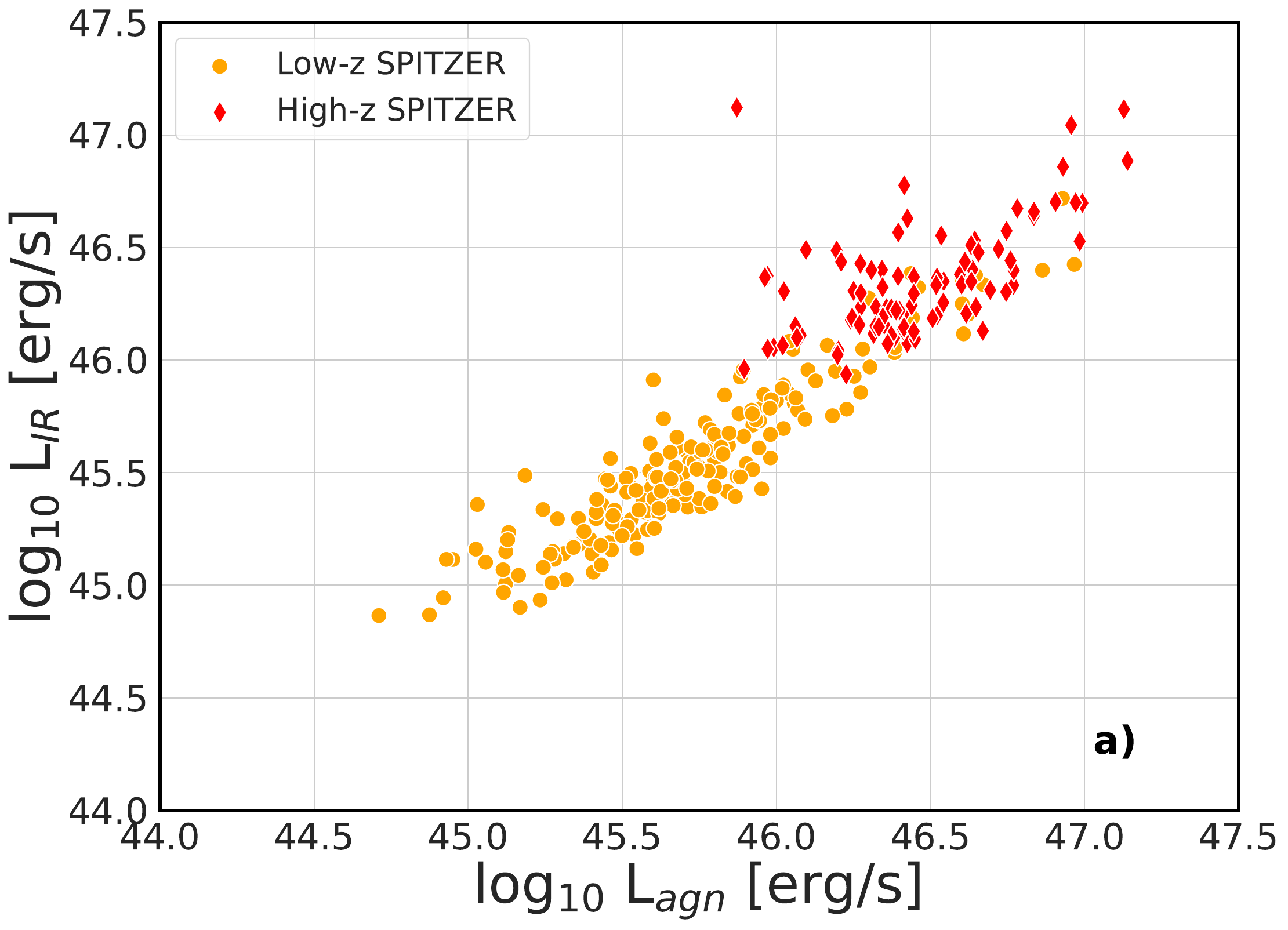}

  \includegraphics[clip,width=1\columnwidth]{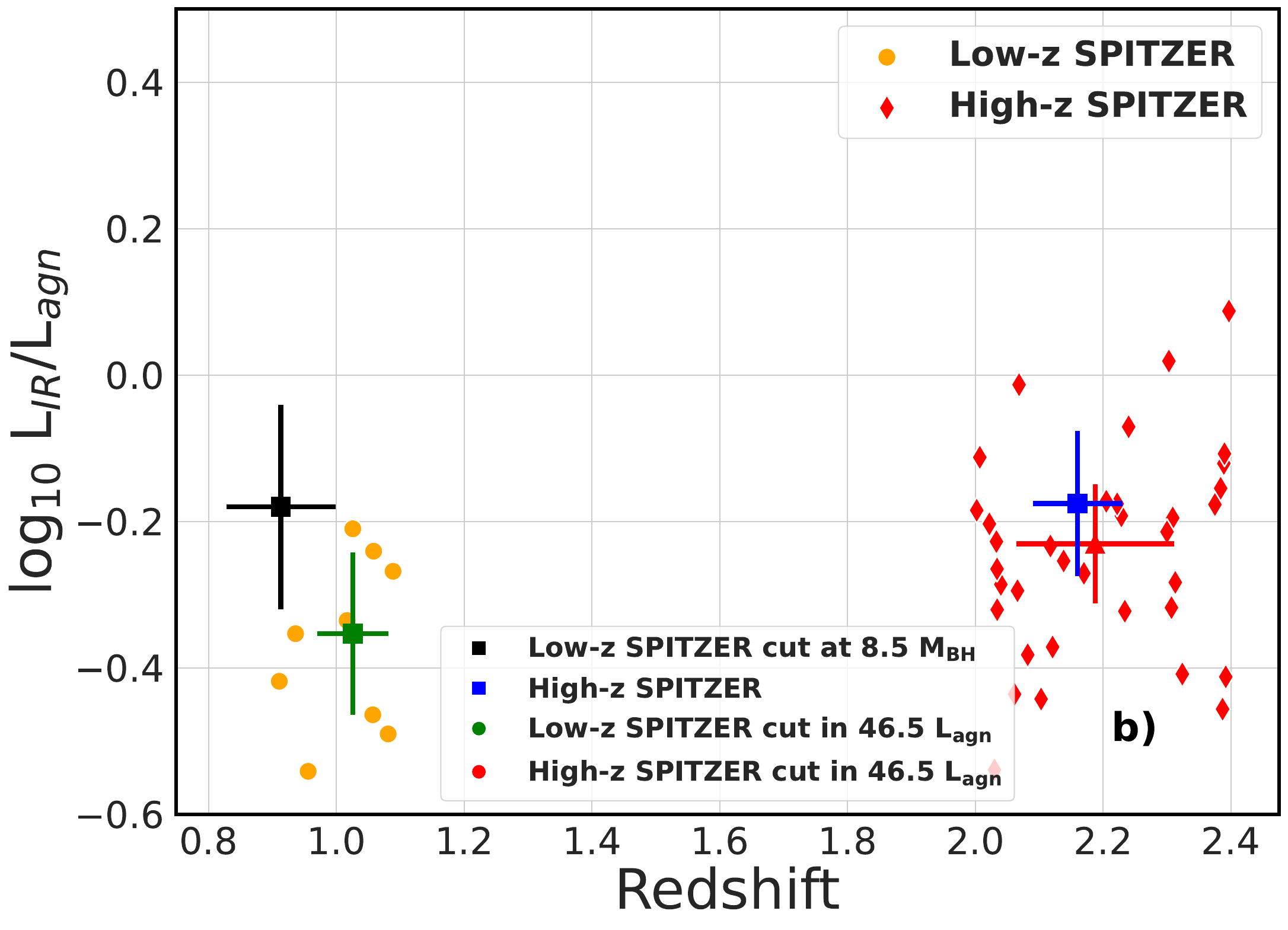}%

\caption{SPITZER data after the SNR$_{\rm W3}>5$, but without the cross-match with VAC. We performed the cut in $\log L_{\rm agn}>46.5$ for both Low-$z$ and High-$z$. The right panel presents the CF distribution for the truncated sample. Black and blue points represents medians for the $M_{BH}$ cut with MAD errors. The green and red points represents the medians and MAD errors for newly cut data.}

\label{fig:test_Luminosity}
\end{figure*}

The $L_{\rm agn}$ luminosity cut was also performed with the selection of both Low-$z$ SPITZER and High-$z$ SPITZER at $\log L_{\rm agn}>46.5$. The lack of large consistent coverage of $L_{\rm IR} - L_{\rm agn}$ space of both datasets is problematic for this method, which is shown in Figure\,\ref{fig:test_Luminosity} a) (left panel). Consequently, the samples after the cut were unevenly sized, and the CF values differed, though they were comparable within the margins of error (Figure\,\ref{fig:test_Luminosity} b)). In the Low-$z$ sample the most luminous and apparent the least obscured objects are present after the cut. For comparison, we also included the $M_{BH}$ cut, which showed consistent results with the luminosity cut within the errors.

\begin{figure*}[!htp]

  \includegraphics[clip,width=1\columnwidth]{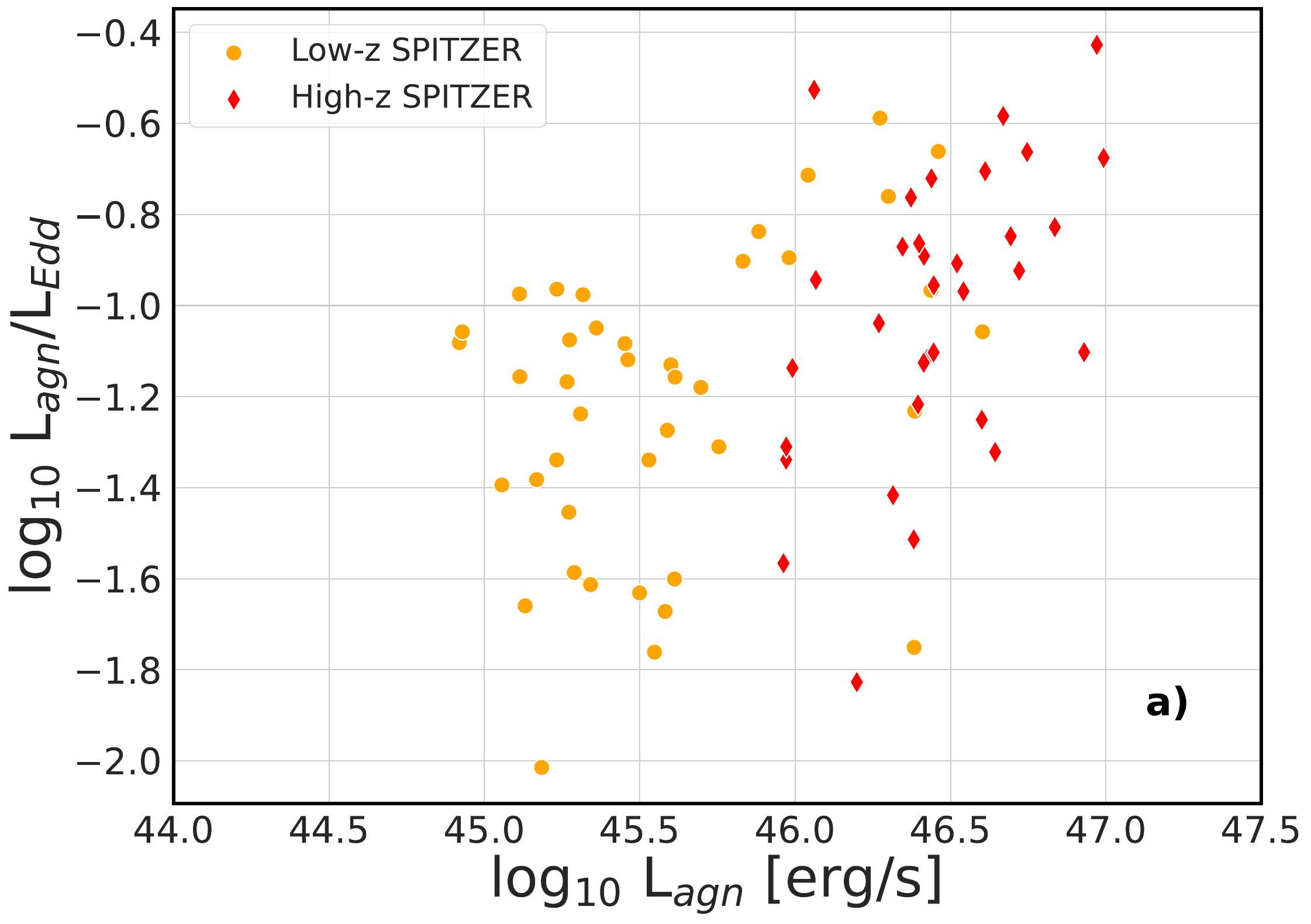}

  \includegraphics[clip,width=1\columnwidth]{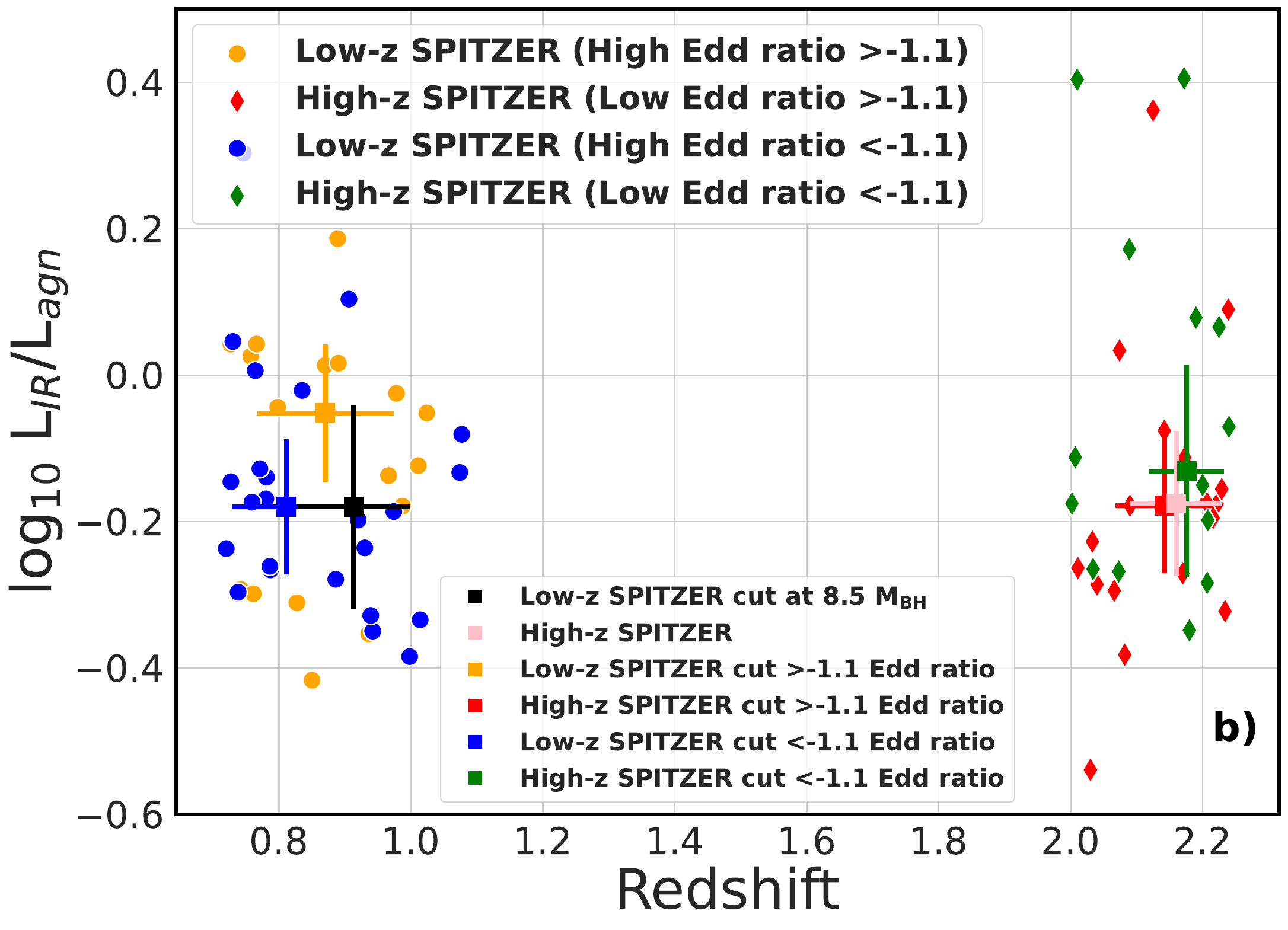}%

\caption{SPITZER data after the SNR$_{\rm W3}>5$, but without the cross-match with VAC. We performed the cut in $\log L_{\rm IR}/L_{\rm agn}$ = -1.1 and created two subsamples for both Low-$z$ and High-$z$. The right panel presents the CF distribution for the truncated sample. Black and pink points represents medians for the $M_{BH}$ cut with MAD errors. The orange and red points represents the medians and MAD errors for low and high sample with the Eddington ratio exceeding --1.1, while blue and green points represents the samples with the Eddington ratio --1.1. The samples are comparable within the error.}

\label{fig:test_EddRatio}
\end{figure*}

Finally, the Eddington ratio cut was performed. The resulting scalling between Eddington ratio and $L_{\rm agn}$ is shown in Figure\,\ref{fig:test_EddRatio}a) (left panel). As one can see the spread in the Eddington ratio is quite large, which can influence this method. We decided to apply a cut at an Eddington ratio value of -1.1. The \ref{fig:test_EddRatio} b) shows data sets with above -1.1 value for both Low-$z$ SPITZER (blue and orange squares) and High-$z$ SPITZER (green and red squares). The distribution of CF is comparable within the error, with the Eddington ratio cut.

\end{appendix}

\end{document}